\documentclass[epj]{svjour}
\usepackage{psfrag}
\usepackage{latexsym}
\usepackage{graphicx}
\usepackage{hyperref}
\newcommand{\intsum}{\makebox[0cm][l]{$\, \sum$} \int} 
\begin{document}
\title{Towards a high precision calculation for the pion-nucleus scattering lengths}

\author{S. Liebig\inst{1} \and V. Baru\inst{1,2} \and  F. Ballout\inst{1}
          \and  C. Hanhart\inst{1} \and A. Nogga\inst{1} 
           }   % end of author               
        
\institute{(1) Institut f\"ur Kernphysik (Theorie), 
Institute for Advanced Simulation and J\"ulich Centre for Hadron Physics, \\
Forschungszentrum J\"ulich, 52425 J\"ulich, Germany\\    
             (2) Institute for Theoretical and Experimental Physics,
 117218, B. Cheremushkinskaya 25, Moscow, Russia
} % end of institute

\date{Received: date / Revised version: 18.03.2010}

\abstract{ We calculate the leading isospin conserving few-nucleon
  contributions to pion scattering on $^2$H, $^3$He, and $^4$He. 
We demonstrate that the strong contributions to the pion-nucleus scattering
lengths can be controlled theoretically to an accuracy of a few percent for
isoscalar nuclei and of 10\% for isovector nuclei. 
In particular, we find the $\pi$-$^3$He scattering length to be $(62 \pm 4\pm 7)\times 10^{-3} m_{\pi}^{-1}$ 
where the uncertainties are due to ambiguities in the $\pi$-N scattering 
lengths and few-nucleon effects, respectively. 
To establish this accuracy we need to identify a suitable power counting for 
pion-nucleus  scattering. For this purpose we study the dependence of the two-nucleon
  contributions to the scattering length on the binding energy of $^2$H. Furthermore,
 we investigate   the relative size of the leading two-, three-, and four-nucleon
  contributions. For the numerical evaluation of the pertinent integrals, a
  Monte Carlo method suitable for momentum space is devised.  Our results
show that in general the power counting suggested by Weinberg is capable to properly
predict the relative importance of $N$-nucleon operators, however, it fails
to capture the relative strength of $N$- and $\left(N+1\right)$-nucleon operators,
where we find a suppression by a factor of 5 compared to the predicted factor
of 50. The relevance for the extraction of the isoscalar $\pi$-N
  scattering length from pionic $^2$H and $^4$He is discussed.  As a side
  result, we show that %physical deuteron binding energy is already beyond
the calculation of the $\pi$-$^2$H scattering length is already beyond
  the range of applicability of heavy pion effective field theory.
\PACS{
      {21.45.-v}{Few-body systems}   \and
      {21.85.+d}{Mesic nuclei} \and 
      {02.70.Tt}{Justifications or modifications of Monte Carlo methods}
     } % end of PACS codes
} %end of abstract

\maketitle
\section{Introduction}
\label{intro}

The probably most fundamental quantities characterizing pion scattering off
nucleons are the pion-nucleon ($\pi$-N) scattering lengths.  Obviously being
low energy observables, one should be able to understand their size based on
the effective field theory (EFT) of Quantum Chromodynamics (QCD), namely
Chiral Perturbation Theory (ChPT). In fact, the impact of symmetries on these
quantities is known for a long time \cite{Weinberg:1966kf,Tomozawa:1966jm}
(for a recent review see Ref.~\cite{Bernard:2007zu}),
but for a quantitative analysis of high accuracy it is of utmost importance to
understand now in detail the higher order corrections, either isospin
symmetric ones or isospin symmetry violating (IV) ones, based on ChPT, for they
point directly at QCD dynamics.

Despite an on-going effort to learn more on the scattering lengths
experimentally, they are still not very accurately known \cite{Gasser:2007zt}.
This is especially true for the isoscalar scattering length which is so
small, that a high control over various other, normally neglected effects,
like IV, is necessary for a reliable extraction of $a^{(+)}$ from
experiment.
 On the theoretical side
important progress for IV $\pi$-N scattering is reported 
in Refs.~\cite{Hoferichter:2009gn,Hoferichter:2009ez,Gasser:2002am}.
A first extraction of the leading charge symmetry breaking (a special case
of IV interactions) 
$\pi$-N scattering
amplitude is reported in Refs.~\cite{vanKolck:2000ip,Bolton:2009rq,Filin:2009yh}
based on an analysis of the recent observation of
the IV forward--backward asymmetry
in $pn\to d\pi^0$~\cite{Opper:2003sb}.

The most promising way to get experimental access to
the $\pi$-N scattering lengths  is the 
investigation of pionic atoms, the simplest system being 
pionic hydrogen. Many efforts have been undertaken to measure 
energy shift and width  of this system, allowing for an extraction of 
the isoscalar and isovector $\pi$-N scattering lengths $a^{(+)}$ and 
$a^{(-)}$ assuming that isospin symmetry is exact
up to standard Coulomb interactions
\cite{Schroder:1999uq,Schroder:2001rc}. 

In principle, it appears to be possible to extract both the
isoscalar and the isovector scattering length from pionic
hydrogen data, for the measurement provides two independent
observables, namely the level shift compared to the QED value
induced by the strong interaction as well as the width,
where the transition $\pi^- p \to \pi^0 n$ gives a significant,
known contribution. However,
in order to improve the extraction accuracy and to better
control systematics, an additional source of $a^{(+)}$ is
desireable. In this context,  it is important
to study more complex systems, like pionic deuterium, $^3$He, or
$^4$He, for the wave functions for those nuclei can be
calculated with high accuracy.
It is easy to see that the pion-nucleus scattering length $a_{\pi -A}$ 
can be written as
\begin{eqnarray}
 a_{\pi -A}& = & \left(\frac{1+\frac{m_\pi}{m_N}}{1+\frac{m_\pi}{A\,m_N}}\right)\left(Aa^{\left(+\right)}-Q_\pi\,
   a^{\left(-\right)} 2T_3\right) \cr 
  && \hspace{3cm}  +\mbox{few-nucleon corrections}\cr
  && \hspace{3cm}  + \mbox{IV corrections}
 \label{eq:piA-scattering}
 \end{eqnarray}
Here, $A$ denotes the mass number of the nucleus, $T_{3}$ the third component
of its isospin, $m_{\pi}$ and $Q_{\pi}$ the mass and charge of the pion and
$m_{N}$ the nucleon mass.  We will not further discuss the IV corrections here,
instead we refer to Refs.~\cite{Meissner:2005ne,Gasser:2007zt} for a review on
the current status of IV in pion-nucleus scattering.  Especially, the isospin
$T=0$ nuclei $^2$H and $^4$He yield important new information since
contributions from $a^{(-)}$ are suppressed and the contribution of $a^{(+)}$
is more prominent.  From Eq.~(\ref{eq:piA-scattering}) it becomes clear that
an understanding of these more complex systems requires to understand with a
high accuracy the few-nucleon contributions to the scattering length.
For $\pi$-$^2$H scattering they have been studied for decades within potential
multiple scattering theory, see
e.g. Refs.\cite{Baru:1997xf,Ericson:2000md,Doring:2004kt} and references
therein. A first attempt to calculate few-body effects systematically and
model-independently, i.e. within ChPT, was done by Weinberg
\cite{Weinberg:1992yk}.  He proposed to organize the transition operator as an
expansion in $m_{\pi}/\Lambda$ with $\Lambda$ being the chiral symmetry
breaking scale of the order of 1 GeV. This transition operator was then convoluted
in Ref.~\cite{Weinberg:1992yk} with phenomenological models for the
nucleon-nucleon (NN) interaction. This work has been refined in
Ref.~\cite{Beane:1997yg}.  The explicit calculations showed that part of the
few-nucleon corrections seem to be suppressed. Maybe even more troublesome for
the extraction of $\pi$-N scattering lengths is the observation that the
few-nucleon corrections were somewhat larger than naive dimensional analysis
predicted. Such issues were addressed in \cite{Beane:2002wk}. Here a refined
power counting, the so-called $Q$--counting, was proposed that takes the small
binding energy of $^2$H explicitly into account. Assuming that typical momenta
of the nucleons are given by the binding momentum $\gamma=\sqrt{E_dm_N}$, with
$E_d$ denoting the deuteron binding energy, it was found that the estimated
magnitudes of various few-body corrections changed drastically and thus the
series needed to be reorganized compared to the original power counting used
in \cite{Weinberg:1992yk}.  Obviously, it is important to investigate the validity
of this assumption, before extending the approach to even more complex systems
or towards higher orders.

As mentioned above, the second issue of the first calculations of
two-nucleon corrections to $\pi$-$^2$H scattering is the relatively
large size of the leading few-nucleon terms. Recently, higher order
contributions to this class of diagrams have been investigated in
Ref.~\cite{Beane:2002wk} and Refs.~\cite{Lensky:2006wd,Baru:2007wf}.
Especially, the results of the latter references indicate that the small 
scale $\gamma$ does not play a role in the higher order terms 
studied in Refs.~\cite{Lensky:2006wd,Baru:2007wf}.
 It is however not clear at this point, whether the
hierarchy of two-, three- and four-nucleon contributions is as
expected by the power counting. An understanding of the systematics of
such higher order corrections is mandatory when analyzing data on
$^3$He and $^4$He.

In this paper, we study the presumably leading two-, three- 
and four-nucleon contributions explicitly.  
In Sec.~\ref{sec:powercount},  we will discuss the implications of using the
different power countings in more detail. 
In Sec.~\ref{sec:fewncontr}, we repeat the well-known 
expressions for two-nucleon operators, derive a complete 
set of leading three-nucleon operators and give expressions 
for the probably most relevant example of a four-nucleon 
operator.  The contribution to the scattering length based on 
these expressions requires the evaluation of high dimensional 
integrals involving the operators and few-nucleon 
wave functions. Such a problem will also appear 
in other applications. Therefore, we explain our
numerical method for such calculations in detail in 
Sec.~\ref{sec:nummeth}. In Sec.~\ref{sec:testcount}, 
we turn to $^2$H and study the dependence of the leading 
two-nucleon contributions on the binding energy based on an
unphysically over- or underbound $^2$H. The results 
will shed light on the validity of $Q$--counting. We then 
introduce the input into our calculations for the 
required few-nucleon wave functions in more detail 
in Sec.~\ref{sec:wavefu}. This is a prerequisite for our 
further studies. Sec.~\ref{sec:3nres}
is devoted to $^3$He where the scattering length 
of $\pi$-$^3$He system is calculated. For the first time, we give explicit 
results for the contribution of three-nucleon corrections 
to pion-nucleus scattering. Finally, in Sec.~\ref{sec:4nres}, 
we estimate four-nucleon contributions to $\pi$-$^4$He 
scattering based on an explicit calculation. We then summarize 
our findings and conclude in Sec.~\ref{sec:concl}. More technical 
aspects of this work are deferred to the appendices.

\section{Power-counting schemes for pion-nucleus scattering}
\label{sec:powercount}

The central goal of this work is to provide the calculations
necessary to extract the isoscalar scattering length 
from data on the pion-nucleus scattering length. The basis for
this is a reliable power counting which
 allows one to identify the relevant operators and
 provides a framework to estimate the diagrams not included.
To reach maximal predictive power we include in each channel all
diagrams that contribute up to one order lower than the contribution
of the leading, unknown counter term.
Especially the estimate of higher order operators requires some care. We here
estimate the contributions from higher order operators,
not calculated explicitly, by two methods:
\begin{itemize}
\item[$\bullet$] We demonstrate that all $N$-nucleon diagrams
  included can be organized in a series in powers of some parameter
  $\chi$. Thus, a calculation performed up to order $\chi^n$ is
  expected to have an uncertainty of order $\chi^{(n+1)}$.
\item[$\bullet$] Using a set of wave functions generated for
  regulators varying over a large range of values, we investigate
  quantitatively the regulator dependence of the various matrix
  elements. This regulator dependence should be absorbed into a
  pertinent counter term. Therefore, we get access to the possible
  counter term contribution.
\end{itemize}
As we will see, for isoscalar nuclei both methods give consistent results supporting
our claim that the scattering of pions off isoscalar targets at threshold can be
calculated with an uncertainty of the order of  5 \% of the leading 
two-nucleon contributions: {$1 \times 10^{-3} \ m_\pi^{-1}$}
for $\pi$-$^2$H and {$2 \times 10^{-3} \ m_\pi^{-1}$} 
for $\pi$-$^4$He scattering. For isovector nuclei the contribution
of the leading contact term is estimated via scaling of the
leading isoscalar contribution. This allows us to estimate
the accuracy of the calculation for  $\pi$-$^3$He scattering
to be of the order of 10\%.

It should also be stressed that up to now we are not able to identify a power
counting that allows one to relate operators with different numbers of active
nucleons, say one-nucleon with two- or three-nucleon operators.
%The origin of this problem might be that bound state wave functions are normalized. As a consequence the momentum independent one--body operators are given just by 
%the pion-nucleon scattering lengths --- c.f. Eq.~(\ref{eq:piA-scattering}). As the number of
%nucleons taking part in the interaction with the 
%external pions increases, the normalization condition
%becomes less and less important while the internal momentum
%gets more and more relevant.  
However, although we do not quantitatively understand the reasons for successive suppression of operators with an increasing number of active nucleons, the suppression observed numerically is quantitatively
sufficiently large to allow for controlled results.
In addition, it turns out that the successive suppression observed
between $N$- and $\left(N+1\right)$-nucleon operators is the same for $N=1, \ 2, \ 3$,
which is important for controlled uncertainty estimates.
We also demonstrate that it is indeed possible to estimate 
reliably the relative importance of various $N$-nucleon operators.

What we need to carry out this program is a reliable power counting.
There are two scales relevant for pion-nucleus scattering at threshold, namely
the pion mass, $m_\pi$, and the nucleus binding momentum, $\gamma$,
which may either be evaluated via
\begin{equation}
\gamma = \sqrt{2\mu \epsilon} \ ,
\label{gammarel}
\end{equation}
with $\mu$ for the reduced mass of a single nucleon with respect to the
remainder and $\epsilon$ for the binding energy with respect
to the first break-up channel, or
via
\begin{equation}
\tilde \gamma = \sqrt{2m_N (E/A)} \ ,
\end{equation}
with $(E/A)$ for the binding energy per nucleon. For the nuclei of
relevance here --- $^2$H, $^3$He, and $^4$He --- both formulas give
similar answers, namely $46$ and $46$ MeV, $82$ and $69$ MeV, and
$167$ and $115$ MeV for $\gamma$ and $\tilde \gamma$ for the three
nuclei in order.  In the following we will therefore only use the
quantity $\gamma$ for the binding momentum.

Let $q$ denote some generic momentum. 
There are two different power-counting schemes
proposed for pion-nucleus scattering at threshold.
These  differ in how 
$q$ is treated relative to $m_\pi$ and $\gamma$.
In Ref.~\cite{Weinberg:1992yk}, referred to 
as Weinberg counting, the assignment
\begin{equation}
q \sim m_\pi
\label{weincount}
\end{equation}
is proposed while in 
Ref.~\cite{Beane:2002wk}, in the following called $Q$--counting,
\begin{equation}
q \sim \gamma \ll m_\pi
\label{beancount}
\end{equation}
is used. In ~\cite{Beane:2002wk} the authors argue that the latter
assignment is to be preferred, since, for $\pi$-$^2$H scattering,
$i)$ it naturally explains why the contribution of Diagram $(a)$ of
Fig.~\ref{fig:2N-ops} is more than an order of magnitude larger than
those of $(b)$ and $(c)$; $ii)$ it quantitatively explains the
magnitude of the contribution of the diagram of
Fig.~\ref{fig:2N-triple}; and $iii)$ it quantitatively explains the
contribution of the boost correction --- this term emerges
as a recoil correction to the leading isoscalar $\pi$-N scattering
amplitude.

As in the
$\pi$N sector, chiral symmetry enforces that operators that are odd
under the exchange of the external pions --- isovector operators ---
appear with at least one derivative acting on the pion field, while
those that are even --- isoscalar --- have at least two derivatives or
one insertion of the quark mass matrix.  As a consequence, isoscalar
operators are one order in $m_\pi/m_N$ suppressed compared to their
isovector counter parts.  This is another argument in favour of using
isoscalar targets to get high accuracy information on the $\pi$-N
scattering lengths: since the contribution of the leading order
counter term sets the limit for the accuracy for the extraction of the
$\pi$-N scattering lengths, the analysis of the scattering off
isovector nuclei provides values for the scattering lengths about an
order of magnitude less accurate.

The pertinent few-body corrections can now be classified in terms 
of $q$ and $m_{\pi}$. 
Because of the arguments above, 
the leading isoscalar 4N$\pi \pi$ counter term contributes to
the $\pi$-$^2$H scattering length at order $(m_\pi/m_N)^2$, where for
simplicity we identified the chiral symmetry breaking scale with the
nucleon mass, while the leading two-nucleon correction --- Diagram $(a)$
of Fig.~\ref{fig:2N-ops} --- scales as $(m_\pi/q)^2$.  Explicit
calculation gives that the leading two-nucleon diagram contributes about
20$\times 10^{-3}\, m_\pi^{-1}$ to the $\pi$-$^2$H scattering length.
Thus in the Weinberg scheme we estimate about 1$\times 10^{-3}\,
m_\pi^{-1}$ for the leading counter term contribution while in
$Q$--counting the leading counter term is expected to contribute only
0.1$\times 10^{-3}\, m_\pi^{-1}$, since the leading two-nucleon term
benefits from an enhancement of order $(m_\pi/q)^2\simeq
(m_\pi/\gamma)^2$ in $Q$--counting.  Thus again, if we aim at a reliable
estimate of the accuracy of the calculations performed we need to
understand which one of the power-counting schemes describes the
hierarchy of diagrams for pion-nucleus scattering at threshold.
This will be done in Sec.~\ref{sec:lamdep}.

It turns out that Weinberg and $Q$--counting predict a very different
binding energy dependence of ratios of few-body corrections. These
relations can be tested empirically: the NN potential at leading
order of the chiral expansion comprises 1$\pi$--exchange and two
counter terms. One of them can be adjusted such that any given
(physical or unphysical) deuteron binding energy can be reproduced.
This has been done to investigate the binding energy dependence of the
relevant ratios of few-nucleon contributions. The main conclusion of
Sec.~\ref{sec:enerdep} will be that the binding energy dependence of
these ratios is in accordance with the prediction from Weinberg --- up
to logarithmic corrections --- and in strong disagreement to the
prediction from $Q$--counting.

In addition to providing some explanation for the apparent suppression
of $a_{\pi -A}^{(1bc)}$ vs. $a_{\pi -A}^{(1a)}$, in
Ref.~\cite{Beane:2002wk} it was argued that boosted $\pi$-N amplitudes
give a significant contribution (number $iii)$ in the list given
above), in discrepancy to what is expected from Weinberg counting. On
the other hand, since the corresponding operator is proportional to
the square of the nucleon momentum, this would imply an observable
effect of the nucleon kinetic energy inside the deuteron, which is in
conflict with general properties of field theories. In
Ref.~\cite{Baru:2007wf} both problems were solved: once the
$\Delta$--isobar is included as an explicit degree of freedom,
simultaneously the residual boost contribution becomes sufficiently
small to be in line with Weinberg counting and appears at an order
where there is also a counter term.

Finally, we analyze the contribution of the triple scattering 
diagram of Fig.~\ref{fig:2N-triple} in Sec.~\ref{i0int}.
In Weinberg counting this diagram is estimated to contribute
only at order $(m_\pi/m_N)^2$ relative to the leading two-nucleon
operator. On the other hand numerical studies revealed that
its actual value exceeds this estimate by about an order
of magnitude.  In Sec.~\ref{i0int} we will demonstrate that
this enhancement is not due to the smallness of the deuteron
binding energy, contrary to the claim in Ref.~\cite{Beane:2002wk},
but due to an integrable singularity in the corresponding
expression that produces an enhancement by a factor of $\pi^2$.
 In view of this, also item $ii)$ 
of the list that was
used to argue in favour of $Q$--counting does not apply 
anymore. 

We  are led to conclude that Weinberg counting gives a more 
reliable power counting scheme than $Q$--counting.

With these remarks, we now want to look at the possible 
few-nucleon contributions in more detail.

\section{Few-nucleon contributions}
\label{sec:fewncontr}

Our main concern here are 
the few-nucleon corrections. For an evaluation of such 
contributions, expectation values of the basic 
pion--few-nucleon ($\pi$-$A$) amplitudes with the few-nucleon wave functions need to be
calculated. Such wave functions can be obtained 
using standard methods to solve the few-nucleon 
Schr\"odinger equation based on nucleon-nucleon (NN) 
interactions \cite{Carlson:1997qn}. Whereas in the first 
attempts to understand such contributions, wave functions based on
phenomenological NN interactions were used 
\cite{Weinberg:1992yk,Beane:1997yg},
it is by now standard to also employ wave functions generated 
with nuclear interactions based on ChPT.  
The basic idea is to use naive dimensional analysis for 
a potential, which is then used to solve a Schr\"odinger 
equation \cite{Weinberg:1990rz,Weinberg:1991um}. Several 
groups have developed NN interactions based on this approach 
\cite{Ordonez:1995rz,Entem:2003ft,Epelbaum:2004fk} and 
also three- (3N) and four-nucleon (4N) interactions have been formulated \cite{vanKolck:1994yi,Bernard:2007sp} and employed
\cite{Epelbaum:2005pn}, see also Ref.~\cite{Epelbaum:2008ga} for a recent
review. Obviously, employing chiral interactions 
is preferable to calculate the pion-nucleus scattering length since 
nuclear interactions and $\pi$-$A$ amplitudes will be consistent.

In this work, we will present results on both kinds of interactions.
We still show results based on phenomenological interactions, namely
AV18 ~\cite{Wiringa:1994wb}, Nijmegen~93 ~\cite{Stoks:1994wp}, and
CD-Bonn ~\cite{Machleidt:2000ge}. For $^3$He and $^4$He, we augment
the nuclear Hamiltonian by a 3N interaction, so that the binding
energies of these nuclei are reasonably well reproduced
\cite{Nogga:2001cz}. These results may serve as benchmark and might
give indications on the size of the model-dependence of older
calculations. We also employ wave functions based on chiral nuclear
interactions between the leading order (LO, order 0 in the chiral
expansion) and next-to-next-to-leading order (N$^2$LO, order 3 in the
chiral expansion).

Chiral interactions require a regularization scheme in order to obtain a
well-defined Schr\"odinger equation. Most realizations use a momentum cutoff
of the order of 500~MeV to this aim. For the leading order, at least when
restricting oneself to $S$-wave interactions, it is possible to obtain fits
for a much larger range of cutoffs \cite{Nogga:2005hy}.  These attempts have
triggered a controversy in the community \cite{Hammer:2006qj}. It should 
be noted that an inconsistency becomes apparent
 for large cutoffs, when higher order NN interactions are
used~\cite{Epelbaum:2009sd}.  Here, we use a wide range of cutoffs only 
for LO interactions, so that also this inconsistency does not apply. This allows 
us to use results for a wide range of cutoffs to estimate the size of leading 
counter terms.  Parameter
sets for the $S$-wave contact interactions are given in
Tables~\ref{tab:potlampara}, \ref{tab:potbepara}, and \ref{tab:potcomplpara}.
Since these calculations are by no means high precision ones, we have
neglected the minor contribution of higher partial waves in this case.

For the higher order chiral forces, we have employed 
order 2 (NLO) and order 3 (N$^2$LO)  ones of 
Ref.~\cite{Epelbaum:2004fk}. In the N$^2$LO case,
we added, as required by power counting, 3N forces, which 
were tuned to reproduce the $^3$He binding energy 
and N-$^2$H scattering lengths. 

Since we restrict ourselves to leading, tree-level 
$\pi$-$A$ operators, we may derive them 
using Feynman diagrams as done below. 
The reduction from four-dimensional quantities to 
three-dimensional ones can be easily 
performed using the on-shell energies of nucleons and pions. 
The pertinent integrals involving the wave functions and 
operators will be given below since their form depends 
on the number of nucleons involved. 

\subsection{Leading two-nucleon contributions}
\label{sec:2ncontr}

\begin{figure*}[t]
  \centering
  \psfrag{k}{\scriptsize ${k}_\pi$}
  \psfrag{k'}{\scriptsize ${k}_\pi^{\,\prime}$}
  \psfrag{q}{\scriptsize ${q}$}
  \psfrag{q0}{\scriptsize ${\tilde q}$}
  \psfrag{q1}{\scriptsize ${\tilde q}$}
  \psfrag{q2}{\scriptsize ${-\tilde q}$}
  \psfrag{a}{\scriptsize $a$}
  \psfrag{b}{\scriptsize $b$}
  \psfrag{c}{\scriptsize $c$}
  \psfrag{d}{\scriptsize $d$}
  \psfrag{e}{\scriptsize $e$}
  \psfrag{tau1}{\scriptsize $\tau_1$}
  \psfrag{tau2}{\scriptsize $\tau_2$}
  \psfrag{tau1d}{\scriptsize $\tau_1^d$}
  \psfrag{tau2e}{\scriptsize $\tau_2^e$}
  \includegraphics[scale=0.55]{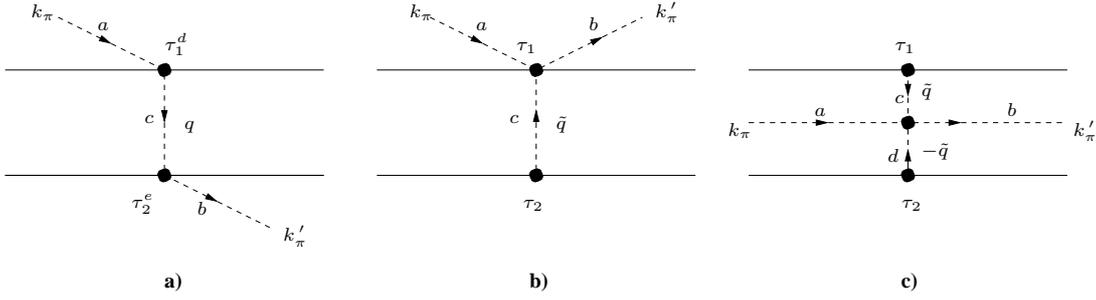}
  \caption{Leading  two-nucleon contributions 
                to $\pi$-$A$ scattering.
           }
  \label{fig:2N-ops}
\end{figure*}

The leading two-nucleon contributions are known 
for many years 
\cite{Weinberg:1992yk,Beane:1997yg,Beane:2002wk}. 
In the following, we call the numerically most important contribution, depicted in Fig.~\ref{fig:2N-ops}(a),  ``Coulombian'' because 
of its $ {1 \over {\vec q}^2}$ pion propagator. The explicit expression 
for the amplitude is
 \begin{eqnarray}
   \label{eq:m2a}
   i\mathcal{M}^{\left(\mbox{\scriptsize \ref{fig:2N-ops}a}\right)} & = & i\frac{m_\pi^2}{4f_\pi^4\;\vec{q}^{\,2}}\big\{2\delta^{ab}\left(\vec{\tau}_1\cdot\vec{\tau}_2\right)-\tau_1^b\tau_2^a-\tau_1^a\tau_2^b\big\}
 \end{eqnarray}
 where $\vec q$ is the momentum transfer between the nucleons,
 the $\vec{\tau}_{i}$ are usual Pauli matrices acting in the isospin 
 space of nucleon $i$ 
 and $f_{\pi}$ the pion decay constant. Throughout this work, we employ $f_{\pi}=92.4$~MeV. The small latin letters refer to isospin 
indices of the pions as given in the figure. 

The amplitudes of 
Figs.~\ref{fig:2N-ops}(b) and \ref{fig:2N-ops}(c) are individually dependent 
on the parametrization of the pion field. The sum of both, however,
is independent of this choice as it should. Therefore, we will only 
show results for the sum of both contributions for which 
the amplitude reads 
\begin{eqnarray}
 \label{eq:m2bc}
  i\mathcal{M}^{\left(\mbox{\scriptsize  \ref{fig:2N-ops}b+\ref{fig:2N-ops}c}\right)} &  = &  -i\frac{g_A^2m_\pi^2}{4f_\pi^4}\frac{1}{\left(\vec{q}^{\,2}+m_\pi^2\right)^2}\left(\vec{\sigma}_1\cdot\vec{q}\right)\left(\vec{\sigma}_2\cdot\vec{q}\right)\nonumber\\
  & &\quad \times\big\{\delta^{ab}\left(\vec{\tau}_1\cdot\vec{\tau}_2\right)-\left(\tau_1^a\tau_2^b+\tau_1^b\tau_2^a\right)\big\} . 
 \end{eqnarray}
Here, we additionally encounter $\vec{\sigma}_{i}$, the Pauli matrices 
acting in the spin space of nucleon $i$. Since the propagators 
contain an additional pion mass, we will refer to this contribution 
as ``non-Coulombian''.   

\begin{figure}[t]
  \centering
    \includegraphics[scale=0.55]{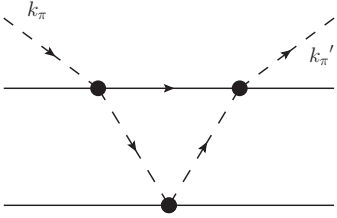}
  \caption{Triple scattering  two-nucleon diagram.}
  \label{fig:2N-triple}
\end{figure}

In addition, in Ref.~\cite{Beane:2002wk}, the diagram shown in
Fig.~\ref{fig:2N-triple} was identified as numerically important,
although in Weinberg counting it appears only at order N$^2$LO as compared
to the leading order diagrams of Fig.~\ref{fig:2N-ops}. This
observation was taken as a support for the modified power counting,
$Q$--counting, as discussed in the previous section. As we will show in
section \ref{i0int}, a part of this amplitude is indeed enhanced,
however, not parametrically but as a consequence of the
special topology of the diagram. Therefore we also include this
triple scattering diagram, although this is not supported
by a power counting yet. Straightforward evaluation
of its most dominant part gives
\begin{eqnarray}
   \label{eq:m2triple}\nonumber
  i\mathcal{M}^{\left(\mbox{\scriptsize \ref{fig:2N-triple}}\right)} &  = &  
-2i\left(\frac{m_\pi}{2f_\pi^2}\right)^3\frac1{8|\vec{q}|} \\ 
& & \times \left\{\tau_1^b\tau_2^a+\tau_1^a\tau_2^b+\frac{i}{2}\epsilon^{abe}\left(\tau_1^e+\tau_2^e\right)\right\} . 
\end{eqnarray}
For a discussion of the full expression for the
amplitude we refer to Sec.~\ref{i0int}.
%Also for we will study the binding energy dependence of 
%its co¥ntribution. This will be specifically interesting
%to discuss the effectiveness of $Q$--counting in 
%Sec.~\ref{sec:enerdep}. 
It is important to note that, although the triple scattering diagram
is enhanced compared to what is expected from dimensional analysis,
this enhancement is not sufficient to fully overcome the parametric
suppression as provided by chiral symmetry. Comparing the double
scattering contribution, Eq.~(\ref{eq:m2a}), to the triple scattering
contribution, Eq.~(\ref{eq:m2triple}), we still find a relative
suppression\footnote{For this comparison we evaluated both operators
  for isoscalar nucleon pairs and under the assumption that
  $\vec{q}\sim m_\pi$, in line with Weinberg counting.} of
order $$m_\pi^2/(16f_\pi^2)\sim 0.14 \ ,$$ which is close to the ratio
found from the exact evaluation of the matrix elements. Thus, there is
a suppression by an order of magnitude between the two contributions.
We conjecture that the related diagram with four $\pi$N interactions
will be suppressed by yet another order of magnitude and is therefore
irrelevant for this study. Note, however, in case of $K$-nucleus
scattering a resummation of the multiple scattering series is
unavoidable \cite{Kamalov:2000iy,Meissner:2006gx,Baru:2009tx}. Also in
Ref.~\cite{Baru:2009tx}, it was argued that a resummation needs to be
done including nucleon recoil effects in the propagators, for the
nucleon recoil in multiple scattering diagrams induces potentially
important corrections of the order of $\sqrt{m_K/m_N}$, with $m_K$
being the kaon mass, as compared to the static contributions.

As was argued above, all diagrams that contribute to lower orders
than the first counter term should be included in this study. In
addition to the diagrams discussed so far, there are potentially also
those that are enhanced due to the presence of either two-nucleon or
$\pi$NN--cuts. Moreover, there are also contributions from the
$\Delta$--isobar.  The effect of the $\pi$NN--cuts was discussed in
detail in Ref.~\cite{Baru:2004kw}, see also Ref.~\cite{Lensky:2005hb}
for a related discussion. 
It was shown in Ref.~\cite{Baru:2004kw} that, although
formally enhanced, the contributions of the $\pi$NN--cuts to the $\pi$-$^2$H scattering length
nearly vanish as a result of a cancellation of %one--body and 
different two-nucleon diagrams enforced by the Pauli principle. The same argument also
applies to the scattering off $^4$He. In
Ref.~\cite{Lensky:2006wd} and \cite{Baru:2007wf} it was shown that
diagrams with an NN--cut, often called dispersive 
terms~\footnote{Note, the group of so--called disconnected diagrams (where the pion is 
absorbed on one nucleon and emitted from another one) are included in
the dispersive corrections.}, and those with an intermediate $\Delta$ start
to contribute at the same order. Although individually sizable, the two
groups of diagrams cancel each other nearly exactly. This cancellation
should also appear in heavier nuclei whenever pions get scattered off an
isoscalar NN pair.
The analogous diagrams for isovector NN pairs were not studied yet,
however, their contribution is intimately connected to the reaction
$pp\to pp\pi^0$. The total cross section near threshold in this pion
production channel, which is the relevant quantity to estimate the
contributions from the NN--cuts, is more than an order of magnitude
smaller than that of $pp\to pn\pi^+$~\cite{Hanhart:2003pg}. In addition,
the $\Delta$--isobar is a lot less important in that channel compared
to the deuteron channel~\cite{Hanhart:1998za}. In total, we
therefore expect a negligible contribution from
intermediate $\Delta$'s and NN--cuts to pion-nucleus scattering.

In Ref.~\cite{Beane:2002wk} also corrections that arise
through the boost of the $\pi$N interaction were
identified as numerically significant. However,
in a formalism that includes the $\Delta$
as explicit degree of freedom this is no longer the case~\cite{Baru:2007wf} ---
see also discussion at the end of Sec.~\ref{sec:powercount}.

Therefore, we may assume that the contributions listed in Eqs.~(\ref{eq:m2a})
to (\ref{eq:m2triple}) are all two-nucleon contributions
that need to be included in this calculation.

With the help of Feynman diagrams, we have obtained 
the amplitude on two free nucleons.  
Following Weinberg, the amplitude resulting for nucleons 
bound can be obtained calculating the expectation value 
of the free amplitude using the $A$-nucleon 
bound state wave function.
E.g. for $A=3$, the wave function  $\psi_{\alpha}\left(\vec{p}_{12},\vec{p}_3\right)$
in momentum space depends on the two momenta $\vec p_{12}$ and $\vec p_{3}$, which
are Jacobi momenta defined by 
\begin{eqnarray}
 \label{eq:jacobi}
  \vec{p}_{12}\:\; & = & \frac{1}{2}\left(\vec{k}_1-\vec{k}_2\right)\nonumber\\
  \vec{p}_{12}^{\,\prime} & = & \frac{1}{2}\left(\vec{k}_1^{\,\prime}-\vec{k}_2^{\,\prime}\right)\nonumber\\
  \vec{p}_3\;\: & = & \frac{2}{3}\vec{k}_3\:-\:\frac{1}{3}\:\left(\vec{k}_1\:+\:\vec{k}_2\right)\nonumber\\
  \vec{p}_3^{\,\prime} & = & \frac{2}{3}\vec{k}_3^{\,\prime}-\frac{1}{3}\left(\vec{k}_1^{\,\prime}+\vec{k}_2^{\,\prime}\right) \ .
 \end{eqnarray}
Here $\vec{k}_i$ and  $\vec{k}_i^{\,\prime}$ are the momenta 
of the nucleons before and after scattering of the pion. 
$\alpha$ denotes an index labeling the different 
spin and isospin states of the three nucleons. 

For the two-nucleon contribution, making use of the conservation 
of momentum of the third nucleon, we obtain in this way
 \begin{eqnarray}
  \label{eq:2bexp}
  & &\langle\hat O^{2-nucleon} \rangle = 
  \intsum_{\alpha^\prime} {d^3p_{12}^{\prime} \over (2\pi)^3 } \intsum_{\alpha} d^3p_{12}\; d^3p_3\; \cr
  &&\cr
  & &\hspace{0.5cm}\times\;\psi_{\alpha'}^\ast\left(\vec{p}_{12}^{\,\prime},\vec{p}_3\right)
  \mathcal{M}_{\alpha' \alpha}\left(\vec{p}_{12}^{\,\prime},\vec{p}_{12}\right)\psi_{\alpha}\left(\vec{p}_{12},\vec{p}_3\right)
 \end{eqnarray}
where the amplitude has also been expressed in terms of 
Jacobi momenta $\vec{p}_{12}$ using the definitions of 
Eq.~(\ref{eq:jacobi}). We note that the factors $(2\pi)^3$ 
take into account that we normalize momentum eigenstates  
to $\delta^3(\vec p - \vec p ')$ 
($(2\pi)^3\, \delta^3(\vec p - \vec p ')$) for the wave functions
(operators) throughout this work.
Eq.~(\ref{eq:2bexp}) is easily generalized to $A=4$ and higher.

The corresponding contribution to the scattering length can in general be 
determined via
\begin{eqnarray}
   &&a_{\pi\scriptsize{-}
    A}^{N{-}nucleon}=\frac{1}{4\pi}\left({A}\atop{N}\right)\left(\frac{1}{1+\frac{m_\pi}{A\,m_N}}\right)\langle
  \hat O^{N{-}nucleon}\rangle \ . \cr 
  & & 
  \label{eq:scattlength}
 \end{eqnarray}
The binomial coefficient in front is introduced
 to take into account the number of nucleon--$N$-tupel
in an $A$-nucleon system.
This factor is three for two-nucleon operators in $^3$He.

\subsection{Leading three-nucleon contributions}
\label{sec:3ncontr}

\begin{figure}[t]
\begin{center}
  \psfrag{k}{\scriptsize ${k}_\pi$}
  \psfrag{k'}{\scriptsize ${k}_\pi^{\,\prime}$}
  \psfrag{q1}{\scriptsize ${q_1}$}
  \psfrag{q3}{\scriptsize ${q_3}$}
  \psfrag{a}{\scriptsize $a$}
  \psfrag{b}{\scriptsize $b$}
  \psfrag{c}{\scriptsize $c$}
  \psfrag{d}{\scriptsize $d$}
  \psfrag{tau1e}{\scriptsize $\tau_1^e$}
  \psfrag{tau2f}{\scriptsize $\tau_2^f$}
  \psfrag{tau3g}{\scriptsize $\tau_3^g$}
\includegraphics[width=5cm]{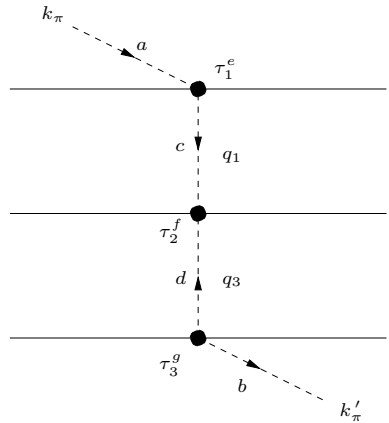}
\caption{Coulombian three-nucleon contribution to 
              $\pi$-$A$ scattering.}
\label{fig:3N-coulcontr}
\end{center}
\end{figure}

One aim of this work is to study the relative 
importance of one-, two-, three-nucleon, etc. contributions 
explicitly. Therefore, we now look at three-nucleon 
contributions to pion-nucleus scattering. 
Based on Weinberg's original power counting,
it is easy to identify the leading ones, 
which are shown in Figs.~\ref{fig:3N-coulcontr}  and 
\ref{fig:3N-contralpha}. Note that we have omitted diagrams that 
vanish in leading order due to threshold kinematics. 
Naively, these diagrams are suppressed by $(m_\pi / m_{N})^2$ 
compared to the leading two-nucleon ones.  
A closer look reveals that the diagram of Fig.~\ref{fig:3N-coulcontr}
is individually independent of the parametrization 
of the pion field. The resulting amplitude reads 
\begin{eqnarray}
\label{eq:m3coul}
  i\mathcal{M}^{\left(\mbox{\scriptsize  \ref{fig:3N-coulcontr}}\right)} & = & \left(\frac{m_\pi}{2f_\pi^2}\right)^3\frac{1}{\vec{q}_1^{\,2}\;\vec{q}_3^{\,2}}\;\nonumber \\[5pt]
  &\times& \big\{\epsilon^{abe}\left[\left(\vec{\tau}_1\cdot\vec{\tau}_2\right)\tau_3^e+\left(\vec{\tau}_2\cdot\vec{\tau}_3\right)\tau_1^e\right]\big\}
 \end{eqnarray}
Again, the propagators resemble Coulombian ones. We therefore 
will refer to this amplitude as ``Coulombian'' three-nucleon 
contribution. The definition of the momentum transfers and 
pion isospin indices can be read off from the figure. 
 Note that due to the isovector structure of this operator it will not contribute 
 to pion scattering on  isoscalar nuclei. 
 
Additionally to this Coulombian contribution, we found a set 
of seven further diagrams, which are individually dependent 
on the parametrization of the pion field. The sum of these 
diagrams is however parametrization-independent as is outlined 
in Appendix~\ref{app:alphadep}, and the explicit expressions are given 
in Appendix~\ref{app:halfcoul3n}. They all have in common 
that at most one of the propagators is Coulombian. In the following,
we will therefore refer to these diagrams as half-Coulombian. 

For the three-nucleon contribution, the momentum of 
the third nucleon is not conserved anymore. 
Therefore, the calculation of the expectation value 
with respect to the $A=3$ wave function reads  
\begin{eqnarray}
\label{eq:3bexp}
  & &\langle\hat O^{3-nucleon} \rangle = \ 
  \intsum_{\alpha^\prime} {d^3p_{12}^{\prime}\over (2\pi)^3} \; {d^3p_3^{\prime}\over (2\pi)^3}  \; \intsum_{\alpha} d^3p_{12}\; d^3p_3\; \cr
  &&\cr
  & &\hspace{0.5cm}\times\;\psi_{\alpha'}^\ast\left(\vec{p}_{12}^{\,\prime},\vec{p}_3^{\, \prime}\right)
  \mathcal{M}_{\alpha' \alpha}\left(\vec{p}_{12}^{\,\prime}\vec{p}_{3}^{\,\prime},\vec{p}_{12}\vec{p}_{3}\right)\psi_{\alpha}\left(\vec{p}_{12},\vec{p}_3\right) \ . \cr & & 
 \end{eqnarray}
Again, the amplitude has been expressed in terms of 
Jacobi momenta.  The contribution to the scattering length due to 
the three-nucleon diagrams is then again found by inserting
Eq.~(\ref{eq:3bexp}) into Eq.~(\ref{eq:scattlength}).

\subsection{Leading four-nucleon contribution}
\label{sec:4ncontr}

\begin{figure}[t]
\begin{center}
  \psfrag{k}{\scriptsize ${k}_\pi$}
  \psfrag{k'}{\scriptsize ${k}_\pi^{\,\prime}$}
  \psfrag{q1}{\scriptsize ${q_1}$}
  \psfrag{q2}{\scriptsize ${q_2}$}
  \psfrag{q3}{\scriptsize ${q_3}$}
  \psfrag{a}{\scriptsize $a$}
  \psfrag{b}{\scriptsize $b$}
  \psfrag{c}{\scriptsize $c$}
  \psfrag{d}{\scriptsize $d$}
   \psfrag{e}{\scriptsize $e$}
  \psfrag{tau1f}{\scriptsize $\tau_1^f$}
  \psfrag{tau2g}{\scriptsize $\tau_2^g$}
  \psfrag{tau3h}{\scriptsize $\tau_3^h$}
   \psfrag{tau4i}{\scriptsize $\tau_4^i$}
\includegraphics[width=5cm]{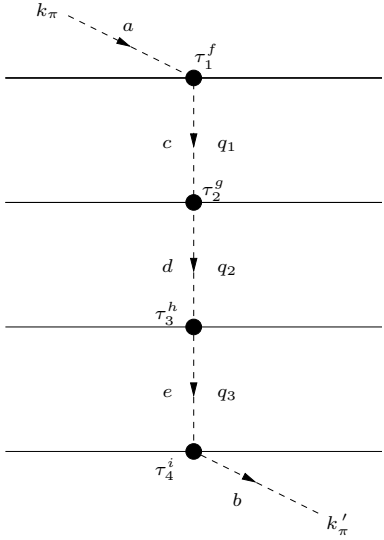}
\caption{Coulombian four-nucleon contribution to 
              $\pi$-$A$ scattering.}
\label{fig:4N-contr}
\end{center}
\end{figure}

In order to also confirm our conclusions on the relative 
importance of higher-body $\pi$-$A$ amplitudes, we 
also need a four-nucleon contribution which, of course,
can only be relevant for scattering on nuclei with $A \ge 4$. 
The probably most important contribution is Coulombian 
and shown in Fig.~\ref{fig:4N-contr}. 
Based on Weinberg's original power counting, this contribution is 
suppressed by $(m_{\pi}/m_{N})^4$ compared to the leading 
two-nucleon contributions. 
Again, it is independent 
of the parametrization of the pion field. Note that this diagram 
is not a complete set of leading four-nucleon terms. 
We will only evaluate its contribution to have an estimate of possible 
higher order contributions to $\pi$-$A$ scattering. 

It is an easy exercise to find the corresponding amplitude
\begin{eqnarray}
\label{eq:m4coul}
  i\mathcal{M}^{\left(\mbox{\scriptsize  \ref{fig:4N-contr}}\right)} & = & 
   i  \left(\frac{m_\pi}{2f_\pi^2}\right)^4\frac{1}{\vec{q}_1^{\,2}\;\vec{q}_2^{\,2}\;\vec{q}_3^{\,2}} \cr 
   & \times & \big\{ 2 \delta^{ab} \left(\vec{\tau}_1\cdot\vec{\tau}_2\right) \left(\vec{\tau}_3\cdot\vec{\tau}_4\right) \cr 
   &  &         + \left(\vec{\tau}_1\cdot\vec{\tau}_4\right) 
                  \left(\vec{\tau}_2^{a} \vec{\tau}_3 ^{b} + \vec{\tau}_2^{b} \vec{\tau}_3 ^{a} \right) \cr 
   &  &       - \left(\vec{\tau}_3\cdot\vec{\tau}_4\right) 
                  \left(\vec{\tau}_1^{a} \vec{\tau}_2 ^{b} + \vec{\tau}_1^{b} \vec{\tau}_2 ^{a} \right) \cr 
   &  &        - \left(\vec{\tau}_1\cdot\vec{\tau}_2\right) 
                  \left(\vec{\tau}_3^{a} \vec{\tau}_4 ^{b} + \vec{\tau}_3^{b} \vec{\tau}_4 ^{a} \right)      \big\}   \  .
 \end{eqnarray}
 This amplitude then enters the evaluation of the expectation value 
 analogously to Eqs.~(\ref{eq:2bexp}) and (\ref{eq:3bexp}) 
 for the two- and three-nucleon operators \pagebreak
 \begin{eqnarray}
\label{eq:4bexp}
  & &\langle\hat O^{4-nucleon} \rangle = \ \nonumber \\[5pt]
& &  \hspace{0.5cm} \intsum_{\alpha^\prime} {d^3p_{12}^{\prime}\over (2\pi)^3} \; {d^3p_3^{\prime}\over (2\pi)^3}\; {d^3q_4^{\prime}\over (2\pi)^3}\;    \intsum_{\alpha} d^3p_{12}\; d^3p_3\; d^3q_4\; \cr
  &&\cr
  & &\hspace{0.5cm}\times\;\psi_{\alpha'}^\ast\left(\vec{p}_{12}^{\,\prime},\vec{p}_3^{\, \prime},\vec{q}_4^{\, \prime}\right)
  \mathcal{M}_{\alpha' \alpha}\left(\vec{p}_{12}^{\,\prime}\vec{p}_{3}^{\,\prime}\vec{q}_{4}^{\,\prime},
                                                      \vec{p}_{12}\vec{p}_{3}\vec{q}_{4}\right) \cr 
  & & \hspace{0.5cm} \times \; \psi_{\alpha}\left(\vec{p}_{12},\vec{p}_3,\vec{q}_4\right) \ .
 \end{eqnarray}
Here, we express the $^4$He wave function $\psi$ in terms 
of Jacobi momenta. $\vec p_{12}$ and $\vec p_{3}$ are the 
same as in the three-nucleon case. $\vec q_{4}$ is the relative 
momentum of nucleon 4 and the cluster of nucleons 1,  2 and 3. 
For details of the representation of the wave functions, we refer to 
\cite{Nogga:2001cz}. 
  
\section{Numerical method}
\label{sec:nummeth}

In this section, we want to introduce briefly the numerical 
method used for the evaluation of the pertinent integrals. 
Whereas expectation values for $^2$H can be easily 
obtained using standard methods of integration or  using 
a partial wave decomposition, this becomes more and more 
tedious for more and more complex nuclei. 
One aim of this work was to establish a scheme to 
evaluate expectation values in momentum space 
based on Monte Carlo (MC) integration without performing a 
partial wave decomposition. 

In this way, we were able to generate the numerical 
expressions of the amplitude reliably with the help 
of a {\em Mathematica} script. The resulting 
FORTRAN code lines could be included in a 
FORTRAN code evaluating the high dimensional 
integrals given above.

For the evaluation, it turned out that 
simple MC integration requires a large
number of trial points to obtain acceptably small standard 
deviations. In our tests, we found that most trial points 
consisted of momenta 
for which the wave functions are nearly zero. 
Therefore, obviously, an importance sampling similar 
to the Metropolis algorithm \cite{Metropolis:1953am}
is required to keep the 
computational needs small and increase the accuracy. 

Usually such an importance sampling is guided by the square of the 
wave function. In configuration space, this quantity is 
perfectly suited as a weight function for the 
Metropolis algorithm since the weight function is then 
automatically normalized to one at least as long as the operators 
are local. For momentum space, the structure is more 
complicated, since the integrals require weight functions 
with higher dimensionality as in configuration space. This implies 
that a simple square of the wave function is not useful 
for the importance sampling anymore. This problem could be solved 
by performing part of the integrals using standard methods
as has been successfully done in  \cite{Wiringa:2008dn}. 
We found this approach less practical in our case, since the 
three- and four-nucleon operators would 
require to perform high dimensional integrations using standard 
integration methods. 

Our solution was to give up weight functions based on the 
wave functions of the system, but choose a rational ansatz 
instead. The parameters of the ansatz were then adjusted so 
that the standard deviation in test cases was minimized. In 
this way, we were able to improve the accuracy sufficiently. 
At the same time, the weight function could be analytically normalized 
to one so that the calculations became feasible. 

E.g. we choose for the importance sampling for integrals 
of the form of Eq.~(\ref{eq:3bexp}) a weight function 
depending on the four integration variables 
$\vec p_{i} = \vec p_{12}$, $\vec p_{3}$, $\vec p_{12}^{\, \prime}$ and 
$\vec p_{3}^{\, \prime}$
 \begin{eqnarray}
   & & w\left(\vec{p}_{12}^{\,\prime},\vec{p}_3^{\,\prime},\vec{p}_{12},\vec{p}_3\right) \equiv  w\left(p_{12}^\prime,p_3^\prime,p_{12},p_3\right)\cr
   & & \quad =  \prod_{i}\frac{ \left( r-3 \right) \left( r-2 \right) \left( r-1 \right) } {8\pi} \: \frac{ C_{p_i}^{ \left(r-3\right)}}{ \left(p_i+C_{p_i}\right)^r }\cr
 &&
 \label{eq:weight-function}
 \end{eqnarray}
For simplicity, the ansatz only depends on the magnitude 
of the momenta. With the parameters $C_{p_{i}}$ and $r$ 
the shape of the weight functions can be influenced. The ansatz guarantees (for large enough $r$) 
that the weight function is normalized to 
\begin{eqnarray}
  \int d^3p_{12}^{\prime} d^3p_3^{\prime}
  \int d^3p_{12} d^3p_3 \,   w\left(\vec{p}_{12}^{\,\prime},\vec{p}_3^{\,  
    \prime},\vec{p}_{12},\vec{p}_3\right) = 1 .
    \label{eq:weight-function-norm}
\end{eqnarray}

\begin{table}
\centering
\caption{Values for the parameters of the weight function
              for various cutoffs of the leading order chiral 
              interaction, AV18 and CD-Bonn wave functions.
              We always chose $C_{p_{12}}=C_{p_{12}^\prime}$
              and $C_{p_{3}}=C_{p_{3}^\prime}$}
\label{tab:wpara}       
\begin{tabular}{llll}
\hline\noalign{\smallskip}
$\Lambda$[fm$^{-1}$] & r & $C_{p_{12}}$[fm$^{-1}$]   & $C_{p_{3}}$[fm$^{-1}$]  \\
\noalign{\smallskip}\hline\noalign{\smallskip}
2.0 & 7.6   & 1.25  &1.0  \\
3.0 & 7.6   & 1.25  &1.0  \\
4.0 & 7.4   & 2.75  &1.25  \\
5.0 & 7.4   & 2.75  &1.25  \\
%6.0 & 7.8   &  3.5   & 1.5  \\
%8.0 & 8.0   &  4.3   &  2.5 \\
10.0 & 8.5  & 5.2   &  3.25 \\
%12.0 & 8.8  & 6.3   &  3.8 \\
%14.0 & 9.0  & 7.5   &  4.5 \\
%16.0 & 9.2  & 8.0   &  5.0 \\
%18.0 & 9.5  & 8.5   &   5.5 \\
20.0 & 9.7  & 8.75 &  6.0 \\
AV18         & 7.6    & 3.0 & 1.25 \\
CD-Bonn   & 7.6    & 3.0  & 1.25  \\
\noalign{\smallskip}\hline
\end{tabular}
\end{table}

A detailed description of our tests of this method can be found in 
\cite{Liebig:2009mth}. Here we only summarize the resulting 
parameters in Table~\ref{tab:wpara}. For the more simple 
integral Eq.~(\ref{eq:2bexp}), the weight function was simplified in
the obvious way just dropping terms depending on $\vec {p'_{3}}$. We used the same parameters in both cases.  

Still, the size of the integrand is driven by the size of the wave functions. 
This is reflected in a strong dependence of the parameters on the wave function
used. In the table, we give our choice for the leading order chiral wave functions 
with different cutoffs $\Lambda$ and the wave functions obtained 
based on AV18 and CD-Bonn. For the higher order chiral wave functions,
we used the same parameters as for the leading order $\Lambda=2$~fm$^{-1}$ 
wave functions, since the momentum dependence is similar in this case. 

\begin{table*}
\centering
\caption{Comparison of PW and MC 
               results. The scattering length contribution 
               of the different two-nucleon operators Figs.\ref{fig:2N-ops} (a), (b)+(c) and  \ref{fig:2N-triple}
               is compared for different cutoffs and phenomenological wave functions.  }
\label{tab:partialmonte}
\begin{center}       
\begin{tabular}{lllllll}
\hline\noalign{\smallskip}
$\Lambda$[fm$^{-1}$] & \multicolumn{2}{c}{$a_{\pi-{^2}{\rm H}}^{\mbox{\scriptsize (\ref{fig:2N-ops}a)}}$ [$10^{-3} \, m_{\pi}^{-1}$]}   &  
\multicolumn{2}{c}{$a_{\pi-^{2}{\rm H}}^{\mbox{\scriptsize (\ref{fig:2N-ops}bc)}}$ [$10^{-3} \, m_{\pi}^{-1}$]}  &  
\multicolumn{2}{c}{$a_{\pi-^{2}{\rm H}}^{\mbox{\scriptsize (\ref{fig:2N-triple})}}$ [$10^{-3} \, m_{\pi}^{-1}$]}  \\
     &   PW  & MC   & PW   & MC  & PW   & MC  \\
\noalign{\smallskip}\hline\noalign{\smallskip} 
3.0 &  -21.37  & -21.16(22)  &  -0.666 & -0.6658(4)    &  2.43 & 2.433(3)    \\
4.0 &  -20.02  & -19.65(9)    &  -0.902 & -0.9017(7)    &  1.77 & 1.767(2)    \\
5.0 &  -19.75  & -20.13(35)  &  -0.889 & -0.8904(6)    &  1.61 & 1.613(3)    \\
%6.0 &  -20.12  & -20.20(20)  &  -0.813 & -0.8151(10)  &  1.89 & 1.908(5)    \\
%8.0 &  -20.78  & -20.75(57)  &  -0.762 & -0.7628(21)  &  2.49 & 2.483(7)    \\
10.0 & -20.72 & -22.68(136)&  -0.824 & -0.8232(14)  &  2.35 & 2.345(7)    \\
%12.0 & -20.49 & -19.78(15)  &  -0.880 & -0.8819(13)  &  1.96 & 1.956(6)     \\
%14.0 & -20.45 & -20.15(30)  &  -0.894 & -0.8920(18)  &  1.89 & 1.860(10)   \\
%16.0 & -20.57 & -19.71(56)  &  -0.882 & -0.8786(29)  &  2.07 & 2.067(15)   \\
%18.0 & -20.71 & -19.93(35)  &  -0.870 & -0.8712(25)  &  2.35 & 2.323(20)  \\
20.0 & -20.78 & -20.60(52)  &  -0.867 & -0.8729(40)  &  2.49 & 2.459(29)   \\
AV18        & -19.62 & -19.45(13)  & -0.749 & -0.7493(6)  & 1.63  & 1.631(3)   \\
Nijm93      & -19.84 & -19.40(11)  & -0.743 & -0.7425(3)  & 1.72  & 1.724(3)  \\
CD-Bonn  & -20.20 & -19.99(8)    & -0.553 & -0.5521(4)  & 1.92  & 1.924(2)  \\
CD-Bonn \cite{Beane:2002wk}  & -20.20 & --- & -0.55 & --- & --- & --- \\
\noalign{\smallskip}\hline
\end{tabular}
\end{center}
\end{table*}

As a first test, we compare results for $^2$H obtained with a 
traditional partial wave (PW) decomposition and with the MC 
method. 
The partial wave decomposed amplitudes are listed in Appendix~\ref{app:2Npw}.
In Table~\ref{tab:partialmonte}, the results are compared. One can see that 
the MC results agree well with the PW values and that these agree well with the 
previously obtained values of \cite{Beane:2002wk}. We note that $2/3$ of the MC
results are within 1$\sigma$ of the PW result. We found a few values with more 
than 2$\sigma$ deviation.  These outliers are generally expected for  a MC calculation
and indicate that the probability distribution is not normal, but has more extended 
shoulders. Note, as usual the numerical accuracy of the MC calculation can 
easily be enhanced by increasing the number of runs --- we stopped our
evaluations as soon as the numerical accuracy was higher than the
theoretical accuracy of the calculation, which is of the order of a few
percent
as discussed below.

\section{Testing the counting schemes}
\label{sec:testcount}

\subsection{Cutoff dependence and estimate of the leading
counter term}
\label{sec:lamdep}

As was argued in the introduction, from the regulator dependence
of the contributions from the various diagrams it is possible
to estimate the size of the leading 4N$\pi\pi$ counter term
contribution. Since the estimated contribution of this term differs
drastically between $Q$--counting and Weinberg counting, a
study of the cutoff dependence provides a non-trivial test
of the counting schemes.

The leading chiral NN interaction is given by 
\begin{equation}
\label{eq:lopot}
V(\vec p^{\, \prime},\vec p) = - \ \left( { g_A \over 2 f_\pi} \right)^2
\vec \tau_1 \cdot \vec \tau_2 \ 
{ \vec q \cdot \sigma_1 \ \vec q \cdot \sigma_2 \over (\vec q)^2 + m_\pi^2 }
+C_{S}+C_{T} \,  \vec \sigma_1 \cdot \vec \sigma_2 \ . 
\end{equation}
In order to obtain a meaningful Schr\"odinger equation, it is necessary 
to introduce a regulator. We here perform regularization by a smooth 
cutoff function given by 
\begin{equation}
f(\vec p) = \exp \left( - \left( { p \over \Lambda } \right)^4  \right)  
\end{equation}
which depends on a cutoff parameter $\Lambda$. 
The potential is then replaced by 
\begin{equation}
V(\vec p^{\, \prime},\vec p)    \longrightarrow f(\vec p^{\, \prime} ) \ V(\vec p^{\, \prime},\vec p) \ f(\vec p)
\end{equation}
In this section, we used $g_{A}=1.2834$, $m_{\pi}=139.57$~MeV, and $f_{\pi}=92.4$~MeV 
for the leading order potential and the pion scattering amplitudes which 
is close to the parameters used in \cite{Beane:2002wk}. Note that we use slightly different 
parameters for studying $^3$He and $^4$He below. 

Since we are only interested in the $^3$S$_{1}$--$^3$D$_{1}$
partial wave, the deuteron channel, we arbitrarily set $C_{T}=0$ 
and fit $C_{S}$ so that the deuteron binding energy is fixed to 
a  experimental value for a given value of the cutoff parameter $\Lambda$. 
The fit results are given in Table~\ref{tab:potlampara}.

\begin{figure}[t]
  \centering
  \psfrag{lambda}{$\Lambda$ [fm$^{-1}$]}
  \psfrag{ascatt}{$a_{\pi-^2 \rm H}^{\left(\mbox{\scriptsize \ref{fig:2N-ops}a}\right)}$ [$10^{-3} m_{\pi}^{-1}$ ]}
  \includegraphics[scale=0.55]{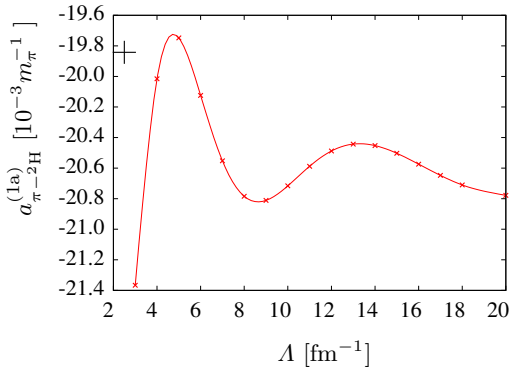}
      
  \caption{Contribution $a_{\pi{\rm -}^2 \rm H}^{\left(\mbox{\scriptsize \ref{fig:2N-ops}a}\right)}$ due to amplitude Eq.~(\ref{eq:m2a}) to 
                the $\pi$-$^2$H scattering length depending on the cutoff $\Lambda$ of the LO potential. 
                 The result for the Nijmegen~93 is also shown ``(+)''. It 
 is independent of $\Lambda$ and is positioned arbitrarily on the plot. }
  \label{fig:lamdepm2a}
\end{figure}

\begin{figure}[t]
  \centering
  \psfrag{lambda}{$\Lambda$ [fm$^{-1}$]}
  \psfrag{ascatt}{$a_{\pi-^2 \rm H}^{\left(\mbox{\scriptsize \ref{fig:2N-ops}bc}\right)}$ [$10^{-3} m_{\pi}^{-1}$ ]}
  
  \includegraphics[scale=0.55]{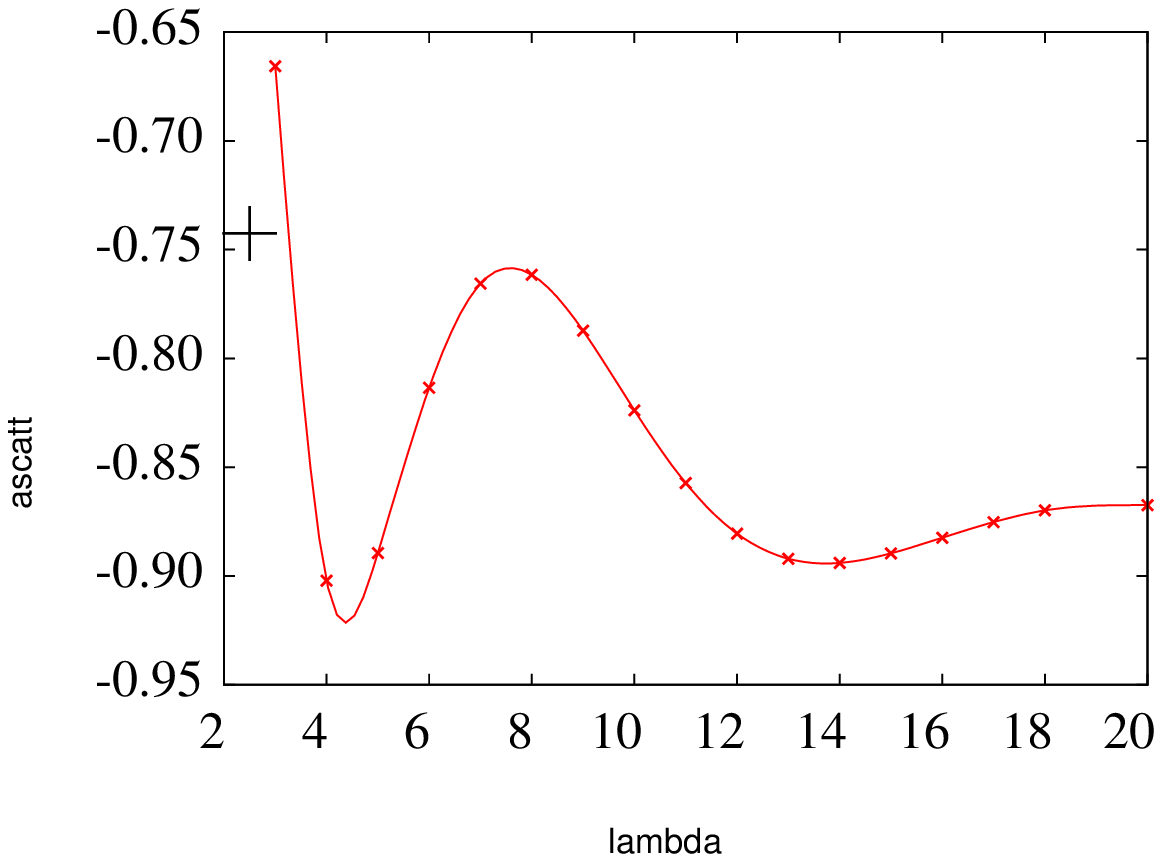}
      
  \caption{Contribution $a_{\pi- ^2 \rm H}^{\left(\mbox{\scriptsize \ref{fig:2N-ops}bc}\right)}$ due to amplitude Eq.~(\ref{eq:m2bc}) to 
                the $\pi$-$^2$H scattering length depending on the cutoff $\Lambda$ of the LO potential. 
                 The result for the Nijmegen~93 is also shown ``(+)''.  }
  \label{fig:lamdepm2bc}
\end{figure}

\begin{figure}[t]
  \centering
  \psfrag{lambda}{$\Lambda$ [fm$^{-1}$]}
  \psfrag{ascatt}{$a_{\pi-^2 \rm H}^{\left(\mbox{\scriptsize\ref{fig:2N-triple}}\right)}$ [$10^{-3} m_{\pi}^{-1}$ ]}
  
  \includegraphics[scale=0.55]{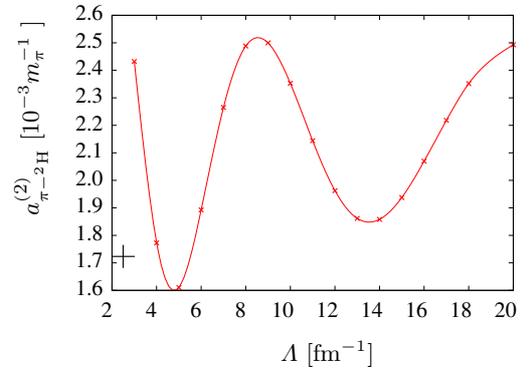}
      
  \caption{Contribution $a_{\pi- ^2 \rm H}^{\left(\mbox{\scriptsize \ref{fig:2N-triple}}\right)}$ due to amplitude 
                Eq.~(\ref{eq:m2triple}) to 
                the $\pi$-$^2$H scattering length depending on the cutoff $\Lambda$ of the LO potential. 
                  The result for the Nijmegen~93 is also shown ``(+)''.  }
  \label{fig:lamdepm2triple}
\end{figure}

We are now in the position to study 
 the $\Lambda$ dependence of the scattering length
contributions due to Eqs.~(\ref{eq:m2a}), (\ref{eq:m2bc}), and
(\ref{eq:m2triple}) based on the LO potential. For this, we have
adjusted the potential so that the deuteron binding energy is close to
the experimental one.  For the Coulombian diagram, the cutoff
dependence has already been studied in \cite{Nogga:2005fv} in momentum
space. This diagram and the triple scattering diagram, as given in Fig.~\ref{fig:2N-triple},
have also been
investigated in Refs.~\cite{Platter:2006pt,PavonValderrama:2006np}
with the result that, for wave functions based on the 1$\pi$--exchange
interaction, the results become independent of the cutoff in the limit
of large $\Lambda$.  In Figs. \ref{fig:lamdepm2a},
\ref{fig:lamdepm2bc}, and \ref{fig:lamdepm2triple}, this result is
seen once more. Note that the scales are entirely different in these figures. 
Additionally, we observe that the same is true for the
contribution due to Eq.~(\ref{eq:m2bc}). We observe a mild oscillation
of the result, the amplitude of which is approximately 5~\% of the
respective contribution --- $1 \times 10^{-3} \, m_{\pi}^{-1}$ in
absolute terms. From the comparison in Table~\ref{tab:partialmonte},
we observe that the leading order results are in good agreement with
the previous potential model results. This is also true for the
amplitude in Eq.~(\ref{eq:m2triple}). From this we conclude that a
theoretical accuracy of $1 \times 10^{-3} \, m_{\pi}^{-1}$ can be
reached at most from a study of pion-nucleus scattering,
which means that a 5\% accuracy can be reached.

As was already argued in Sec.~\ref{sec:powercount}, the dependence
of the few-body operators on the regulator used for the deuteron
wave function can give us an idea on the numerical size of the
leading counter term contribution. In Weinberg counting the leading isoscalar 
counter term is expected to be suppressed by a factor $(m_\pi/m_N)^2 \sim 2 \%$
compared to the numerically leading double scattering term.
This is fully in line 
 with the observed  amount of $\Lambda$ dependence, as described
in the previous paragraph, but in gross disagreement
to the expectations of $Q$--counting, where a counter term contribution
of the order of $0.2$ \% would be expected.
Thus, from the study of the cutoff dependence we conclude that
Weinberg counting provides the more accurate estimate for 
the leading counter term contribution.

\begin{table}
\centering
\caption{Values for $C_{S}$  depending on the cutoff 
               $\Lambda$.   For these fits, the experimental binding 
               energy $E_{d}=2.225$~MeV is reproduced. We fixed 
               $C_{T}=0$ in all cases.}
\label{tab:potlampara}       
\begin{tabular}{lr}
\hline\noalign{\smallskip}
$\Lambda$[fm$^{-1}$] &  $C_{S}$[GeV$^{-2}$]  \\
\noalign{\smallskip}\hline\noalign{\smallskip}
3.0 &       -34.2225   \\
4.0 &       48.1751  \\
5.0 &         562.089 \\
%6.0 &        -304.164 \\
%8.0 &        -103.476 \\
10.0 &      -50.9683 \\
%12.0 &      2.37148 \\
%14.0 &      170.776 \\
%16.0 &       -357.634 \\
%18.0 &      -140.205 \\
20.0 &      -92.1179  \\
\noalign{\smallskip}\hline
\end{tabular}
\end{table}

\subsection{Dependence of the $\pi$-$^2$H scattering length on the deuteron binding energy}
\label{sec:enerdep}

Following up on the discussion in Sec.~\ref{sec:powercount}, 
we now study the energy dependence of the ratios 
of few-nucleon contributions. If $Q$--counting were operative, the relative scaling of Diagrams 
Figs.~\ref{fig:2N-ops}(a) and \ref{fig:2N-ops}(b)+\ref{fig:2N-ops}(c) would strongly depend on the 
binding momentum. A straightforward analysis of Eqs.~(\ref{eq:m2a}) and
(\ref{eq:m2bc}) shows that 
Diagram $(a)$ of
Fig.~\ref{fig:2N-ops} scales as
$$
a_{\pi A}^{(1a)} \sim m_\pi^2/q^2 \ ,
$$
while the sum of Diagram $(b)$ and $(c)$ 
scales as
$$
a_{\pi A}^{(1bc)} \sim q^2/m_\pi^2 \ .
$$
Thus, for the binding energy dependence of the ratio of these two contributions
we find
\begin{equation}
\frac{a_{\pi A}^{(1bc)}}{a_{\pi A}^{(1a)}}
\propto  \left\{{\mbox{const. \ (Weinberg)}}\atop
{\hspace{0.5cm} {E_{d}}^2 \hspace{0.6cm}  \mbox{($Q$--counting)}}   \right. \ ,
\label{epsdepestimates1}
\end{equation}
where we used the $Q$--counting relation $q\sim \gamma$ together with Eq.~(\ref{gammarel}).
Analogously one gets
\begin{equation}
\frac{a_{\pi A}^{(2)}}{a_{\pi A}^{(1a)}}
\propto  \left\{{\mbox{const. \ (Weinberg)}}\atop
{\hspace{0.5cm} \sqrt{E_{d}} \hspace{0.6cm}  \mbox{($Q$--counting)}}   \right. \ ,
\label{epsdepestimates2}\end{equation}
using the explicit expression for the contributions of the
individual diagrams.

We have studied the binding energy 
dependence of both classes of diagrams based on the leading 
chiral NN interaction. To this aim, we have adjusted the contact interaction 
acting in the deuteron channel so that the deuteron was bound with 
unphysically large and small binding energy (see Table \ref{tab:potbepara}). 
It was then an easy exercise to 
calculate the contributions of Figs.~\ref{fig:2N-ops}(a), \ref{fig:2N-ops}(b)+\ref{fig:2N-ops}(c)
and \ref{fig:2N-triple} to the scattering length depending on the binding energy
and explicitly compare to the expectations from $Q$-- and Weinberg counting.

\begin{table}
\centering
\caption{Values for $C_{S}$  depending on the chosen binding energy of the 
               deuteron $E_{d}$. For these fits, we fixed 
               $C_{T}=0$ and $\Lambda = 20$~fm$^{-1}$.}
\label{tab:potbepara}       
\begin{tabular}{lr}
\hline\noalign{\smallskip}
$E_{d}$[MeV] &  $C_{S}$[GeV$^{-2}$]  \\
\noalign{\smallskip}\hline\noalign{\smallskip}
0.002 &    -71.7689   \\
0.005 &    -72.1400   \\
0.01 &      -72.5574  \\
0.02 &      -73.1460  \\
0.05 &      -74.3088  \\
0.1 &       -75.6126 \\
0.2 &      -77.4471 \\
0.5 &     -81.0703  \\
1.0 &     -85.1564  \\
5.0 &      -103.225 \\
10.0 &      -118.793 \\
20.0 &      -147.032\\
30.0 &  -178.149 \\
40.0 &  -217.147 \\
50.0 &  -271.056  \\
\noalign{\smallskip}\hline
\end{tabular}
\end{table}

\begin{figure*}[t]
  \centering

  \psfrag{ener}{$E_{d}$ [MeV]}
  \psfrag{ascatt}{$a_{\pi-^2 \rm H}^{\left(\mbox{\scriptsize \ref{fig:2N-ops}a}\right)}$ [$10^{-3} m_{\pi}^{-1}$ ]}
 \includegraphics[scale=0.55]{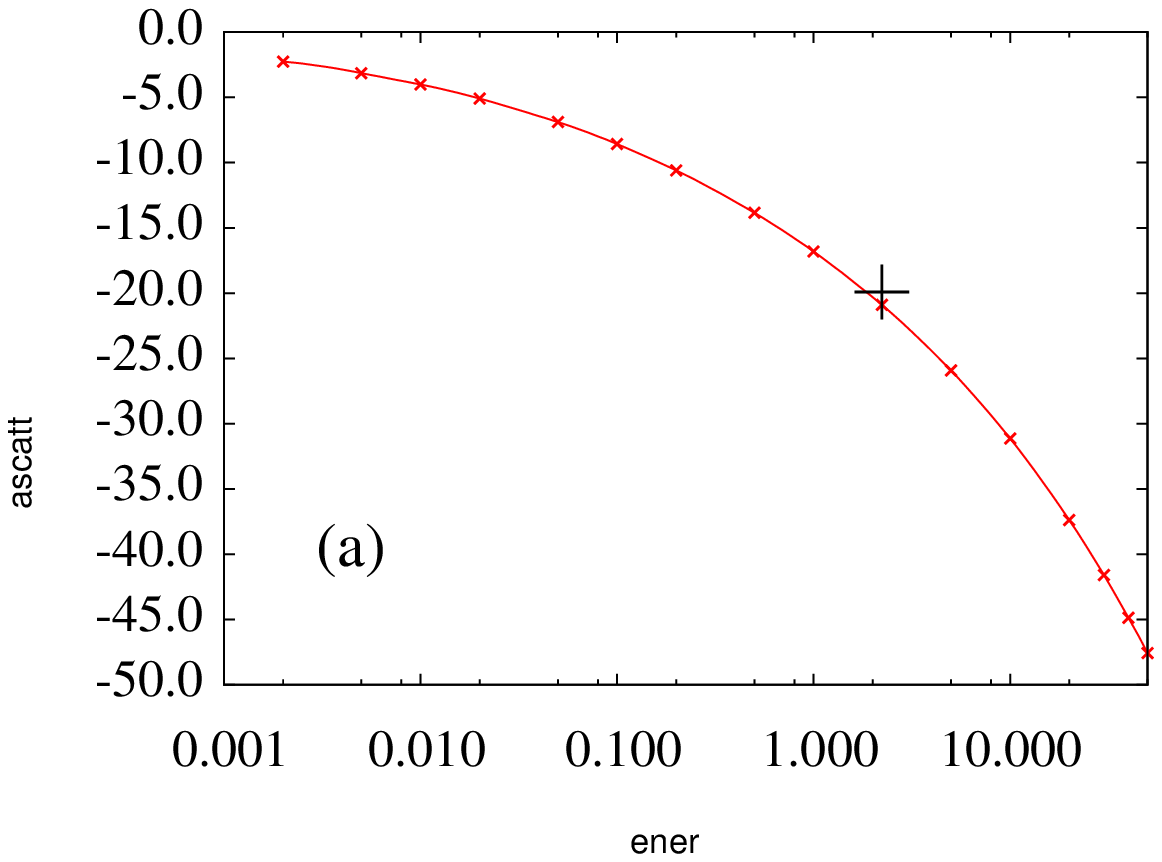}
 \psfrag{ener}{$E_{d}$ [MeV]}
 \psfrag{ascatt}{$a_{\pi-^2 \rm H}^{\left(\mbox{\scriptsize \ref{fig:2N-ops}bc}\right)}$ [$10^{-3} m_{\pi}^{-1}$ ]}   
  \includegraphics[scale=0.55]{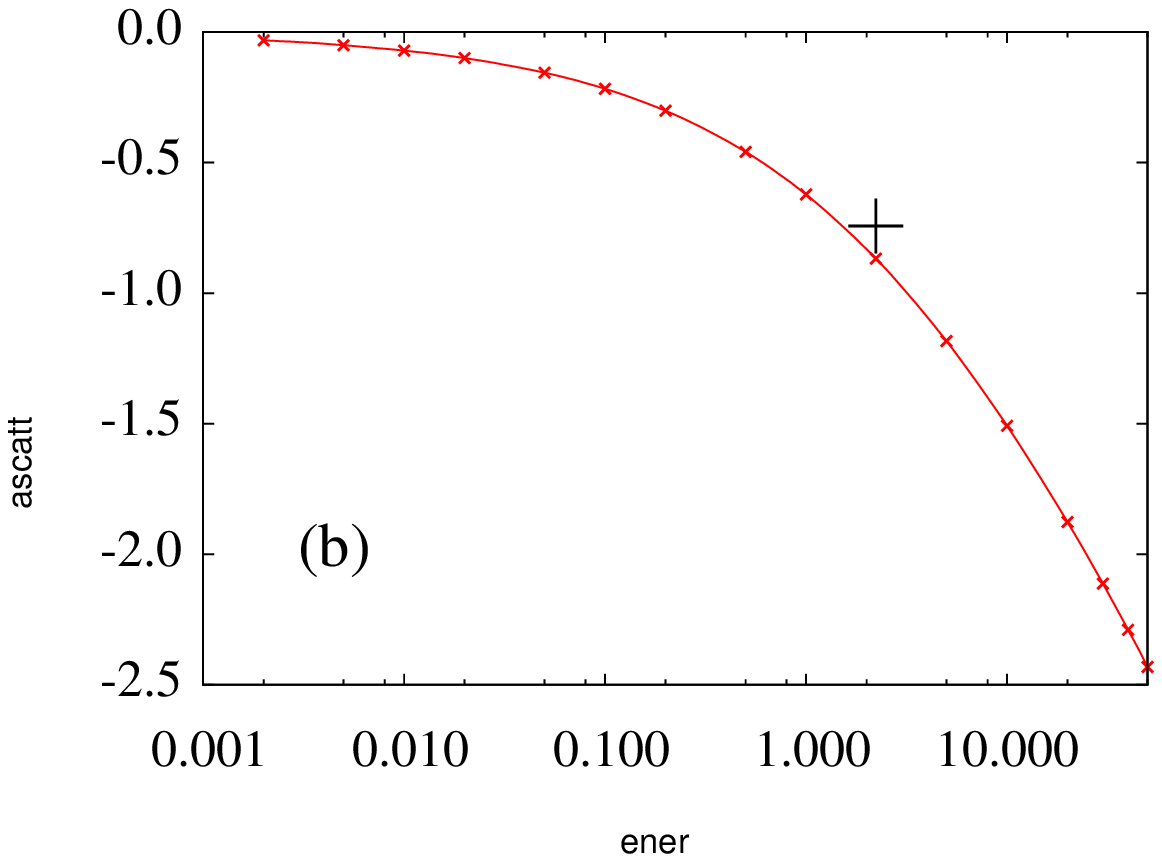}
  
 \psfrag{ener}{$E_{d}$ [MeV]}
  \psfrag{ascatt}{$a_{\pi-^2 \rm H}^{\left(\mbox{\scriptsize \ref{fig:2N-triple}}\right)}$ [$10^{-3} m_{\pi}^{-1}$ ]}
 \includegraphics[scale=0.55]{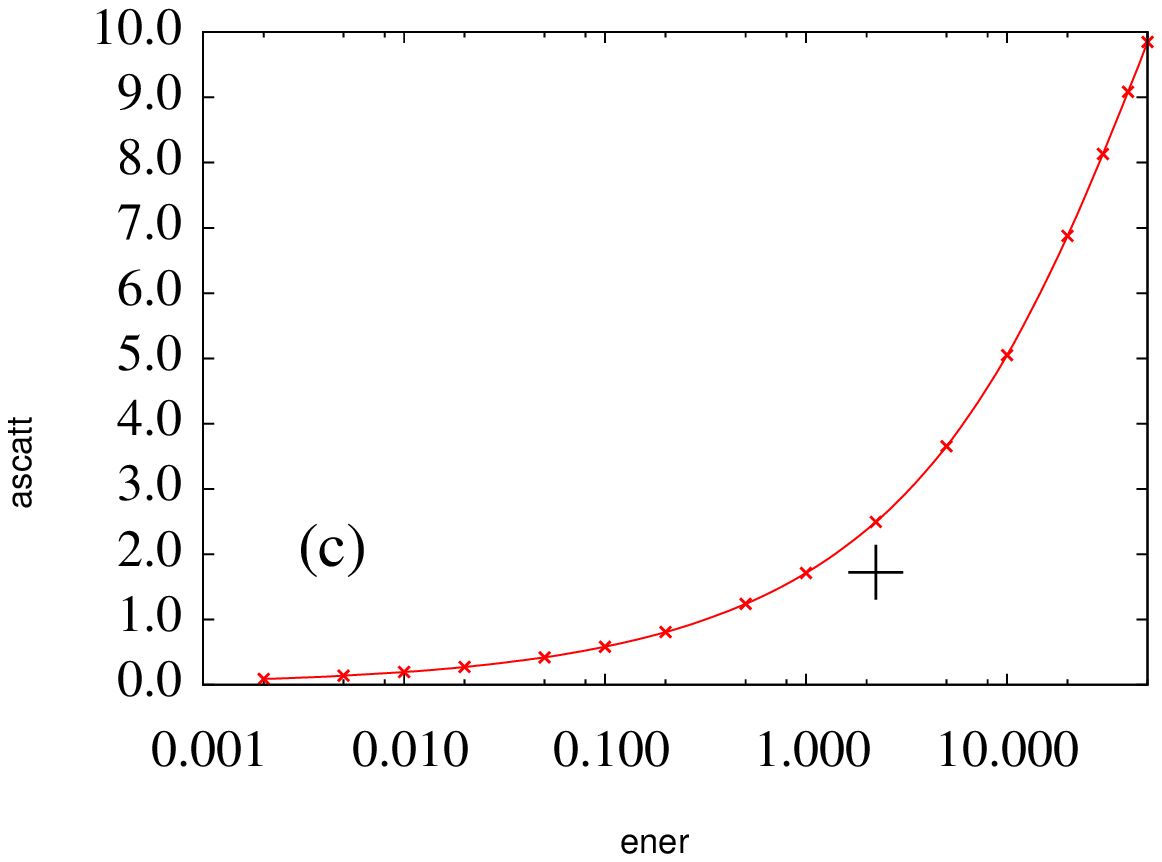}
 \psfrag{ener}{$E_{d}$ [MeV]}
  \psfrag{ratio}{$r$}  
  \includegraphics[scale=0.55]{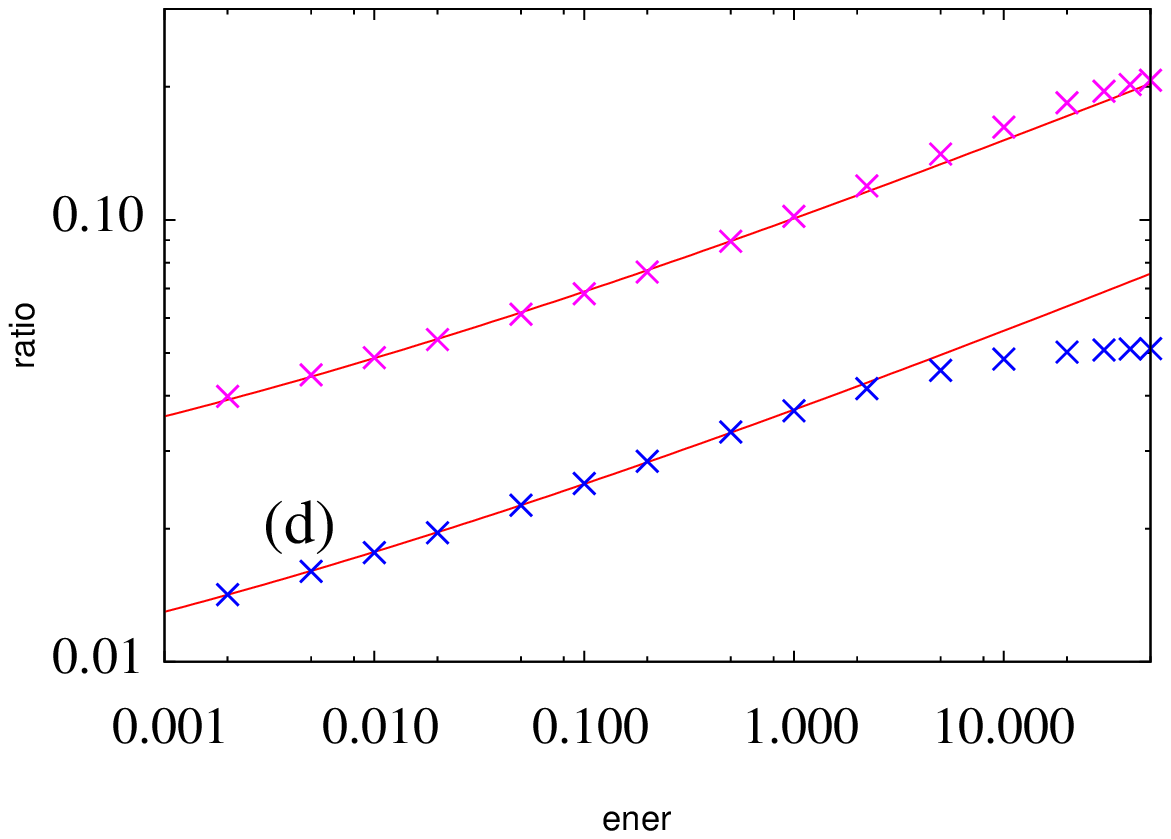}

  \caption{Contributions $a_{\pi-^2 \rm H}^{\left(\mbox{\scriptsize \ref{fig:2N-ops}a}\right)}$ (a), 
       $a_{\pi-^2 \rm H}^{\left(\mbox{\scriptsize \ref{fig:2N-ops}bc}\right)}$ (b), and 
       $a_{\pi-^2 \rm H}^{\left(\mbox{\scriptsize \ref{fig:2N-triple}}\right)}$ (c)   
  due to the amplitudes of Eqs.~(\ref{eq:m2a}), (\ref{eq:m2bc}), and (\ref{eq:m2triple}) to 
                the $\pi$-$^2$H scattering length depending on the binding 
               energy $E_{d}$ of $^2$H. 
                The results for Nijmegen~93 are also shown ``(+)''.  (d) shows 
                the ratios $r=\left| {a_{\pi-^2 \rm H}^{\left(\mbox{\scriptsize \ref{fig:2N-ops}bc}\right)}  \big/ 
                            a_{\pi-^2 \rm H}^{\left(\mbox{\scriptsize \ref{fig:2N-ops}a}\right)}  } \right| $ (lower ``x'')
                and  $r= \left| {a_{\pi-^2 \rm H}^{\left(\mbox{\scriptsize \ref{fig:2N-triple}}\right)}  \big/
                            a_{\pi-^2 \rm H}^{\left(\mbox{\scriptsize \ref{fig:2N-ops}a}\right)}  } \right| $ (upper ``x''). 
                            The lines are fits to the low energy results 
                             assuming that the ratio is $\propto E^{0.2}$. }

  \label{fig:enerdepm2}
 
\end{figure*}

Based on this observation, we are now in the position to look in more
detail at the dependence of the various contributions on the binding
energy. For this, we arbitrarily choose $\Lambda=20$~fm$^{-1}$, which
is in the region where the results are almost independent of the
cutoff. We start displaying the binding energy dependence of the
individual contributions to the scattering length in
Fig.~\ref{fig:enerdepm2}. We observe that, independent of the binding
energy, the contribution of the Coulombian diagram is the most
important two-nucleon contribution.  In contrast to the naive power
counting estimates, the amplitude of Eq.~(\ref{eq:m2triple}) --- 
the triple scattering diagram, depicted in Fig.~\ref{fig:2N-triple} ---  is the
next important one. Still, it is suppressed by one order of magnitude
compared to the leading Coulombian two-nucleon diagram.  The
contribution of this diagram will be discussed in detail in
Sec.~\ref{i0int}.  The amplitude Eq.~(\ref{eq:m2bc}) ---
shown in Fig.~\ref{fig:2N-ops}(b)+\ref{fig:2N-ops}(c) --- gives an
extraordinarily small shift of the scattering length. From the
observation that this suppression is not strongly depending  on the binding
energy, we conclude that this suppression is unrelated to the binding
momentum as suggested in Ref.~\cite{Beane:2002wk}, but probably
accidental.

In order to be more quantitative on the relative suppression of these contributions, 
we show in Fig.~\ref{fig:enerdepm2}(d) the ratios of the shifts of the scattering lengths due 
to Eqs.~(\ref{eq:m2bc}) and (\ref{eq:m2triple}) and the one due to 
Eq.~(\ref{eq:m2a}).  Based on $Q$--counting, the ratio for  Eq.~(\ref{eq:m2bc})
should scale like $E_{d}^2$ (c.f.
Eqs.~(\ref{epsdepestimates1}) for small energies, the one for Eq.~(\ref{eq:m2triple})
should scale like  $\sqrt{{E_{d}}}$ (c.f.
Eqs.~(\ref{epsdepestimates2})). 
Our explicit calculation, however,
 shows a much weaker dependence on the binding energy. This is in strong contrast 
to the expectation from $Q$--counting. 
Therefore, we conclude that $Q$--counting is not realized for pion scattering 
on light nuclei, i.e. $^2$H.

\begin{figure*}[t]
  \centering
  \psfrag{contact}[cr][cr]{contact}
  \psfrag{Hulthen}[cr][cr]{Hulth\'{e}n}
  \psfrag{lam=4}[cr][cr]{$\Lambda=4$ fm$^{-1}$}
  \psfrag{lam=10}[cr][cr]{$\Lambda=10$ fm$^{-1}$}
  \psfrag{lam=20}[cr][cr]{$\Lambda=20$ fm$^{-1}$}  
  \psfrag{u(r) [fm^{-1/2}]}{$u(r) [fm^{-1/2}]$}
  \psfrag{r [fm]}{$r$ [fm]}
 \includegraphics[scale=0.55]{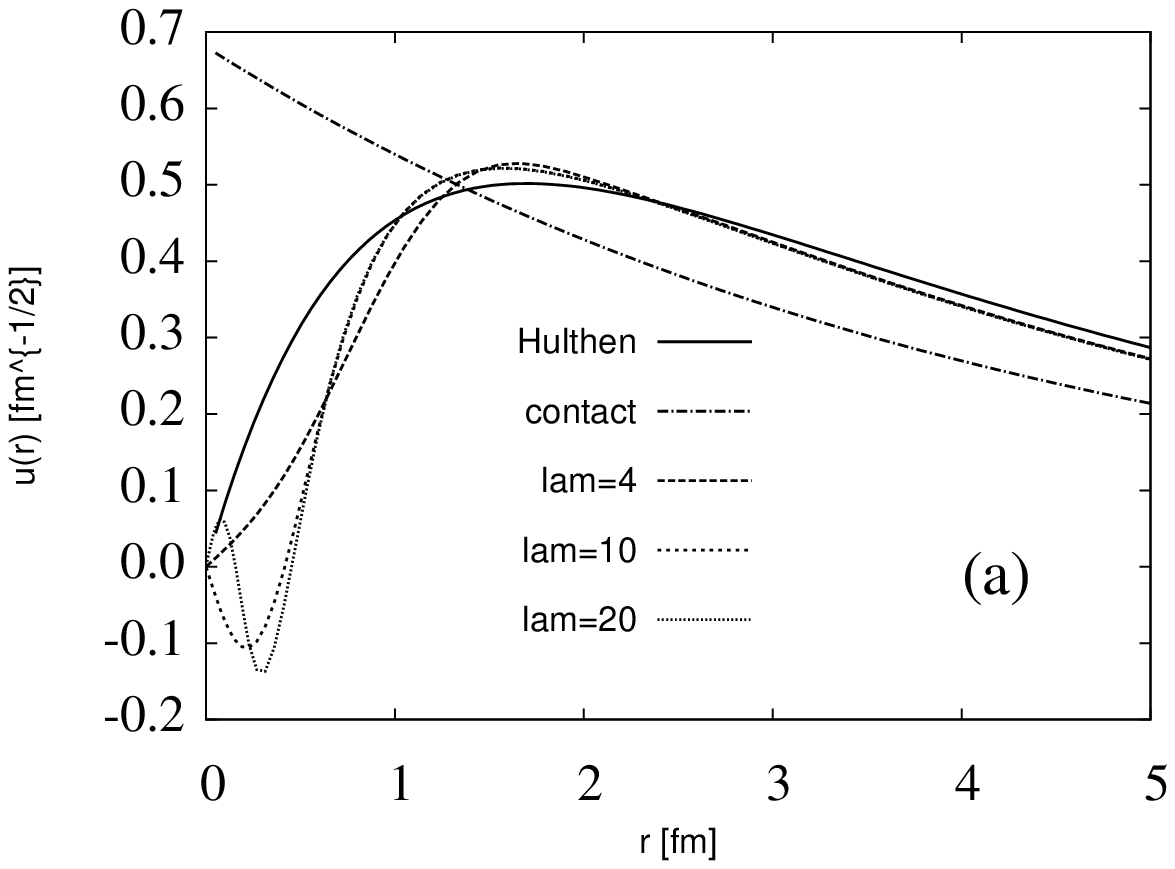}
  \psfrag{contact}[cr][cr]{contact}
   \psfrag{Hulthen}[cr][cr]{Hulth\'{e}n}
  \psfrag{E=0.01}[cr][cr]{$\Lambda=20$ fm$^{-1}$}
   \psfrag{u(r) [fm^{-1/2}]}{$u(r) [fm^{-1/2}]$}
   \psfrag{r [fm]}{$r$ [fm]}
 \includegraphics[scale=0.55]{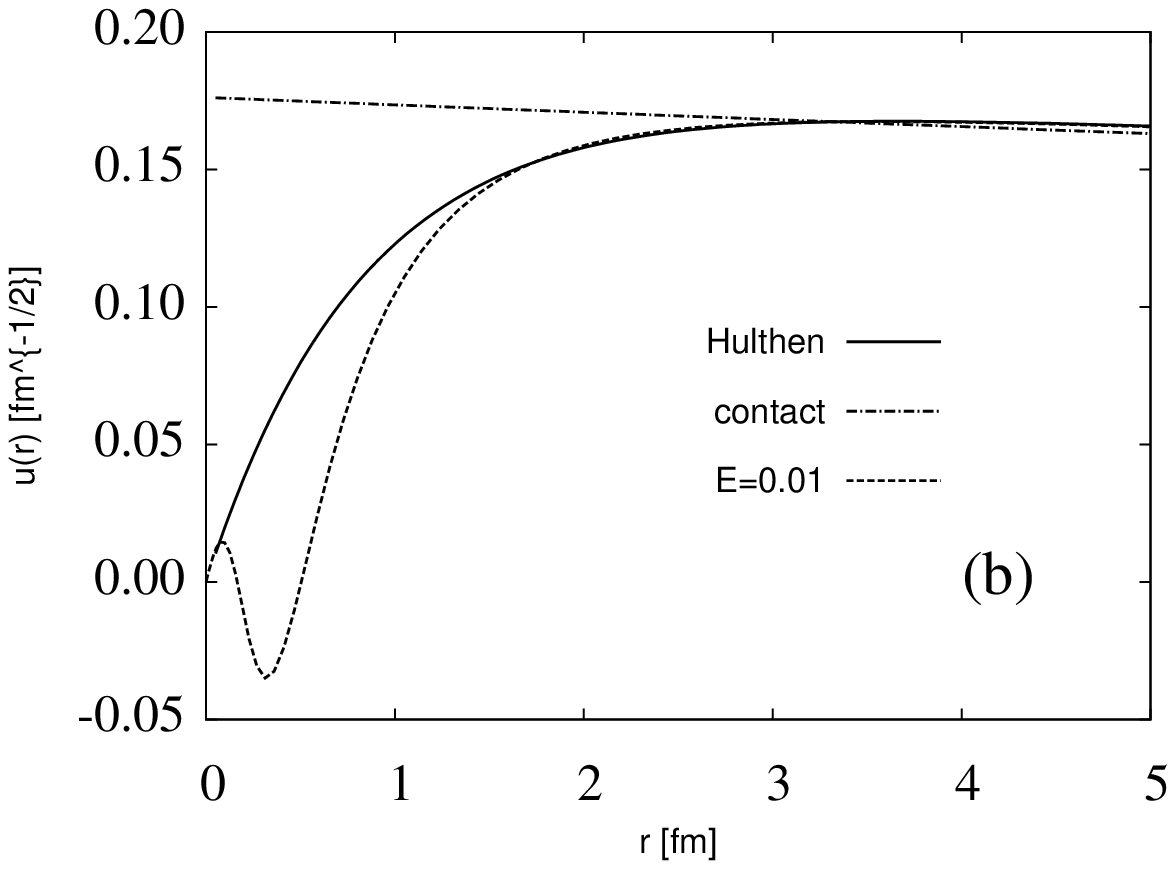} 
      
  \caption{Comparison of the Hulth\'{e}n deuteron wave function to 
               the $S$-wave part of several LO wave functions at the physical 
                deuteron binding energy (a) and  at $E_{d}=0.01$~MeV (b).                     
                The 
                Hulth\'{e}n wave function is shown for $\beta = 1.7 m_{\pi}$.
                We also show the deuteron wave functions based on contact 
                interactions ($\beta \to \infty$).}

   \label{fig:deutwf}
 
\end{figure*}

It is instructive to investigate further the source of the
residual binding energy dependence reported above. 
From this we will find that the ratios of contributions
to the scattering length has a logarithmic and not a 
power law dependence. 
In addition, we will be able to
show that the physical deuteron binding energy is
already beyond the
 range of applicability of heavy pion
effective field theory.
To this aim, we study the ratios within an analytical model of the 
deuteron wave function. 
To get there we start from pionless EFT to obtain the deuteron 
based on a zero range approximation of the NN interaction.
However, from different investigations it became clear that the binding
momentum cannot be the only scale affecting the wave functions of the
deuteron in a model independent way:
  $\pi$-$^2$H scattering at
threshold was studied based on the perturbative treatment of pions
in \cite{Borasoy:2003gf} and using 
heavy pion EFT in Refs.~\cite{Beane:2002aw,Meissner:2005bz} 
with the result that a counter
term is required in leading order where two-nucleon diagrams
contribute. In practice, this would imply that an extraction of
$a^{(+)}$ is unfeasible based on $\pi$-$^2$H atoms.  However, it was
realized that this problem is tamed once pions are treated
non-perturbatively
\cite{Nogga:2005fv,Platter:2006pt,%
PavonValderrama:2006np,PavonValderrama:2005gu}.
In this case, the short distance $^2$H wave function is affected by
1$\pi$--exchange in such a way, that counter terms are not required
to obtain cutoff independent results. No contradiction to the naive
power counting of Weinberg is observed.  Obviously, the pion
introduced scales into the wave function beyond the binding momentum. Therefore, we 
add a range factor to the vertex function, so that we are able to 
introduce the intrinsic
non-perturbative scale that enters through pion exchange.
In momentum space, the wave function
then reads
\begin{equation}
\Psi (p) = N(\gamma,\beta)\frac{1}{\vec p\, ^2 + \beta^2}\frac{1}{\vec p\, ^2 + \gamma^2} \ ,
\label{eq:hulthen}
\end{equation}
where the normalization factor $N$ is fixed to
$$
N(\gamma,\beta)^2 = 8\pi \gamma \beta (\gamma + \beta)^{3}
$$
by the normalization condition for the deuteron wave function.
This is a wave function of the Hulth\'{e}n type \cite{Hulthen:1957hb}.
In Refs.~\cite{PavonValderrama:2005gu,Nogga:2005fv,%
Valderrama:2007ja}, it was shown that for radii
larger than 0.6 fm the LO chiral wave functions are
basically independent of the regulator used for their construction.
We therefore fix $\beta$ by fitting to the tail of the wave function
at $E_{d}=0.01$~MeV.  This
way we find
\begin{equation}
\beta = 1.7 m_\pi \ .
\label{eq:betavalue}
\end{equation}

It is reassuring that $\beta$ turns out to be of the 
order of the pion mass. If the proposed picture is
correct, the scale $\beta$ should be independent
of the deuteron binding energy. We confirmed that
this is indeed the case as long as we do not go to very large 
binding energies above the physical one. Fig.~\ref{fig:deutwf} 
shows the $S$-wave deuteron wave functions 
compared to the Hulth\'{e}n ansatz. For both binding energies, we use 
the same $\beta$. As one can see, the Hulth\'{e}n and chiral wave functions 
nicely agree for larger distances. In Fig.~\ref{fig:deutwf}(a) one can 
also see that this long range part is independent of the cutoff used. 

The limit of a point like vertex is achieved 
by the limit $\beta\to \infty$. In this limit
the wave function of Eq.~(\ref{eq:hulthen})
becomes identical to the one used in theories
with perturbative pions~\cite{Borasoy:2003gf} as well
as the one used when treating the pion as heavy field~\cite{Beane:2002aw,Meissner:2005bz}.
For completeness, we also show this wave function in Fig.~\ref{fig:deutwf}(a). 
It is obvious that such a simplistic wave function is not a good approximation 
to complete chiral wave functions. The most important effect is a reduction of the 
long range part, which becomes necessary to insure the correct normalization 
of the wave function.

%\begin{figure}[t]
%  \centering
%   \psfrag{Hulthen}[cr][cr]{Hulth\'{e}n}
%   \psfrag{swave}[cr][cr]{LO (only $S$-wave)}
%   \psfrag{ascatt}{$a^{\left(\mbox{\scriptsize \ref{fig:2N-ops}a}\right)}$ [$10^{-3}\, m_{\pi}^{-1}$]}
%   \psfrag{ed}{$E_{d}$ [MeV]}
%  \includegraphics[scale=0.55]{hulthen/aa.eps}
%      
%  \caption{Comparison of $a^{\left(\mbox{\scriptsize \ref{fig:2N-ops}a}\right)}$ obtained from the Hulth\'{e}n deuteron wave function and 
%                the $S$-wave part of the LO wave 
%                function for $\Lambda=20$~fm$^{-1}$ depending on the 
%                deuteron binding energy $E_{d}$. The
%                Hulth\'{e}n result is shown for $\beta = 1.7 m_{\pi}$}
%  \label{fig:aaedepanaly}
%\end{figure}

%\begin{figure}[t]
%  \centering
%   \psfrag{Hulthen}[cr][cr]{Hulth\'{e}n}
%   \psfrag{swave}[cr][cr]{LO (only $S$-wave)}
%   \psfrag{ascatt}{$a^{\left(\mbox{\scriptsize \ref{fig:2N-ops}bc}\right)}$ [$10^{-3}\, m_{\pi}^{-1}$]}
%   \psfrag{ed}{$E_{d}$ [MeV]}
%  \includegraphics[scale=0.55]{hulthen/abc.eps}
%      
%  \caption{Comparison of $a^{\left(\mbox{\scriptsize \ref{fig:2N-ops}bc}\right)}$ obtained from the Hulth\'{e}n deuteron wave function and 
%                the $S$-wave part of the LO wave 
%                function for $\Lambda=20$~fm$^{-1}$ depending on the 
%                deuteron binding energy $E_{d}$. The
%                Hulth\'{e}n result is shown for $\beta = 1.7 m_{\pi}$}
%  \label{fig:abcedepanaly}
%\end{figure}

\begin{figure*}[t]
  \centering
   \psfrag{Hulthen}[cr][cr]{Hulth\'{e}n}
   \psfrag{swave}[cr][cr]{LO (only $S$-wave)}
   \psfrag{ascatt}{r}
   \psfrag{ed}[cr][rr]{$E_{d}$ [MeV]}
  \includegraphics[scale=0.55]{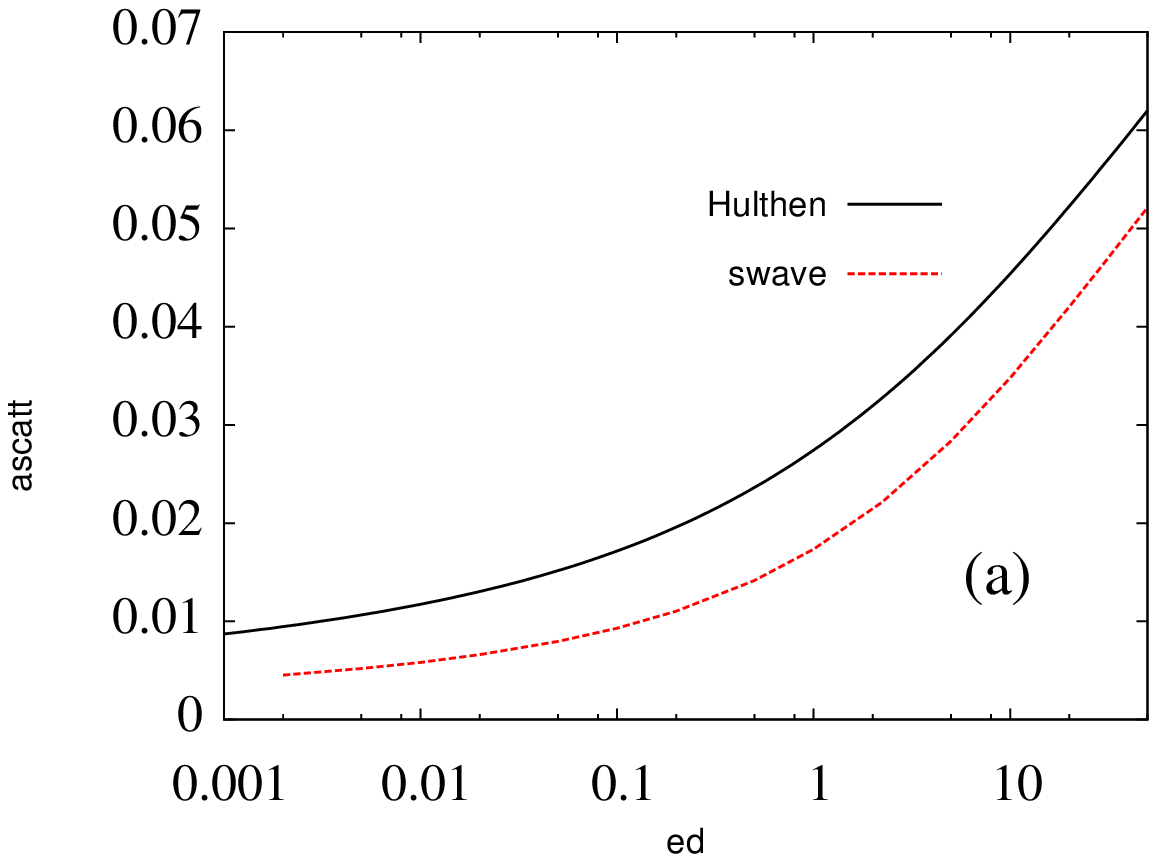}
  \psfrag{Hulthen}[cr][cr]{Hulth\'{e}n}
   \psfrag{expan}[cr][cr]{expansion of Hulth\'{e}n}
   \psfrag{swave}[cr][cr]{LO (only $S$-wave)}
  \psfrag{ascatt}{$a_{\pi-^2 \rm H}^{\left(\mbox{\scriptsize \ref{fig:2N-ops}a}\right)}$ [$10^{-3} m_{\pi}^{-1}$ ]}
   \psfrag{ed}[cr][rr]{$E_{d}$ [MeV]}
  \includegraphics[scale=0.55]{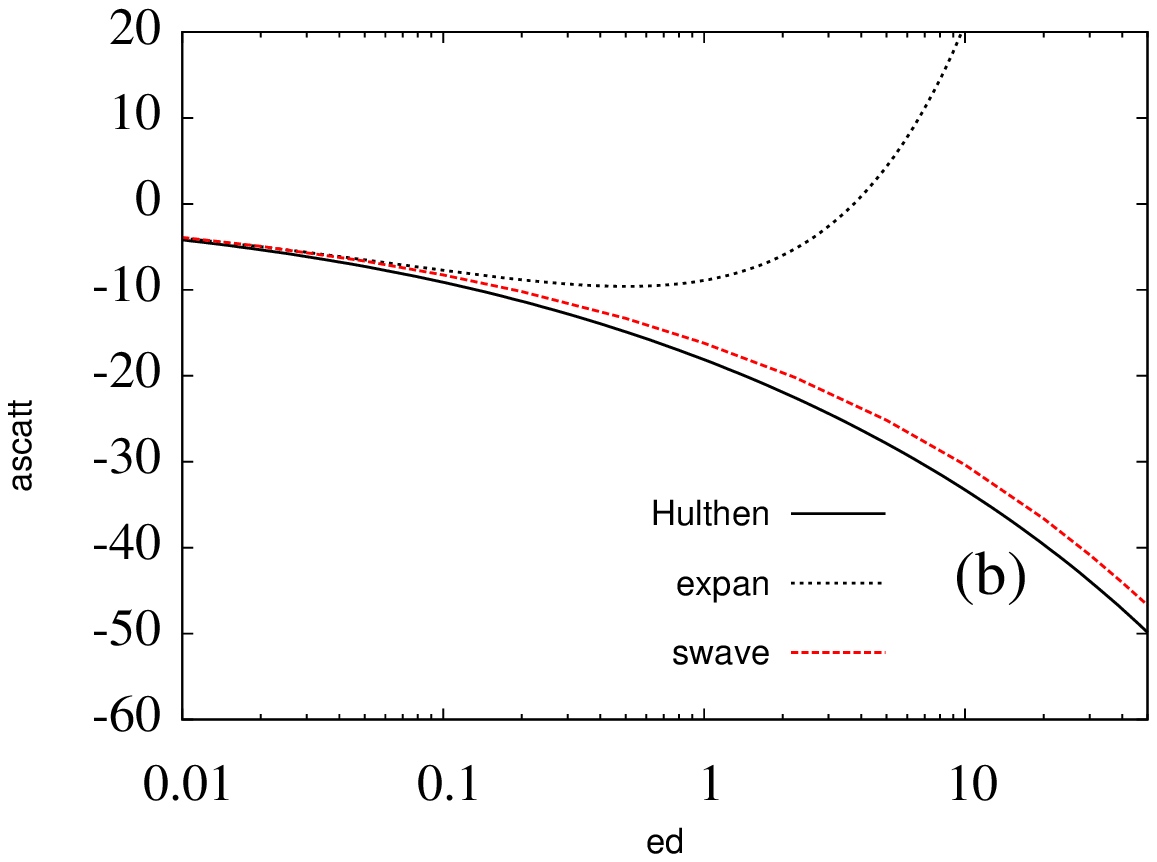}

 \caption{Comparison of the ratio 
    $r=\left| { a_{\pi-^2 \rm H}^{\left(\mbox{\scriptsize \ref{fig:2N-ops}bc}\right)}
    / a_{\pi-^2 \rm H}^{\left(\mbox{\scriptsize \ref{fig:2N-ops}a}\right)}}  \right| $  (a) 
    and of  $ a_{\pi-^2 \rm H}^{\left(\mbox{\scriptsize \ref{fig:2N-ops}a}\right)} $ (b) 
     obtained from the Hulth\'{e}n deuteron wave function and 
                the $S$-wave part of the LO wave 
                function for $\Lambda=20$~fm$^{-1}$ depending on the 
                deuteron binding energy $E_{d}$. The
               Hulth\'{e}n result is shown for $\beta = 1.7 m_{\pi}$. In (b) the results 
                are also compared to the expansion of Eq.~(\ref{eq:aaanalyt_lo}).}
                
  \label{fig:edepanaly}                        

\end{figure*}

The wave function of Eq.~(\ref{eq:hulthen}) is still
sufficiently simple that analytic calculations can be
performed for the various matrix elements discussed in
this section. In particular we find
\begin{eqnarray}
\label{eq:aaanalyt}
a_{\pi -^2{\rm H}}^{(1a)}= \kappa \, x \, \frac{(1+x)}{(1-x)^2}
\ln \left(\frac{4x}{(1+x)^2}\right) \ ,
\end{eqnarray}
where we introduced the dimensionless parameter $x=\gamma/\beta$
and
$$
\kappa = \beta\frac{1}{8\pi^2\left(1+m_\pi/2m_N\right)}\frac{m_\pi^2}{f_\pi^4} \ .
$$
In addition, we find
\begin{eqnarray} \nonumber
\label{eq:abcanalyt}
a_{\pi -^2{\rm H}}^{(1bc)}&=& \kappa \, x \, \frac{(1+x)}{(1-x)^2}\frac{g_A^2}{12} \\ \nonumber
&\times&  \left\{ \ln \left(\frac{(1+x+\bar m_\pi)^2}{(2x+\bar m_\pi)(2+\bar m_\pi)}\right) \right. \\
& &    \left.- \frac{\bar m_\pi (1-x)^2}{(1+x+\bar m_\pi)(2x+\bar m_\pi)(2+\bar m_\pi)}
\right\} \ ,
\end{eqnarray}
with $\bar m_\pi=m_\pi/\beta$ 
and
\begin{eqnarray} \nonumber
\label{eq:a2analytb} 
a_{\pi -^2{\rm H}}^{(2)}&=& \kappa \, x \, \frac{(1+x)}{(1-x)^2}
\\
&\times&
\left(\frac{m_\pi \beta}{4\pi f_\pi^2}\right)
 \left\{ \ln \left(\frac{2}{x+1}\right) 
+x \ln \left(\frac{2x}{x+1}\right)\right\} \ ,
\end{eqnarray}

We used these analytic results, found with the Hulth\'{e}n wave
functions, 
to predict the ratio 
\begin{equation}
r=\left| { a_{\pi-^2 \rm H}^{\left(\mbox{\scriptsize \ref{fig:2N-ops}bc}\right)}
    \over a_{\pi-^2 \rm H}^{\left(\mbox{\scriptsize \ref{fig:2N-ops}a}\right)}}  \right| .
\end{equation}    
In Fig.~\ref{fig:edepanaly}(a), this analytical result is compared 
to a LO calculation for which we neglect the $D$-wave 
contribution of the wave function. Although the results do not 
agree quantitatively, it is obvious that the simplified calculation 
based on the Hulth\'{e}n ansatz is able to describe the energy dependence qualitatively.
The inclusion of the $D$-wave changes the result
for Diagram \ref{fig:2N-ops}(a) only marginally, however, the
full contribution for Diagram \ref{fig:2N-ops}(b)+\ref{fig:2N-ops}(c) even changes its sign.
Let us now focus on the contributions from only the deuteron $S$-wave.
From Eqs.~(\ref{eq:aaanalyt}) and (\ref{eq:abcanalyt})
it follows directly that the suppression of Diagram
\ref{fig:2N-ops}(b)+\ref{fig:2N-ops}(c) compared to the Coulombian \ref{fig:2N-ops}(a) is
only logarithmic, in line with Weinberg counting,
and not power law ($\gamma^4$) as predicted by Q--counting.
Close inspection reveals that the bulk of
the suppression of (\ref{fig:2N-ops}$bc$) with respect to 
(\ref{fig:2N-ops}$a$) comes from spin-isospin factors leading to the factor of
$1/12$ in $a_{\pi -^2{\rm H}}^{(1bc)}$. Such kind
of accidental suppression, a power counting can not capture.

It is also very interesting to investigate various limits
of  Eqs.~(\ref{eq:aaanalyt}) and (\ref{eq:abcanalyt}).
As mentioned above the expressions for the wave functions 
relevant for theories with perturbative pions, as used
e.g. in Ref.~\cite{Borasoy:2003gf}, are recovered when
taking the limit $\beta\to \infty$. Then the above
expressions collaps to
\begin{eqnarray}
\label{eq:aaanalyt_lo}
a_{\pi -^2{\rm H}, \mbox{LO}}^{(1a)}= \kappa \, x \, 
\ln (4x) \ ,
\end{eqnarray}
and
\begin{eqnarray} \nonumber
\label{eq:abcanalyt_lo}
a_{\pi  -^2{\rm H}, \mbox{LO}}^{(\rm 1bc)}&=& -\kappa \, x \, \frac{g_A^2}{12} \\ 
%\nonumber
&\times&  \left\{ \ln \left(2(2x+\bar m_\pi)\right)
% \right. \\
%& &    \left.
+ \frac{\bar m_\pi}{(2x+\bar m_\pi)}
\right\} \ .
\end{eqnarray}
Identifying $\beta/2$ with $\Lambda^*$, the given equations
agree with those of Ref.~\cite{Borasoy:2003gf}.
In addition we find
\begin{eqnarray} \nonumber
\label{eq:a2analyta} 
a_{\pi -^2{\rm H}, \mbox{LO}}^{(2)}&=& \kappa \, x \left(\frac{m_\pi \beta}{4\pi f_\pi^2}\right)
 \ln \left(2\right) 
 \ .
\end{eqnarray}
Contrary to the theory with perturbative pions, in the heavy pion effective field
theory the Diagrams \ref{fig:2N-ops}(b)+\ref{fig:2N-ops}(c) do not appear explicitly but are 
absorbed in local counter terms.
A comparison of the given leading order expressions with
the full result reveals that, for Diagrams \ref{fig:2N-ops}(b)+\ref{fig:2N-ops}(c), there
is no regime of binding energies where the given formula
well represents the full result. Contrary to this,
the leading order expression for Diagram \ref{fig:2N-ops}(a) works
well for very small values of the binding energy 
(see Fig.~\ref{fig:edepanaly}(b)). This
is to be expected since there must be a kinematical regime
where the heavy pion effective field theory is applicable. However,
we find that the expressions break down already for very low
values of the binding energy. Already for a binding energy
of 0.1 MeV the leading order expressions show a significant
deviation from the full result. We therefore conclude that
for accurate calculations of $\pi$-$^2$H scattering neither
the treatment with perturbative pions nor heavy pion effective
field theory are applicable for the physical deuteron.

\subsection{Analysis of the triple scattering diagram}
\label{i0int}

The observation that the diagram shown in
Fig.~\ref{fig:2N-triple} is significantly enhanced
compared to the expectation based on Weinberg
counting was taken as a further support for
$Q$--counting~\cite{Beane:2002wk}.
In this subsection, we will demonstrate that
the triple scattering diagram is not enhanced
as a result of the smallness of the deuteron
binding momentum but because of the special
topology of its loop diagram.

The loop that appears in the triple scattering diagram
may be written as
\begin{eqnarray}\nonumber
I_0(\omega, v \cdot Q, Q^2) &=&
\frac1{i}\int \frac{d^dl}{(2\pi)^d}\frac1{(v\cdot l-\omega-i\epsilon)} \\
& & \hspace{-1.5cm} \times \frac1{(m_\pi^2-l^2-i\epsilon)(m_\pi^2-(l-Q)^2-i\epsilon)} \ ,
\end{eqnarray}
where $v=(1,0,0,0)$.
In Ref.~\cite{Dmitrasinovic:1999cu} a general solution for this integral
is given. In the kinematics relevant for $\pi$-$A$ scattering at threshold
we find
\begin{eqnarray}
\label{i0res}
I_0(m_\pi, 0 , -\vec{q}^2) = \frac1{8|\vec{q}|}+\delta I_0  ,
\end{eqnarray}
with
\begin{equation}
\delta I_0 = \frac1{8\pi^2|\vec{q}|}\int_0^\pi dx\left(
 \arctan \left(\frac{2m_\pi}{\sin (x)|\vec{q}| }\right) -\frac{\pi}{2}\right) \ .
\end{equation}
In Ref.~\cite{Beane:2002wk} only the first term on the right hand
side was included.
Note that in the limit of heavy pions,
as used in Ref.~\cite{Meissner:2005bz}, the contribution 
of $\delta I_0$ vanishes.
Dimensional analysis, which is the basis of
Weinberg counting, allows us to estimate
integrals. In case of $I_0$ this analysis gives
(assuming $q\sim m_\pi$)
\begin{equation}
I_0 \sim 1/((4\pi)^2 m_\pi) \ .
\label{weinestimate}
\end{equation}
Clearly, the first term on the right hand side of Eq.~(\ref{i0res})
is enhanced by a factor of $2\pi^2$ compared to this estimate. 
The remainder, $\delta I_0$, on the other hand is numerically
fully in line with the estimate (\ref{weinestimate}). The
power counting can only capture parametric suppressions. 
We therefore conclude that the fact that $\delta I_0$ behaves
in accordance with Weinbergs power counting is a further
support for its applicability. However, in a high accuracy
calculation for pion-nucleus scattering the first
term of Eq.~(\ref{i0res}) is to be kept and it is this
piece that we have in the list of operators to 
be included in the calculation --- c.f. Eq.~(\ref{eq:m2triple}).

How can we understand the large enhancement of
a part of $I_0$? To see this, we first observe that
the enhanced part of $I_0$ can be directly calculated
from the full Feynman integral by only keeping the
term that corresponds to the nucleon pole. For this
piece we get
\begin{eqnarray} \nonumber
I_0^{\rm nucl. \ pole} &=& \int \frac{d^3l}{(2\pi)^3}
\frac1{\vec{l}^2 (\vec{l}-\vec{q})^2} \\
&=& \frac1{8\pi^2 |\vec{q}|}
\int_0^\infty \frac{dx}{x}\ln \left( \left(\frac{x+1}{x-1}\right)^2\right) \ .
\end{eqnarray}
The implicit assumption behind any dimensional
analysis is that integrals, once converted into
dimensionless expressions, are of order 1. And
indeed, if the given integral were of order 1, the
full expression would be perfectly in line with power
counting. However, one finds
$$
\int_0^\infty \frac{dx}{x}\ln \left( \left(\frac{x+1}{x-1}\right)^2\right) = \pi^2 \ .
$$
We trace the appearance of this large result to 
the presence of an integrable singularity at $x=1$ in
the expression given above. Indeed, 80 \% of
the exact result originate from values of $x\le 2$.

It is interesting to note that enhancements
by factors of $\pi$ were already observed to emerge
also in pion loop contributions to the NN potential~\cite{Friar:2003yv},
the scalar nucleon form--factor~\cite{Becher:1999he}
and the $\pi^0$ photoproduction amplitude~\cite{Bernard:1991rt}
from similar topologies as those discussed here.
A deeper understanding, when these dimensionless factors
appear, would be very desireable.

It is also important to note that integrals
of the same topology are also relevant for the reaction
${\rm NN} \to {\rm NN}\pi$, though in different kinematics.
For a discussion of how to treat these large momentum
transfer reactions in ChPT, see Refs.~\cite{Hanhart:2000gp,Hanhart:2003pg}.
Loops for this reaction were studied in 
Refs.~\cite{Dmitrasinovic:1999cu,Hanhart:2002bu,Lensky:2005jc,Kim:2008qha}.
It is conceivable  that the problems in understanding
quantitatively especially the reaction $pp\to pp\pi^0$ 
are connected to the same enhancement of loops as discussed
in this section. 

\section{Few-nucleon wave functions}
\label{sec:wavefu}

\begin{table}
\centering
\caption{Values for $C_{S}$ and $C_{T}$ depending on the cutoff 
               $\Lambda$ of the LO potential.  }
\label{tab:potcomplpara}       
\begin{tabular}{lrr}
\hline\noalign{\smallskip}
$\Lambda$[fm$^{-1}$] &  $C_{S}$[GeV$^{-2}$] &  $C_{T}$[GeV$^{-2}$]  \\
\noalign{\smallskip}\hline\noalign{\smallskip}
2.0 &         -83.6941  &    2.63787   \\
3.0 &         -29.0931  &    16.3942 \\
4.0 &          86.3303     &    52.6427  \\
5.0 &         -435.354   &   -122.611  \\
%6.0 &        -146.303   &    -27.1868 \\
%8.0 &        -75.6971   &    -4.84401  \\
10.0 &      -39.8356   &   6.36715   \\
%12.0 &       20.8571     &  26.0876  \\
%14.0 &     -970.194    & -304.637   \\
%16.0 &     -131.357   & -25.3109   \\
%18.0 &      -87.2204    & -10.8255   \\
20.0 &      -66.2861    & -4.03158    \\
\noalign{\smallskip}\hline
\end{tabular}
\end{table}

Before we can give results for the scattering lengths, we need to specify the 
input going into our calculations of the few-nucleon wave functions in more 
detail. 
We will again study the results for LO wave functions. In contrast to $^2$H, we restrict 
to interactions in $^3$S$_{1}$-$^3$D$_{1}$ and $^1$S$_{0}$ partial 
waves requiring fits of the strength of both contact interactions, $C_{S}$ and 
$C_{T}$.  We determined these constants fitting to the binding energy of 
$^2$H and the $^1$S$_{0}$  neutron-proton phase shift at a laboratory 
energy of 1~MeV. The potential was given already in Eq.~(\ref{eq:lopot}). 
For the new fits, we used for historical reasons a different, more sharp 
regulator 
\begin{equation}
f(\vec p) = \exp \left( - \left( { p \over \Lambda } \right)^8  \right)  \ ,
\end{equation}
$g_{A}=1.29$ (utilizing the Goldberger-Treiman relation \cite{Goldberger:1958tr}) and $m_{\pi}=138.0$~MeV. 
The results for $C_{S}$ and $C_{T}$
are summarized in Table~\ref{tab:potcomplpara}. 

\begin{table}
\centering
\caption{Summary of the $^3$He and $^4$He binding energy results for 
              the LO, NLO, and N$^2$LO chiral interactions, AV18 and CD-Bonn. 
              For the LO interaction, the cutoff $\Lambda$ is given in [fm$^{-1}$]. 
              For the chiral interaction, the Lippmann-Schwinger cutoff $\Lambda$ 
              and spectral function cutoff $\tilde \Lambda$ is given in MeV  
              \cite{Epelbaum:2004fk}.
              The binding energies are given in MeV. For $^{4}$He, we have not 
              performed calculations for all cutoffs in LO. }
\label{tab:3n4nbind}       
\begin{tabular}{llrr}
\hline\noalign{\smallskip}
 & $\Lambda$ / $\tilde \Lambda$ &  $B(^3{\rm He})$ &  $B(^4{\rm He})$  \\
\noalign{\smallskip}\hline\noalign{\smallskip}
LO & 2.0 / -- &          11.042   & 39.88   \\
LO & 3.0 / -- &          6.878   & 20.25   \\
LO & 4.0 / -- &          6.068    & 17.08    \\
LO & 5.0 / -- &      5.987     & 16.48    \\
%LO & 6.0 / -- &      5.987     & 16.38   \\
%LO & 8.0 / -- &      5.810    & --- \\
LO & 10.0 / -- &    5.611    & 15.05     \\
%LO & 12.0 / -- &    5.530      & --- \\
%LO & 14.0 / -- &   5.486      & --- \\
%LO & 16.0 / -- &   5.459     & --- \\
%LO & 18.0 / -- &   5.441      & ---  \\
LO & 20.0 / -- &   5.429    & --- \\
\noalign{\smallskip}\hline
NLO &  400/500          &   7.678   &   28.57  \\
NLO &  550/500         &    6.991   &   24.38  \\
NLO &  550/600          &   7.051   &   24.72    \\
NLO &  400/700          &   7.699   &   28.77  \\
NLO &  550/700         &    7.090   &   24.94   \\
\noalign{\smallskip}\hline
N$^2$LO & 450/500           &   7.717    & 28.04  \\
N$^2$LO & 600/500           &   7.740    & 28.11  \\
N$^2$LO & 550/600           &   7.722    & 28.28  \\
N$^2$LO & 450/700           &   7.726    & 27.65  \\
N$^2$LO & 600/700           &   7.808    & 28.57  \\
\noalign{\smallskip}\hline
CD-Bonn & ---        &    7.719     & 28.28  \\ 
AV18       &  ---       &    7.736     & 28.36  \\
\noalign{\smallskip}\hline
Expt.       &  ---       &  7.718 & 28.30  \\
\noalign{\smallskip}\hline
\end{tabular}
\end{table}

Based on these fits, it is a straightforward task to calculate 
the binding energies of $^3$He and $^4$He. To this aim, we have 
solved Faddeev/Yakubovsky equations in momentum space following 
\cite{Nogga:2001cz}. Thereby, we subtracted the poles 
of unphysical spurious NN bound states from the 2N $t$-matrix
as outlined in \cite{Nogga:2005hy}. As was already shown 
in the same reference, the binding energies become rather independent of 
$\Lambda$ for large $\Lambda$. However, in LO, the binding energies 
of $^3$He and $^4$He do not well reproduce the experimental values. 
Table~\ref{tab:3n4nbind} shows our results for the various NN potentials 
used. In LO, 
the $^3$He binding energy is varying for the different cutoffs by almost 6~MeV. 
Such a large variation can be expected in low orders, since 
the binding energies are specifically sensitive to changes 
of the potential \cite{Nogga:2006ir}. Note that we did not include the 
Coulomb interaction in these LO calculations, whereas 
we did include the Coulomb interaction for the other orders and the 
phenomenological NN forces. 
In NLO, for a smaller 
range of cutoffs, the variation is reduced, but still visible. In NLO, again, there is an 
appreciable deviation from the experimental values. It is only in N$^{2}$LO, that 
three-nucleon forces (3NF's) contribute  \cite{vanKolck:1994yi,Epelbaum:2002vt},
which ensure by construction that the binding energies are close to the 
experimental results. 
For the phenomenological interactions, it is by now standard to augment 
the Hamiltonians by phenomenological 3NF's \cite{Nogga:1997mr,Nogga:2001cz,Pudliner:1997ck}
mostly based on the Urbana \cite{Carlson:1983kq} or Tucson-Melbourne \cite{Coon:2001pv}
models. Also here, the 3NF's are then adjusted so that the binding energies 
of $^3$He and $^4$He are close to the experimental values. 
Based on these adjustments, we are now in the position to calculate 
shifts of the pion-nucleus scattering length due to the few-nucleon 
contributions based on LO, NLO, and N$^2$LO chiral wave functions and 
on state-of-the-art phenomenological ones.

\section{Two- and three-nucleon contributions to $\pi$-$^3$He scattering}
\label{sec:3nres}

We now study the few-nucleon contributions to  $\pi$-$^3$He scattering 
in more detail with the goal to get a better, quantitative understanding of the
relative importance of $N$-  and $\left(N+1\right)$-nucleon operators.

\begin{table*}[t]
\centering
  \caption{Summary of the shifts of the $\pi$-$^3$He scattering length due to the few-nucleon corrections  
                  $a_{\pi-^3 \rm He}^{\left(\mbox{\scriptsize \ref{fig:2N-ops}a}\right)}$, 
                  $a_{\pi-^3 \rm He}^{\left(\mbox{\scriptsize \ref{fig:2N-ops}bc}\right)}$,
                  $a_{\pi-^3 \rm He}^{\left(\mbox{\scriptsize \ref{fig:2N-triple}is}\right)}$,
                  $a_{\pi-^3 \rm He}^{\left(\mbox{\scriptsize \ref{fig:2N-triple}iv}\right)}$,
                  $a_{\pi-^3 \rm He}^{\left(\mbox{\scriptsize \ref{fig:3N-coulcontr}}\right)}$, and 
                  $a_{\pi-^3 \rm He}^{\left(\mbox{\scriptsize \ref{fig:3N-contralpha} }\right)}$. 
                  For LO, the cutoff $\Lambda$ is  given in fm$^{-1}$, for NLO and N$^2$LO, both cutoffs ($\Lambda$/$\tilde \Lambda$) 
                  are given in MeV.
                  Central values and standard deviation for the
                  scattering length results are given in units of
                  $\left[10^{-3}\,m_\pi^{-1}\right]$.}
\label{tab:3he-res}

   \begin{center}
   \begin{tabular}{rlrrrrrr}
   \hline\noalign{\smallskip}  \\
     & $\Lambda$/$\tilde \Lambda$   &
    \multicolumn{1}{c}{$a_{\pi-^3 \rm He}^{\left(\mbox{\scriptsize \ref{fig:2N-ops}a}\right)}$}&
    \multicolumn{1}{c}{$a_{\pi-^3 \rm He}^{\left(\mbox{\scriptsize \ref{fig:2N-ops}bc}\right)}$}&
    \multicolumn{1}{c}{$a_{\pi-^3 \rm He}^{\left(\mbox{\scriptsize \ref{fig:2N-triple}is}\right)}$}&
    \multicolumn{1}{c}{$a_{\pi-^3 \rm He}^{\left(\mbox{\scriptsize \ref{fig:2N-triple}iv}\right)}$}&
    \multicolumn{1}{c}{$a_{\pi-^3 \rm He}^{\left(\mbox{\scriptsize \ref{fig:3N-coulcontr}}\right)}$}&
    \multicolumn{1}{c}{$a_{\pi-^3 \rm He}^{\left(\mbox{\scriptsize \ref{fig:3N-contralpha} }\right)}$}    \\[5pt]     
   \hline\noalign{\smallskip}    
    CD Bonn & --- & \hspace{0.4cm}$-25.08(6)$  & $\hspace{0.5cm}-0.329(1)$ & $2.769(2)$& $0.890(1)$
                             & \hspace{0.7cm}$-4.020(73)$ & \hspace{0.5cm}$-0.789(6)$\\
    AV18   & --- & $-24.13(9)$ & $-0.884(1)$ & $2.286(3)$ & $0.788(1)$& $-3.536(36)$ & $-0.728(4)$\\
  \hline\noalign{\smallskip}
      LO  & $2.0$/ --  & $-30.20(5) $ & $ 1.711(1) $ &$4.418 (1)$ &$1.935(1)$ & $-6.755(123)$ & $-2.339(6)$ \\
      LO  & $3.0$/ --  & $-23.76(13)$ & $-0.053(1) $ &$2.445 (1)$ &$1.553(1)$ & $-4.493(38) $ & $-1.762(4)  $ \\
      LO  & $4.0$/ --  & $-20.70(25)$ & $-0.327(1) $ &$1.077 (3)$ &$1.581(1)$ & $-3.620(25) $ & $-1.556(15) $ \\
      LO  & $5.0$/ --  & $-20.33(13)$ & $-0.120(1) $ &$0.868 (3)$ &$1.812(1)$ & $-3.830(93) $ & $-1.740(15) $ \\
%      LO  & $6.0$/ --  & $-20.66(10)$ & $ 0.099(1) $ &$1.147 (5)$ &$2.061(1)$ & $-3.941(67) $ & $-1.960(20) $ \\
 %     LO  & $8.0$/ --  & $-21.45(35)$ & $ 0.215(2) $ &$1.419 (8)$ &$2.371(2)$ & $-4.160(149)$ & $-2.067(21) $ \\
      LO  & $10.0$/ -- & $-21.92(99)$ & $ 0.128(3) $ &$0.770 (11)$&$2.576(5)$ & $-4.283(302)$ & $-1.872(57) $ \\
%      LO  & $12.0$/ -- & $-19.55(70)$ & $ 0.100(5) $ &$0.022 (16)$&$2.830(5)$ & $-3.744(391)$ & $-1.741(15) $ \\
%      LO  & $14.0$/ -- & $-20.46(87)$ & $ 0.167(7) $ &$-0.144(30)$&$3.138(11)$& $-3.364(159)$ & $-1.652(57) $ \\
%      LO  & $16.0$/ -- & $-18.94(76)$ & $ 0.235(10)$ &$-0.163(40)$&$3.404(13)$& $-3.939(701)$ & $-1.635(43) $ \\
%      LO  & $18.0$/ -- & $-19.56(99)$ & $ 0.295(10)$ &$-0.112(57)$&$3.714(24)$& $-4.359(133)$ & $-1.712(72) $ \\
      LO  & $20.0$/ -- & $-19.22(146)$& $ 0.296(18)$ &$-0.392(98)$&$4.005(76)$& $-3.835(894)$ & $-1.743(158)$ \\
       \hline\noalign{\smallskip}
     NLO & 400/500  & $-25.36(4)$ & $ 0.828(1)$  &$3.117(1)$ &$0.984(1)$ & $-3.934(29) $ & $-0.695(1)$ \\
     NLO & 550/500  & $-24.33(4)$ & $-0.061(1)$  &$2.714(1)$ &$0.645(1)$ & $-3.431(141)$ & $-0.374(1)$ \\
     NLO & 550/600  & $-24.05(2)$ & $-0.037(1)$  &$2.637(1)$ &$0.671(1)$ & $-3.245(22) $ & $-0.397(4)$ \\
     NLO & 400/700  & $-25.23(3)$ & $ 0.847(1)$  &$3.085(2)$ &$1.002(1)$ & $-3.898(21) $ & $-0.720(2)$ \\
     NLO & 550/700  & $-24.05(5)$ & $-0.020(1)$  &$2.564(1)$ &$0.692(1)$ & $-3.311(15) $ & $-0.435(4)$ \\
       \hline\noalign{\smallskip}
     N$^2$LO & 450/500  & $-25.76(4)  $ & $ 0.642(1)$  &$3.189(2)$ &$0.987(1)$& $-3.979(7) $ & $-0.721(2)$\\
     N$^2$LO & 600/500  & $-25.60(5)  $ & $-0.021(1)$  &$3.039(3)$ &$0.779(1)$& $-3.826(58)$ & $-0.496(4)$\\
     N$^2$LO & 550/600  & $-25.55(2)  $ & $ 0.233(1)$  &$3.104(3)$ &$0.952(1)$& $-4.057(75)$ & $-0.708(4)$ \\
     N$^2$LO & 450/700  & $-25.25(4)  $ & $ 0.611(1)$  &$3.104(1)$ &$1.052(1)$& $-4.038(15)$ & $-0.806(1)$ \\
     N$^2$LO & 600/700  & $-25.51(7)  $ & $ 0.094(1)$  &$3.022(2)$ &$0.985(1)$& $-4.056(53)$ & $-0.734(2)$\\
       \hline\noalign{\smallskip}
   \end{tabular}
  \end{center}
\end{table*}

The results for $^3$He are summarized in Table~\ref{tab:3he-res}. To
obtain these values, we have used the Monte Carlo scheme introduced in
Sec.~\ref{sec:nummeth}. The table gives the averaged results together
with the estimate of the standard deviation. In all cases, we have
performed several independent runs and checked that the spread of the
different results is in reasonable agreement with the expectations
from our estimates of the standard deviation. The table distinguishes
the shifts of the $\pi$-$^3$He scattering due to Eqs.~(\ref{eq:m2a})
($a_{\pi-^3 \rm He}^{\left(\mbox{\scriptsize
      \ref{fig:2N-ops}a}\right)}$), (\ref{eq:m2bc}) ($a_{\pi-^3 \rm
  He}^{\left(\mbox{\scriptsize \ref{fig:2N-ops}bc}\right)}$),
(\ref{eq:m2triple}) (isoscalar part $a_{\pi-^3 \rm
  He}^{\left(\mbox{\scriptsize \ref{fig:2N-triple}is}\right)}$ and
isovector part $a_{\pi-^3 \rm He}^{\left(\mbox{\scriptsize
      \ref{fig:2N-triple}iv}\right)}$, respectively), and
(\ref{eq:m3coul}) ($a_{\pi-^3 \rm He}^{\left(\mbox{\scriptsize
      \ref{fig:3N-coulcontr}}\right)}$), and due to the sum of the
contributions listed in Appendix~\ref{app:halfcoul3n} ($a_{\pi-^3 \rm
  He}^{\left(\mbox{\scriptsize \ref{fig:3N-contralpha} }\right)}$).

Based on reasonable assumptions for $a^{( \pm )}$, the one-nucleon
contribution to the scattering length was found to be $a_{\pi-^3 \rm
  He}^{\left(\mbox{\scriptsize 1N}\right)}=(92 \pm 15) \times 10^{-3} \,
m_{\pi}^{-1}$ \cite{Baru:2002cg}.  The uncertainty in this result is mainly
due to the uncertainty in $a^{\left(+\right)}$ multiplied by 3 as follows from
Eq.~(\ref{eq:piA-scattering}). On the other hand, it was realized in
Refs.\cite{Gasser:2007zt,Meissner:2005ne,Baru:2007ca} that the inclusion of
the leading IV effects  in $\pi$-N scattering leads to
the replacement of $a^{\left(+\right)}$ by $\tilde a^{\left(+\right)}$ in
 Eq.(\ref{eq:piA-scattering}). 
%To be more precise, 
In addition to this, at the same order there is also an isospin
 violating electromagnetic correction to $\pi$-$^3$He scattering:
  $-\alpha f_2/2/(1+\frac{m_\pi}{3\,m_N})$
 with $\alpha=1/137$ and LEC $f_2=(-0.97\pm 0.38 ) $GeV$^{-1}$
 \cite{Gasser:2002am}.
 The latter, however, gives a relatively small shift of the scattering length by
 $ (0.5\pm 0.2)\times 10^{-3} m_{\pi}^{-1}$.
%To
%be more precise, at LO there is also an isospin violating correction to the
%one--body term of $\pi ^3$He scattering.
A recent systematic analysis of isospin violating effects in $\pi$-N
scattering up to NLO \cite{Hoferichter:2009gn,Hoferichter:2009ez} resulted in
updated values for $\tilde a^{\left(+\right)}$ and $a^{\left(-\right)}$ \cite{Hoferichter:2009cq} from a
combined analysis of pionic hydrogen and deuterium data:
\begin{equation}
\tilde a^{\left(+\right)}=( 1    \pm 1)   \times  10^{-3} m_{\pi}^{-1},\ \
a^{\left(-\right)}= (86.5    \pm 1.2)   \times 10^{-3} m_{\pi}^{-1} \ .
\end{equation}
In the same works numerically significant subleading IV corrections were 
identified. Those may be included here by changing 
$$
\tilde a^{\left(+\right)}\to \tilde a^{\left(+\right)}+\Delta \tilde
a^{\left(+\right)} \ \mbox{and} \ a^{\left(-\right)}\to
a^{\left(-\right)}+\Delta a^{\left(-\right)} \ , 
$$
with 
$\Delta \tilde
a^{\left(+\right)}=(-3.35\pm 0.28)$ and $\Delta a^{\left(-\right)}=(1.39\pm
1.33)$, both values again in units of $ 10^{-3} \, m_{\pi}^{-1}$.
Equipped with these numbers we get an  updated value for the one-nucleon contribution 
$a_{\pi-^3 \rm He}^{\left(\mbox{\scriptsize  1N}\right)}=(88 \pm 4) \times 10^{-3} \, m_{\pi}^{-1}$.

Also in Ref.~\cite{Baru:2002cg}, the contribution of the two-nucleon diagrams
was estimated based on approximated wave functions for CD-Bonn. Their result
is $a_{\pi-^3 \rm He}^{\left(\mbox{\scriptsize 2N}\right)}=-26 \times 10^{-3}
\, m_{\pi}^{-1}$ which is in acceptable agreement with our full calculation.
In this work, we aim at the theoretical improvement of the result of
Ref.~\cite{Baru:2002cg} in several aspects. First, using chiral nuclear wave
functions up to N$^2$LO allows us to analyze systematically the model
dependence of our results. Second, the empirical enhancement of the
triple scattering diagram, discussed in subsection~\ref{i0int}, calls for an
inclusion of this two-nucleon operator also in $\pi$-$^3$He scattering.
We find that the isoscalar part of the
triple scattering diagram reduces the leading double scattering contribution
by about 12\% which is fully in line with the corresponding contribution to
$\pi$-$^2$H scattering. Moreover, the triple scattering diagram is even further
enhanced to a quite sizable 15\% contribution once the isovector part is
included.
 In
addition, for the first time, we investigate the role of the leading
three-nucleon contributions.  From Table \ref{sec:3nres} it becomes clear
that three-nucleon contributions are suppressed compared to the
two-nucleon ones.  Again, the Coulombian contributions $a^{(1a)}_{\pi -^3{\rm
    He}}$, $a^{(2)}_{\pi -^3\rm He}$ and $a^{(3)}_{\pi -^3\rm He}$ are a lot
more important than the non- or half-Coulombian ones although the binding
momentum is larger for $^3$He than for $^2$H. 

It was already discussed in Sec.~\ref{sec:powercount} that, in case of isovector
nuclei, counter terms start to contribute from lower orders than in isoscalar
nuclei.  Due to this fact the theoretical accuracy of the extraction of the
$\pi$N low energy parameters from $\pi$-$^3$He is, unfortunately,
significantly lower than from $\pi$-$^2$He or $\pi$-$^4$He. The contribution of
the isovector counter term in $\pi$-$^3$He scattering can be estimated using
dimensional analysis to be $\sim m_{\pi}/m_N\,\cdot a^{(1a)}_{\pi -^3\rm He}
\approx 4\,\times 10^{-3}\,m_\pi^{-1}$. On the other hand, the isovector
contact term is expected to be enhanced by $m_N/m_\pi$ compared to its
isoscalar counter part, which was estimated above (see 
sec.~\ref{sec:lamdep}) to be of order $ 1\,\times
10^{-3}\,m_\pi^{-1}$. This would provide us with an estimate of $7\times
10^{-3}$ $m_{\pi}^{-1}$. Both numbers are consistent and we use the latter
uncertainty below.

\begin{figure}[t]
  \centering
   \psfrag{two-body contributions}{$\hspace{1cm}{a^{(1a)}_{\pi -^3\rm He}}$}
   \psfrag{three-body contributions}{$\hspace{1cm}{a^{(3)}_{\pi -^3\rm He}}$}
   \psfrag{NLO}{\footnotesize{NLO}}
   \psfrag{N2LO}{\footnotesize{N$^2$LO}}
   \psfrag{Lambda}{\large $\mathbf{\Lambda\,\left[\mbox{\textbf{fm}}^{-1}\right]}$}
   \psfrag{scatlength}{\large $\mathbf{a_{\pi - ^3{{{He}}}}\,\left[10^{-3}\,m_\pi^{-1}\right]}$}
   \includegraphics[scale=0.30]{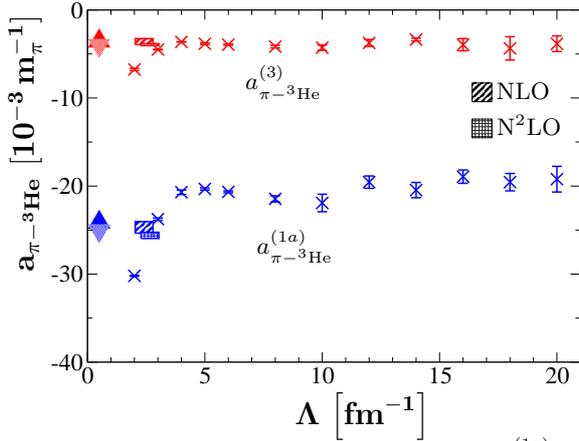}
   \caption{Leading few-nucleon contributions $a^{(1a)}_{\pi -^3\rm He}$
     and $a^{(3)}_{\pi -^3\rm He}$ to the $\pi$-$^3$He
     scattering length. The crosses denote the results obtained by
     using the LO chiral wave function (triangle up: AV18, triangle
     down: CD-Bonn).}
  \label{fig:cutoff-dep_all_wfs_I}
\end{figure}

\begin{figure}[t]
  \centering
   \psfrag{two-body contributions}{$\hspace{1cm}a^{(1bc)}_{\pi -^3He}$}
   \psfrag{three-body contributions}{$\hspace{1cm}a^{(19)}_{\pi -^3He}$}
   \psfrag{2b}{\scriptsize{2b}}
   \psfrag{3b}{\scriptsize{3b}}
   \psfrag{NLO,2b}{\footnotesize{NLO, 2b}}
   \psfrag{N2LO,2b}{\footnotesize{N$^2$LO, 2b}}
   \psfrag{NLO,3b}{\footnotesize{NLO, 3b}}
   \psfrag{N2LO,3b}{\footnotesize{N$^2$LO, 3b}}
   \psfrag{Lambda}{\large $\mathbf{\Lambda\,\left[\mbox{\textbf{fm}}^{-1}\right]}$}
   \psfrag{scatlength}{\large $\mathbf{a_{\pi -^3{{{\rm He}}}}\,\left[10^{-3}\,m_\pi^{-1}\right]}$}
   \includegraphics[scale=0.30]{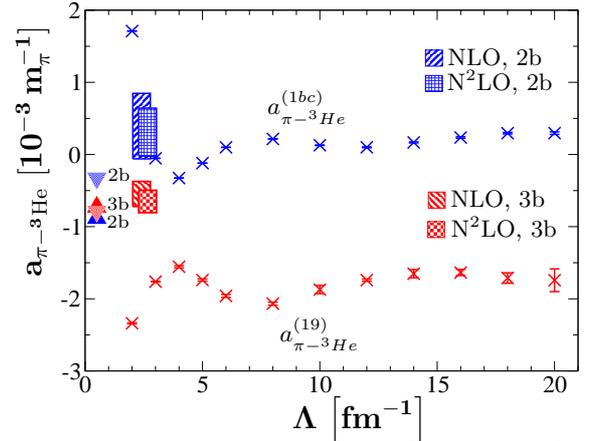}
   \caption{Few-nucleon contributions $a^{(1bc)}_{\pi -^3He}$ and $a^{(19)}_{\pi -^3He}$ 
     as a function of  $\Lambda$. The crosses, the triangle up and the triangle
     down mark again the LO chiral wave functions, AV18 and CD-Bonn,
     respectively. }
  \label{fig:cutoff-dep_all_wfs_II}
\end{figure}

To support our uncertainty estimates it is important to study the cutoff
dependence of the various scattering length shifts and to compare it to the
counter term contribution. The cutoff dependence is shown in
Figs.~\ref{fig:cutoff-dep_all_wfs_I} and \ref{fig:cutoff-dep_all_wfs_II}.  We
noticed that the results depend strongest on the cutoff of the
Lippmann-Schwinger equation $\Lambda$. Therefore, we plotted the results
depending on this cutoff for NLO and N$^2$LO neglecting the mild dependence on
the spectral function cutoff $\tilde \Lambda$ (see \cite{Epelbaum:2004fk} for
more details on the definition of this cutoff).  We also included the results
for CD-Bonn and AV18 in these figures, which we arbitrarily positioned left of
the other data.

Fig.~\ref{fig:cutoff-dep_all_wfs_I} shows the most important two- and three-nucleon 
contributions, the Coulombian ones. It becomes clear that  most of the dependence on the 
cutoff is for cutoffs below 5~fm$^{-1}$.  
%For cutoffs larger than 5~fm$^{-1}$, a mild 
%cutoff dependence is remaining. 
%, which is however of the order which can be 
%expected from higher order short range counter terms. 
The suppressed contributions 
for the non- or half-Coulombian diagrams are shown in  Fig.~\ref{fig:cutoff-dep_all_wfs_II}. 
Also here, we observe that the cutoff dependence becomes mild for larger $\Lambda$. But the 
results based on the LO wave functions are not always in good agreement with the ones  
for phenomenological  and higher order wave functions. Apparently, the cutoff dependence 
is not as strong as other higher order contributions for these diagrams. Obviously, the small size of the contributions amplifies small effects. Especially, 
the counter term contribution is significantly larger than the few-nucleon 
terms presented in Fig.\ref{fig:cutoff-dep_all_wfs_II}.
%we can expect 
%counter term contributions which are quantitatively comparable to these 
%few-nucleon contributions.  

\begin{figure}[t]
  \centering
   \psfrag{E_bind}{\large $\mathbf{|B_{^3\mbox{{\textbf{He}}}}|\,\left[MeV\right]}$}
   \psfrag{scatlength}{\large
     $\mathbf{a_{\pi-^3\mbox{\scriptsize\textbf{{He}}}}\,\left[10^{-3}\,m_\pi^{-1}\right]}$}
   \psfrag{2N-triple, IS}{$\hspace{3cm}{a_{\pi-^3 \rm He}^{\left(\mbox{\tiny \ref{fig:2N-triple}is}\right)}}$}
   \psfrag{three-body contributions}{$\hspace{0.4cm}{a_{\pi-^3 \rm He}^{\left(\mbox{\tiny \ref{fig:3N-coulcontr}}\right)}}$}
   \psfrag{two-body contributions}{$\hspace{0.65cm}{a_{\pi-^3 \rm He}^{\left(\mbox{\tiny \ref{fig:2N-ops}a}\right)}}$}
   \includegraphics[scale=0.30]{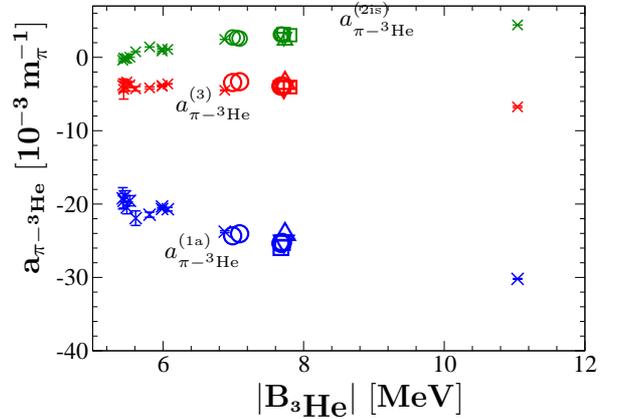}
   \caption{Results for the leading two- and three-nucleon contributions
     and for the isoscalar part of Fig.~\ref{fig:2N-triple},
     respectively. Crosses: LO chiral wave functions, circles: NLO,
     squares: N$^2$LO, triangle down: CD-Bonn, triangle up: AV18. For
     the NLO (N$^2$LO) four (two) results were selected since they are
     very close and therefore not distinguishable.}
  \label{fig:E_bind-dep_all_wfs_I}
\end{figure}

Quantitatively more significant is the cutoff dependence of the Coulombian 
diagrams for small $\Lambda$. We found that for these contributions the 
cutoff dependence is driven by the binding energy of $^3$He. This is shown 
explicitly in Fig.~\ref{fig:E_bind-dep_all_wfs_I}, 
where the results of Fig.~\ref{fig:cutoff-dep_all_wfs_I}
are plotted again, this time depending on the binding energy.  
%Clearly, 
%the dominant two- and three-nucleon shifts of the scattering length 
%are correlated with the binding energy predicted for the interactions. 
Basically, the smaller is the cutoff the larger is the binding energy 
(see Table~\ref{tab:3n4nbind}) and the larger is the scattering length 
for the dominant two- and three-nucleon contributions. 
Whereas this fact does not change our conclusion on the power counting 
below, it will be interesting for the extraction of $\pi$-N scattering 
lengths from data on light nuclei. Obviously, the uncertainty due to higher 
order contributions to the wave function can be reduced by the 
requirement that the binding energy is correctly described. 
In this way, the effective dependence of the scattering length shifts 
on the order of the interaction is reduced to approximately 1~$\times 10^{-3} \, m_{\pi}^{-1}$. 
This uncertainty is smaller than  the estimated contribution of the contact term 
and therefore irrelevant for the estimation of total theoretical accuracy.  
%{\bf May be remove the next paragraph? It is already discussed above that the CT
%exceeds some small few body effects and thus they are irrelevant}
%We note that the correlation discussed above is 
%not effective for the much smaller non- and half-Coulombian contributions. These 
%are more sensitive to short range parts of the wave function, which are not 
%constrained by the binding energy. Quantitatively, these contributions 
%will probably not be relevant in a final evaluation of data on light nuclei
%since, again, contact terms  omitted will be more significant. 

The net  contribution of two- and three-nucleon terms to $\pi$-$ ^3$He scattering
length can be read off from Table \ref{sec:3nres} (the numbers are in $ 10^{-3} m_{\pi}^{-1}$)
\begin{eqnarray}\nonumber
\!a^{(2N+3N)}_{\pi -^3\rm He}\!&\! =\!&\!(-25.6+4.0+0.3)+(-4.0-0.7) \pm 7\\ % 10^{-3} m_{\pi}^{-1}\\
&=&- 26.0    \pm 7 \ . % 10^{-3} m_{\pi}^{-1}.
\end{eqnarray}
Here the numbers in the first(second) bracket correspond to the average
results for two(three)-nucleon contributions calculated with N$^2$LO wave
functions.  The uncertainty due to the use of different wave functions is not
shown, for it is much smaller than the estimated contact term contribution.
In addition, we argued in Sec.~\ref{sec:2ncontr} that we do not expect large
corrections to the $\pi$-$^3$He scattering length due to the net effect of the
dispersive and the $\Delta$ contributions.  Thus, our prediction for the
$\pi$-$^3$He scattering length is
\begin{equation}
a^{(1N+2N+3N)}_{\pi -^3\rm He}=(62 \pm 4\pm 7)\times  10^{-3} m_{\pi}^{-1}
\label{apiheth}
\end{equation}
where the first uncertainty is due to ambiguities in $\pi$-N scattering lengths
whereas the second one represents the uncertainty in the few-nucleon effect. 
This result does not include  isospin violating few-nucleon effects. Those are found 
to be sizable for $\pi$-$^2$H scattering~\cite{Baru:2010tmp} although 
significantly smaller than the theoretical uncertainty of $\pi$-$^3$He calculation. 
\begin{table}[ht]
\caption{Experimental results for the $\pi$-$^3$He scattering length.
The  entries contain measured energy level shifts of
$\pi$-$^3$He atomic bound states together with corresponding
scattering lengths.}
\label{piHeexp}
\begin{center}
\begin{tabular}{lcc}
&$\epsilon_{1s}$ [eV] & $a_{\pi-^3 \rm He}\left[10^{-3}\,m_\pi^{-1}\right]$\\
\hline
R. Abela {\it et al.} \protect{\cite{Abela:1977mj}} &
$44 \pm 5$ & $56 \pm 6$ \\
G. R. Mason {\it et al.} \protect{\cite{Mason:1980vg}} &
$34 \pm 4$ & $43 \pm 5$ \\
I. Schwanner {\it et al.} \protect{\cite{Schwanner:1984sg}} &
$32 \pm 3$ & $41 \pm 4$ \\
\hline
\end{tabular} \end{center}
\end{table}
The result of Eq.(\ref{apiheth}) 
is to be compared to the experimental results for the $\pi$-$^3$He scattering lengths given 
in Table~\ref{piHeexp}. 
Those are extracted from the measurements of the 1s level shifts in $\pi^-$-$^3$He atom due to the strong
interactions \cite{Abela:1977mj,Mason:1980vg,Schwanner:1984sg} 
by using DGBT-type formulae \cite{Deser:1954vq} and including logarithmic corrections of 
Ref.~\cite{Lyubovitskij:2000kk}. The table demonstrates that the results of the
first measurement are in contradiction with the others even within the large experimental uncertainties. 
It is getting even more intriguing because it is only the first measurement that agrees 
with our
theoretical prediction (\ref{apiheth}) 
based on ChPT. Clearly, a new measurement of this quantity 
recently performed at PSI \cite{Gotta:2010prop} is of high importance to resolve the 
existing discrepancies.

\begin{table}[t]
\centering

 \caption{Numerical results for the relative scalings of the
     few-nucleon contributions.}
% compared to the estimates based on
%     Weinberg- and $Q$--counting}
  \label{tab:3hescal}

 \begin{center}
  \begin{tabular}{llcc}
    \hline\noalign{\smallskip} 
     & $\Lambda$/$\tilde \Lambda$
        & $a_{\pi-^3 \rm He}^{\left(\mbox{\scriptsize 2N}\right)}
          /  a_{\pi-^3 \rm He}^{\left(\mbox{\scriptsize  1N}\right)}$
        & $a_{\pi-^3 \rm He}^{\left(\mbox{\scriptsize 3N}\right)}
          /  a_{\pi-^3 \rm He}^{\left(\mbox{\scriptsize 2N}\right)}$
%        & $a_{\pi-^3 \rm He}^{\left(\mbox{\scriptsize \ref{fig:2N-ops}bc}\right)}
%             / a_{\pi-^3 \rm He}^{\left(\mbox{\scriptsize \ref{fig:2N-ops}a}\right)} $
%        & $a_{\pi-^3 \rm He}^{\left(\mbox{\scriptsize \ref{fig:3N-contralpha} }\right)}
%            / a_{\pi-^3 \rm He}^{\left(\mbox{\scriptsize \ref{fig:3N-coulcontr}}\right)} $
      \\[3pt]
     \hline\noalign{\smallskip}  \hline\noalign{\smallskip} 
     CD-Bonn & ---  & 0.220  & 0.221\\
     AV18    & ---  & 0.222  & 0.194 \\
     \hline\noalign{\smallskip} 
   NLO & 400/500  & 0.206 & 0.227 \\
   NLO & 550/500  & 0.212 & 0.181 \\
   NLO & 550/600  & 0.210 & 0.175 \\
   NLO & 400/700  & 0.205 & 0.228 \\
   NLO & 550/700  & 0.210 & 0.180 \\
  \hline\noalign{\smallskip}                                             
N$^2$LO & 450/500 & 0.212 & 0.222 \\ 
N$^2$LO & 600/500 & 0.220 & 0.194 \\
N$^2$LO & 550/600 & 0.215 & 0.224 \\ 
N$^2$LO & 450/700 & 0.207 & 0.236 \\
N$^2$LO & 600/700 & 0.216 & 0.224 \\
   \hline
   \end{tabular}
   \end{center}
\end{table}

In Table~\ref{tab:3hescal}, we have compiled the relative
contributions of one-nucleon, two-nucleon, and three-nucleon diagrams
($a_{\pi - ^3 \rm He}^{\left(\mbox{\scriptsize 1N}\right)}$, $a_{\pi
  -^3 \rm He}^{\left(\mbox{\scriptsize 2N}\right)}$, and $a_{\pi -^3
  \rm He}^{\left(\mbox{\scriptsize 3N}\right)}$). 
We omitted the results of the LO
wave functions here, since their description of the binding energies
is generally poor. It sticks out that, numerically, the suppression of
few-nucleon corrections is less than expected by Weinberg's
power counting. Based on these findings, we are led to the conclusion
that the power counting gives reasonable guidance in
identifying the most important contributions, however, for a
quantitative understanding, explicit calculations for the leading
few-nucleon contributions are necessary to estimate the contribution
of the class of $N$-nucleon contributions. Qualitatively,
more-nucleon diagrams are still sufficiently suppressed so that the
series of one- , two-, ... nucleon contributions can be truncated at
sufficiently low complexity of the problem --- we find a factor of 5
suppression when going from an $N$-nucleon operator to an
$\left(N+1\right)$-nucleon operator. It is important to note in this
context that we find the same suppression factor for $N=1$, $N=2$, and, as
will be shown in the next section, $N=3$. Due to this, the four-nucleon
diagrams turn out to be already insignificant.

In summary, we have calculated $\pi$-$^3$He scattering length including
leading three-nucleon terms and  two-nucleon operators. Due to the presence of the 
large contact term in the isovector channel the present calculation
basically reaches the edge of the theoretical accuracy. 
We also find that the results are in qualitative agreement with Weinberg's counting.

\section{Two- and four-nucleon contributions to $\pi$-$^4$He}
\label{sec:4nres}

\begin{table*}[t]
\centering
  \caption{Summary of the shifts of the $\pi$-$^4$He scattering length 
                due to the few-nucleon corrections  
                  $a_{\pi -^4 \rm He}^{\left(\mbox{\scriptsize \ref{fig:2N-ops}a}\right)}$, 
                  $a_{\pi -^4 \rm He}^{\left(\mbox{\scriptsize \ref{fig:2N-ops}bc}\right)}$,
                  $a_{\pi -^4 \rm He}^{\left(\mbox{\scriptsize \ref{fig:2N-triple}}\right)}$, and 
                  $a_{\pi -^4 \rm He}^{\left(\mbox{\scriptsize \ref{fig:4N-contr}}\right)}$. 
                  For NLO and N$^2$LO, both cutoffs ($\Lambda$/$\tilde \Lambda$) 
                  are given in MeV.
                  Central values and standard deviation for the scattering length results are given
                   in $10^{-3} m_{\pi}^{-1}$.  }
 \label{tab:4he-res}
 \begin{center}
  \begin{tabular}{lccccc}  
     \hline 
               &    $\Lambda$/$\tilde \Lambda$
                                 &      $a_{\pi -^4 \rm He}^{\left(\mbox{\scriptsize \ref{fig:2N-ops}a}\right)}$
                                 &       $a_{\pi -^4 \rm He}^{\left(\mbox{\scriptsize \ref{fig:2N-ops}bc}\right)}$
                                 &       $a_{\pi -^4 \rm He}^{\left(\mbox{\scriptsize \ref{fig:2N-triple}}\right)}$
                                 &       $a_{\pi -^4 \rm He}^{\left(\mbox{\scriptsize \ref{fig:4N-contr}}\right)}$    \\[3pt] 
\hline
AV18       & ---           & -49.5(7)     &    -1.29(2)  &   5.00(5)   &   2.73(84)  \cr 
\hline
NLO        & 400/500  & -56.1(15)   &     3.02(1)  &   7.18(1)   &   2.79(11)   \cr 
NLO        & 550/500  & -51.0(8)     &      0.41(1) &   6.04(2)   &   2.18(22)   \cr  
NLO        & 550/600  & -51.4(5)     &      0.53(1) &   5.95(1)   &   2.02(22)   \cr
NLO        & 400/700  & -54.5(3)     &      3.10(1) &   7.13(2)   &   3.92(42)   \cr  
NLO        & 550/700  & -51.6(12)   &      0.58(1) &   5.72(1)   &   2.54(59)   \cr  
\hline 
N$^2$LO & 450/500  & -54.4(4)     &     1.92(1) &   6.98(2)   &   3.00(20)   \cr 
N$^2$LO & 600/500  & -52.0(8)    &     -0.09(2) &   6.16(3)   &   2.13(11) \cr 
N$^2$LO & 550/600  & -52.7(6)    &      0.50(1) &   6.42(3)   &   2.31(31)   \cr 
N$^2$LO & 450/700  & -52.7(7)    &      1.81(1) &   6.68(3)   &   2.56(10)   \cr 
N$^2$LO & 600/700  & -53.9(8)    &      0.36(1) &   6.34(2)   &   2.81(17)   \cr 
    \hline
   \end{tabular}
   \end{center}
\end{table*}

Finally, we want to discuss the few-nucleon contributions to
$\pi$-$^4$He scattering. Because of the isovector character of the
leading three-nucleon contributions, we here only need to study two-
and four-nucleon ones. Our results are summarized in
Table~\ref{tab:4he-res}. Qualitatively, the results for the
two-nucleon operators are similar to the ones for $\pi$-$^2$H and for $\pi$-$^3$He scattering. 
The leading two-nucleon term $a_{\pi - ^4   \rm He}^{\left(\mbox{\scriptsize \ref{fig:2N-ops}a}\right)}$ is the
by far most important contribution depending on $a^{(-)}$. 
Just based on  the symmetry arguments 
$a_{\pi - ^4  \rm He}^{\left(\mbox{\scriptsize \ref{fig:2N-ops}a}\right)}$ should be approximately twice 
as large as the one in $\pi$-$^3$He scattering. 
As one can see from Tables \ref{tab:3he-res} and  \ref{tab:4he-res}, this is indeed supported by our
results.
In addition, the contribution $a_{\pi - ^4  \rm He}^{\left(\mbox{\scriptsize \ref{fig:2N-ops}a}\right)}$ is approximately twice as
large as for the deuteron although the number of NN pairs is six. This 
can be expected from the isospin structure of this amplitude leading
to opposite signs for the contributions of neutron-proton pairs and
proton-proton (or neutron-neutron) pairs \cite{Kubis:priv}.
Trivially, the one-nucleon terms in $\pi$-$^4$He and in $\pi$-$^2$H  scattering also scale 
with a factor 2, as seen from Eq.~(\ref{eq:piA-scattering}).
Unfortunately,  this implies that the correlation
of $a^{(-)}$ and $a^{(+)}$ due to experimental results for
$\pi$-$^4$He scattering is very similar to the one based on a
$\pi$-$^2$H analysis.
The most important correction to the leading two-nucleon term originates 
from the triple scattering diagram. Again, it 
is enhanced compared to the estimate of naive dimensional 
analysis. In full analogy with $\pi$-$^2$He and $\pi$-$^3$He scattering  the ratio 
$a_{\pi -^4 \rm He}^{\left(\mbox{\scriptsize \ref{fig:2N-triple}}\right)}
/a_{\pi - ^4   \rm He}^{\left(\mbox{\scriptsize \ref{fig:2N-ops}a}\right)}$ is about 0.1-0.13
and is only smoothly dependent on the cutoff.

From the cutoff variation of the N$^2$LO
results, we estimate that missing counter terms should contribute of
the order of $2\times 10^{-3} m_{\pi}^{-1}$. This is approximately 4~\%
and in line with our expectations from Weinberg counting.
$a_{\pi - ^4 \rm He}^{\left(\mbox{\scriptsize \ref{fig:2N-ops}bc}\right)}$
is again suppressed and negligible compared to the cutoff 
dependence of $a_{\pi -^4 \rm He}^{\left(\mbox{\scriptsize \ref{fig:2N-ops}a}\right)}$. 
Similar to the results for $\pi$-$^3$He, even the sign of 
this contribution is not fixed clearly indicating the sensitivity 
to the short distance part of the wave functions. 

Let us now turn to the contribution of the four-nucleon 
operator $a_{\pi -^4 \rm He}^{\left(\mbox{\scriptsize \ref{fig:4N-contr}}\right)}$. Its contribution is opposite in sign to the leading 
two-nucleon contribution, but  similar in size as 
the cutoff variation of $a_{\pi -^4 \rm He}^{\left(\mbox{\scriptsize \ref{fig:2N-ops}a}\right)}$ and, therefore,  
comparable to the contribution of the
first isoscalar two-nucleon counter terms. The relative 
suppression of the $a_{\pi -^4 \rm He}^{\left(\mbox{\scriptsize \ref{fig:4N-contr}}\right)}$ compared to 
$a_{\pi -^4 \rm He}^{\left(\mbox{\scriptsize \ref{fig:2N-ops}a}\right)}$ is 
approximately 0.06.  From Weinberg counting, however,
we naively  expect a much larger suppression of 
$(m_{\pi}/m_N)^4 \simeq 4 \times 10^{-4}$. On the other hand, this deviation 
was expected and  in line with the results
of our calculations for $\pi$-$^3$He where we also observed a significant overestimation 
of the calculated ratio  $a_{\pi -^3 \rm He}^{\left(\mbox{\scriptsize \ref{fig:3N-coulcontr}}\right)}
/a_{\pi -^3 \rm He}^{\left(\mbox{\scriptsize \ref{fig:2N-ops}a}\right)}$
compared to the  dimensional analysis.
In fact we find numerically a suppression of about $1/5^2$ for the four-nucleon
operator
compared to the two-nucleon operator, completely in line with the pattern of 
successive suppression found in case of  $\pi$-$^3$He scattering.
As a result of this, the contribution of the four-nucleon  operators can be, in principle, omitted, for 
it coincides in the magnitude with  the estimated counter term contribution. This is the reason
why we do not take into account the other four-nucleon topologies analogous
 to those given in Fig.~\ref{fig:3N-contralpha}
for $\pi$-$^3{\rm He}$ scattering: Based on the study of  the $\pi$-$^3{\rm He}$ process,
 we expect their contribution to be suppressed as compared to the 
leading four-nucleon operator.

Thus, we conclude that pion scattering on $^4$He can not provide any additional constraints on $a^{\left(+\right)}$ and $a^{\left(-\right)}$
although this process  will be helpful  to study the systematic uncertainties of experiment and theoretical 
calculations. 
Again we find  that Weinberg 
counting is not able to capture quantitatively 
the relative suppression of operators with a different number 
of active nucleons.

\section{Conclusions and Outlook}
\label{sec:concl}

In this paper, we have systematically studied the few-nucleon contributions to
pion-nucleus scattering. We compared our numerical findings to power counting
estimates of two power-counting schemes, Weinberg counting and $Q$--counting.
Our numerical results do qualitatively support Weinberg's power counting but
not $Q$--counting. An evidence for this was provided by studying the
dependence of different contributions on the binding energy of the
deuteron. This dependence is expected to be significantly different in these
two counting schemes varying from a constant (up to logarithmic corrections)
for the ratio of relevant two-nucleon contributions in Weinberg counting to a
power law behavior in $Q$--counting. It is shown both numerically and
analytically that the ratio is indeed very weakly depending on the binding
energy in clear favour of Weinberg counting. In the course of the analytic
analysis, we found that wave functions based on a perturbative treatment of
pions should not be used for the calculation of pion-nucleus
scattering. Especially, a treatment of the pion as a heavy field would be
justified only for unphysically small deuteron binding energies below 0.1~MeV.

%\\ There\-fore, we analyzed 
%the evidence in favor of $Q$-counting in more detail and On the other hand,
Numerical results for some particular diagrams are not in line with the
estimates based on Weinberg counting. This was resolved when we
could trace back enhancement or suppression of the pertinent diagrams
either to accidental spin-isospin factors or to specific properties of the
loop function of the triple scattering diagram, which can not be, of course,
taken into account in the counting scheme. Especially, the last insight might
be of relevance for the treatment of pion production and also NN forces.

Generally, our results point to a smaller suppression of more-nucleon terms
than expected by naive dimensional analysis.  To be specific, our results
demonstrate that the ratio of one-nucleon to two-nucleon to three-nucleon to
four-nucleon contributions scales roughly as 100:25:5:1 as compared to the
pattern $50^3$:$50^2$:50:1 which can be expected from Weinberg counting.
Apparently, the additional integral measures that enter when nucleons are
added and that provide most of the suppression in the dimensional analysis are
partially canceled by other mechanisms. It should be stressed that even the
relatively mild suppression found is still quantitatively sufficient to allow
for controlled calculations for pion-nucleus scattering.  Especially, for the
$\pi$-$^4$He scattering length the four-nucleon operator is already numerically
insignificant.
%Practically, in order 
%to be more quantitative, it is necessary to evaluate leading 
%more-nucleon contributions to get a reliable estimate of the 
%whole class of $A$-nucleon diagrams. 

We performed explicit calculations up to four-nucleon contributions to
pion-nucleus scattering.  Based on these calculations, we conclude that
three-nucleon contributions might still be relevant.  On the other hand, it
turned out that the most important three-nucleon diagram is
isovector. Therefore, for the important case of $^4$He, only one-nucleon and
two-nucleon diagrams will contribute significantly.  It turned out that the
relative contribution of one- and two-nucleon diagrams is very similar for
$\pi$-$^2$H and $\pi$-$^4$He. Therefore we expect that the corresponding bands
$a^{\left(+\right)}$ vs $a^{\left(-\right)}$ for $\pi$-$^2$H and $\pi$-$^4$He
will be almost on top of each other.  Thus, we tend to conclude that pion
scattering on $^4$He can not provide any additional constraints on
$a^{\left(+\right)}$ and $a^{\left(-\right)}$, however, an improved
measurement of the $\pi^-$-$^4$He atom will provide an additional cross check
of the systematics of the analysis.  On the other hand, $\pi$-$^3$He does
contain a new non-trivial dependence on $a^{\left(-\right)}$. That is why we
have studied this process in more detail. For the first time three-nucleon
contributions were included. Based on the latest numbers for $\tilde
a^{\left(+\right)}$ --- which denotes the isoscalar scattering length
including the leading, isoscalar isospin violating corrections --- and
$a^{\left(-\right)}$ we obtained $a_{\pi -^3 {\rm He}}=(62 \pm 4 \pm 7)\times
10^{-3} m_{\pi}^{-1}$ where the uncertainties are due to ambiguities in
$\pi$-N scattering lengths and in the unknown isovector two-nucleon contact
operator, respectively. Unfortunately, the large theoretical uncertainty
related to the contact term in the isovector channel precludes one from
considering $\pi$-$^3$He scattering on the same footing with $\pi $-H and 
$\pi$-$^2$H processes to extract $\tilde a^{\left(+\right)}$ and
$a^{\left(-\right)}$. On the other hand, the results of a new measurement of
$\pi$-$^3$He at PSI \cite{Gotta:2010prop} which are currently analyzed, are of
high interest to check the predictions of ChPT, especially since the
theoretical value lies above the most recent experimental value (we find a
2$\sigma$ discrepancy), while it is consistent with an earlier measurement.

    Our results also showed that we are able to predict
    few-nucleon contributions for pion scattering off isoscalar (isovector)
nuclei up to a level of 5~\% (30~\%\footnote{This is 10\% of the full
  scattering length, since for isovector targets, contrary to their isoscalar
counter terms, the scattering length is
dominated by the one-nucleon term.}) of the leading 
    two-nucleon contributions 
    before additional counter terms enter.  Since isospin violation is
    predicted to contribute at the few-percent level as well, it will be required to
    further work on IV contributions, before we finally extract
    $a^{(+)}$ and $a^{(-)}$ from the available data on pionic atoms.
    Work in this direction is in progress~\cite{Baru:2010tmp}.

\begin{acknowledgement} 
We thank E. Epelbaum, D. Gotta, M.  Hoferichter, B.  Kubis, A. Kudryavtsev, 
U.-G. Mei{\ss}ner, and D.  R. Phillips for useful discussion and 
valuable comments on the manuscript.  
The work was supported in parts by funds provided 
from the Helmholtz Association (grants VH-NG-222, VH-VI-231) 
and  by the EU HadronPhysics2 project. The work of V.B.  was 
supported by the State Corporation of the Russian Federation ``Rosatom''.  
Part of the numerical calculations have been performed 
on the super computer cluster of the NIC, J\"ulich, Germany.
\end{acknowledgement}

\appendix

\section{Independence of three-nucleon diagrams on the parameterization of the 
              pion field}
\label{app:alphadep}

\begin{figure*}[t]
  \centering
  \psfrag{tau1}{$\tau_1$}
  \psfrag{tau3}{ $\tau_3$}
  \psfrag{q1}{ $\vec{q}_1$}
  \psfrag{q3}{ $\vec{q}_3$}   
  \psfrag{k}{ \hspace{-0.3cm}$\vec{k}_\pi$}
  \psfrag{k'}{ $\vec{k}_\pi^{\,\prime}$}
  \psfrag{a}{ $a$}
  \psfrag{b}{ $b$}
  \psfrag{c}{ $c$}
  \psfrag{d}{ $d$}
  \psfrag{e}{ $e$}
  \psfrag{f}{ $f$}
  \psfrag{tau2f}{ $\tau_2^f$}
  {\large (a)}
    \parbox[t]{4.5cm}{\vspace{0cm}
        \includegraphics[scale=0.55]{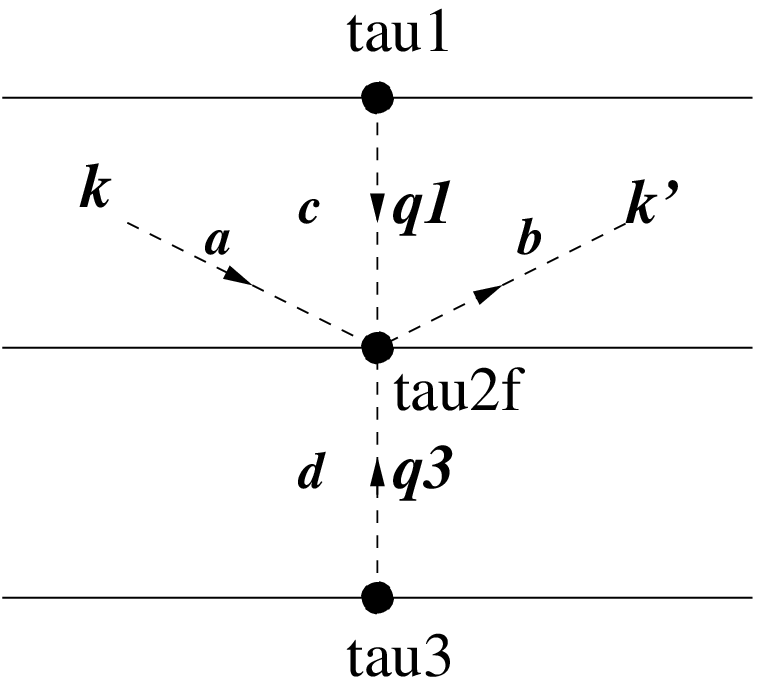}}
  \hspace{0.5cm}
  \psfrag{q2}{$\vec{q}_2$}
  \psfrag{q3}{$\vec{q}_3$}
  {\large (b)}
    \parbox[t]{4.5cm}{\vspace{0cm}
        \includegraphics[scale=0.55]{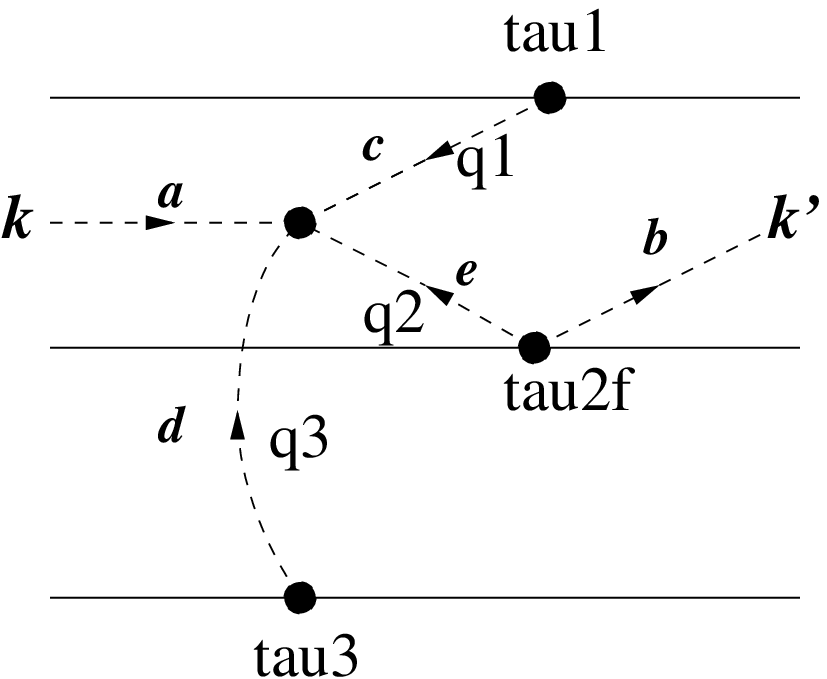}}
  \hspace{0.5cm}
  {\large (c)}
    \parbox[t]{4.5cm}{\vspace{0cm}
        \includegraphics[scale=0.55]{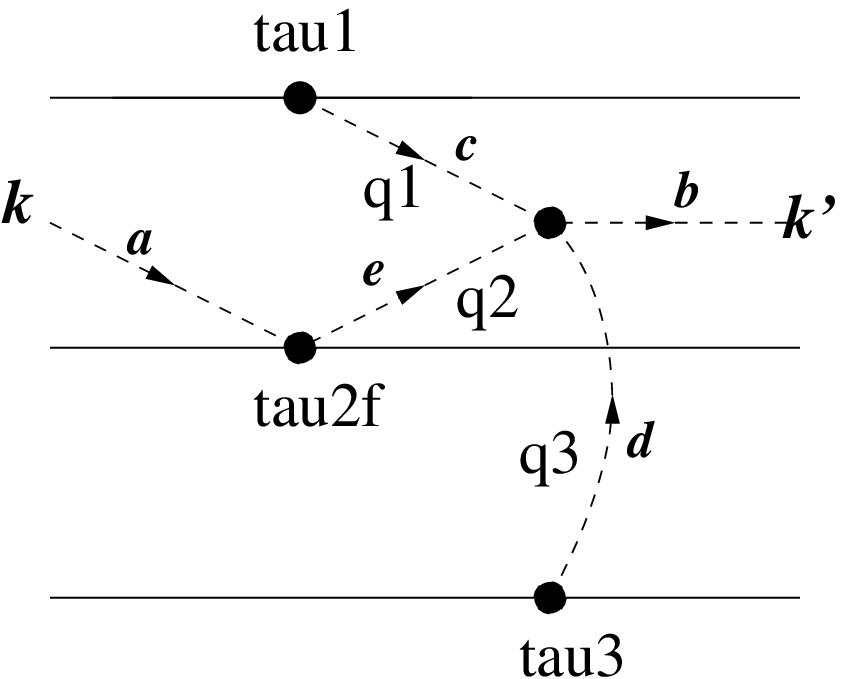}} \\[15pt]

  {\large (d)}
    \parbox[t]{4.5cm}{\vspace{0cm}
        \includegraphics[scale=0.55]{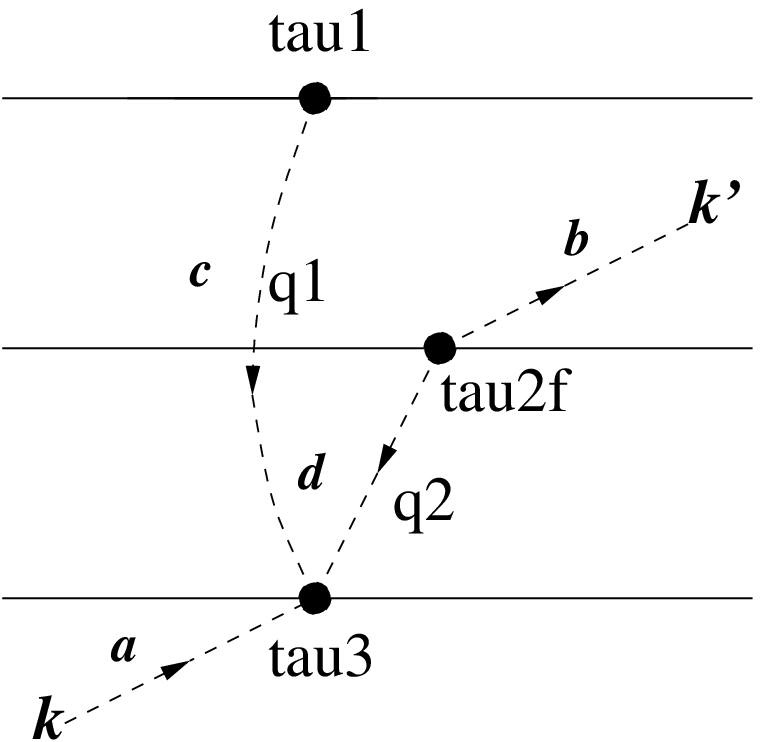}}
  \hspace{0.5cm}
  {\large (e)}
    \parbox[t]{4.5cm}{\vspace{0cm}
        \includegraphics[scale=0.55]{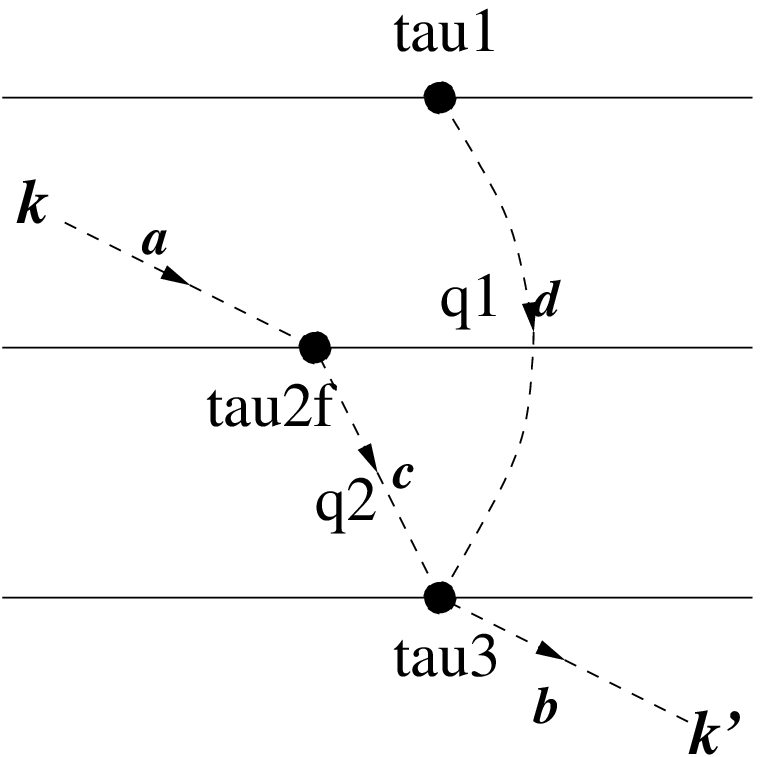}}
  {\large (f)}
    \parbox[t]{4.5cm}{\vspace{0cm}
        \includegraphics[scale=0.55]{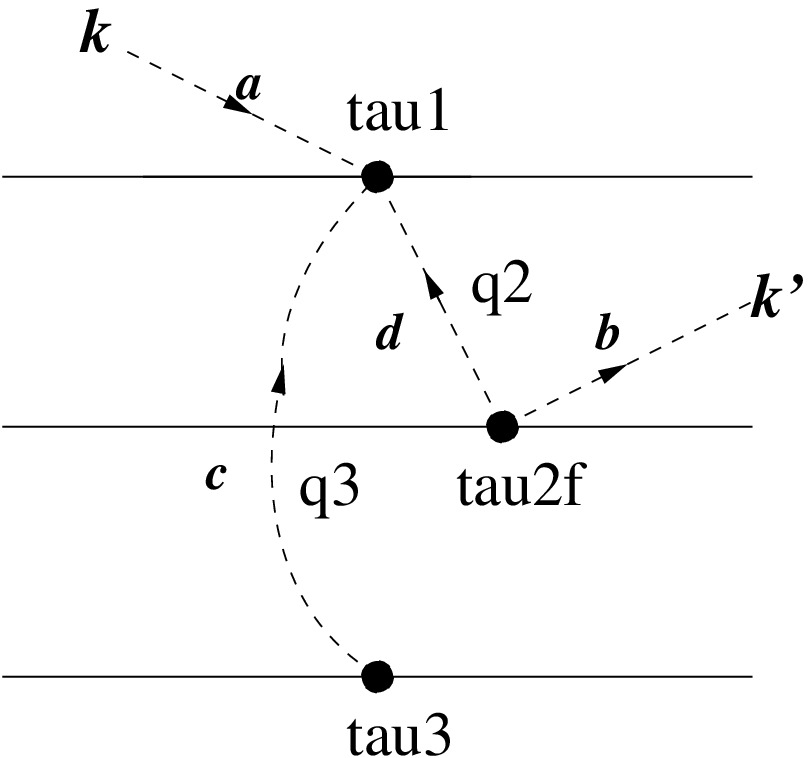}}
  \hspace{0.5cm}

  {\large (g)}
    \parbox[t]{4.5cm}{\vspace{0cm}
        \includegraphics[scale=0.55]{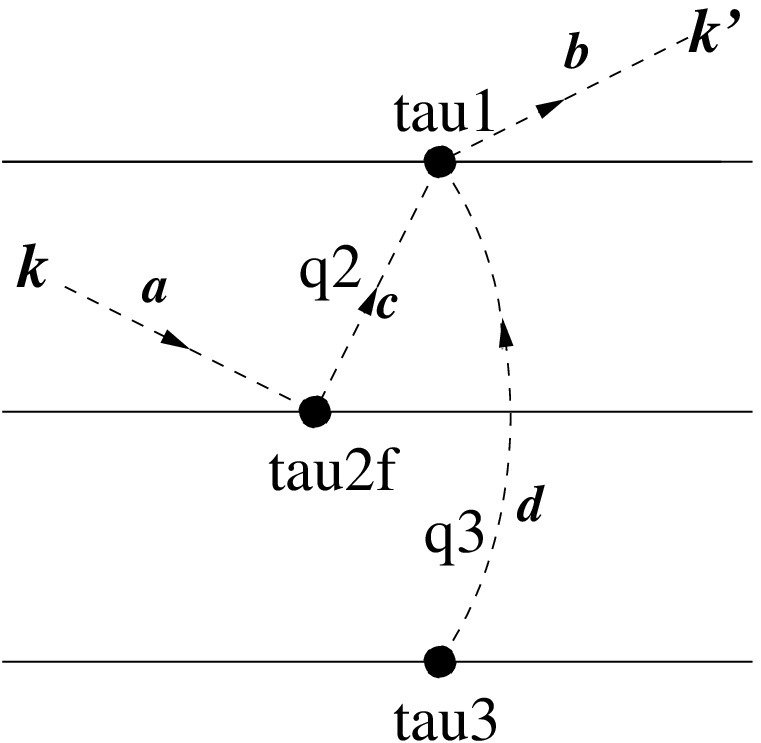}}
    
  \caption{Individually $\alpha$-dependent leading three-nucleon 
                contributions to pion-nucleus scattering.}
  \label{fig:3N-contralpha}
\end{figure*}

In this appendix, we explicitly show the independence of 
the sum of Feynman diagrams depicted in
Fig.~\ref{fig:3N-contralpha} on the parametrization of the pion field. 
For this, we closely follow the recipe outlined in 
\cite{Hanhart:2007mu}. 
We start with the most general expression for the
chiral matrix $U\left(\vec{\pi}\right)$ 
\begin{eqnarray} 
U\left(\vec{\pi}\right) & = & \exp\left(\frac{i}{f_\pi}\left(\vec{\tau}\cdot\vec{\pi}\right)g\left(\vec{\pi}^{\,2}/ f_\pi^2\right)\right).
\end{eqnarray}
where the function $g$ is an arbitrary regular function 
with the property $g\left(0\right)=1$. 
After expansion up to second order in the pion
fields it can be written as
\begin{eqnarray} 
g\left(\vec{\pi}^{\,2}/f_\pi^2\right) & = & 1+\left(\alpha+\frac{1}{6}\right)\frac{\vec{\pi}^{\,2}}{f_\pi^2}+ \ldots .
\end{eqnarray}
For $\alpha=-\frac{1}{6}$ this equals the expression for the chiral
matrix $U$ in the so-called exponential gauge,
$U=\exp\left(\frac{i}{f_\pi}\left(\vec{\tau}\cdot\vec{\pi}\right)\right)$.
In the $\sigma$-gauge we have
$U=\sqrt{1-\frac{\vec{\pi}^2}{f_\pi^2}}+\frac{i}{f_\pi}\vec{\tau}\cdot\vec{\pi}$,
which is reproduced up to terms of fourth order in the pion fields by
using $\alpha=0$ (for details see \cite{Hanhart:2007mu,Liebig:2009mth}).

Based on this ansatz for the chiral matrix $U$, it is easy to derive
the corresponding Feynman rules for the NN$\pi$, NN$2\pi$, 
NN$3\pi$, NN$4\pi$, and $4\pi$ vertices. The first two of the vertices,
turn out to be independent of $\alpha$:
\begin{eqnarray}
\label{eq:V_NNPi}
V_{NN\pi}  & = & -\frac{g_A}{2f_\pi}\left(\vec{\sigma}\cdot\vec{q}\right)\tau^a \\
\label{eq:V_NN2Pi}
V_{NN2\pi} & =&  \frac{1}{4f_\pi^2}v\cdot\left(q_2-q_1\right)\epsilon^{abc}\tau^c 
\end{eqnarray}
The third, fourth, and fifth one depend on $\alpha$. They 
are given by 
\begin{eqnarray}
\label{eq:V_NN3Pi}
V_{NN3\pi} & =&  -\frac{g_A}{4f_\pi^3}\vec{\sigma}\cr 
  & &\cdot\big\{\delta^{ab}\tau^c\left[\vec{q}_1+\vec{q}_2+4\alpha\left(\vec{q}_1+\vec{q}_2+\vec{q}_3\right)\right]\cr
 & &+\delta^{ac}\tau^b\left[\vec{q}_1+\vec{q}_3+4\alpha\left(\vec{q}_1+\vec{q}_2+\vec{q}_3\right)\right]\nonumber\\
 & &+\delta^{bc}\tau^a\left[\vec{q}_2+\vec{q}_3+4\alpha\left(\vec{q}_1+\vec{q}_2+\vec{q}_3\right)\right]\big\} \\
\label{eq:V_NN4Pi}
 V_{NN4\pi} & =&  \frac{1}{8f_\pi^4}\left(1+8\alpha\right)v \cr 
   & & \cdot\{\delta^{ab}\epsilon^{fcd}\left(q_4-q_3\right)+\delta^{bc}\epsilon^{fad}\left(q_4-q_1\right)\cr
 &  &+\delta^{cd}\epsilon^{fab}\left(q_2-q_1\right)+\delta^{ac}\epsilon^{fbd}\left(q_4-q_2\right)\nonumber\\
 &  &+\delta^{ad}\epsilon^{fbc}\left(q_3-q_2\right)+\delta^{bd}\epsilon^{fac}\left(q_3-q_1\right)\}\tau^f   \\
 \label{eq:V_4Pi}
V_{4\pi} & = & \frac{i}{f_\pi^2}\Bigg\{\cr 
 & & \left[\left(q_1+q_2\right)^2-m_\pi^2+2\alpha\sum_{i=1}^4\left(q_i^2-m_\pi^2\right)\right] \cr 
 & & \qquad \delta^{ab}\delta^{cd}\cr
 & &+\left[\left(q_1+q_3\right)^2-m_\pi^2+2\alpha\sum_{i=1}^4\left(q_i^2-m_\pi^2\right)\right] \cr 
  & & \qquad  \delta^{ac}\delta^{bd}\nonumber\\
 & &+\left[\left(q_1+q_4\right)^2-m_\pi^2+2\alpha\sum_{i=1}^4\left(q_i^2-m_\pi^2\right)\right] \cr 
 & & \qquad \delta^{ad}\delta^{bc}\Bigg\}
\end{eqnarray}
For these expressions, we assume that all momenta 
are going out of the vertices. 

Here, we are not interested in the expressions for the amplitudes, 
but only in the $\alpha$ dependence. Therefore, for simplicity,
we only keep the $\alpha$-dependent terms and show that
these vanish for the sum of the diagrams in 
Fig.~\ref{fig:3N-contralpha}. 

We identified seven $\alpha$-dependent diagrams 
in leading order, which we summarize in Fig.~\ref{fig:3N-contralpha}. 
Choosing $v=\left(1,0,0,0\right)$, it is easy to derive for the 
$\alpha$-dependent part of the 
threshold amplitude shown in Diagram (a) 
\begin{eqnarray}
\label{eq:3balphaa}
i\mathcal{M}^{\left( \mbox{\scriptsize \ref{fig:3N-contralpha}a} \right)} & = &  -\alpha\;\frac{g_A^2m_\pi}{4f_\pi^6}\frac{1}{\left(q_1^2-m_\pi^2\right)\left(q_3^2-m_\pi^2\right)} \cr 
& & \left(\vec{\sigma}_1\cdot\vec{q}_1\right)\left(\vec{\sigma}_3\cdot\vec{q}_3\right)\tau_2^f\nonumber\\
 &&\cr
& & \times\big\{\epsilon^{adf}\left(\tau_1^b\tau_3^d+\tau_1^d\tau_3^b\right)+\epsilon^{dbf}\left(\tau_1^a\tau_3^d+\tau_1^d\tau_3^a\right) \cr 
& & +2\epsilon^{abf}\left(\vec{\tau}_1\cdot\vec{\tau}_3\right)\}
\end{eqnarray}
Similarly, for Diagram (b)-(g) one finds 

\begin{eqnarray}
i\mathcal{M}^{\left(\mbox{\scriptsize \ref{fig:3N-contralpha}b}\right)} & = &   \alpha\;\frac{g_A^2m_\pi}{4f_\pi^6}\left(\vec{\sigma}_1\cdot\vec{q}_1\right)\left(\vec{\sigma}_3\cdot\vec{q}_3\right)\tau_2^f \cr 
& & \big\{\epsilon^{dbf}\left(\tau_1^a\tau_3^d+\tau_1^d\tau_3^a\right)+\epsilon^{abf}\left(\vec{\tau}_1\cdot\vec{\tau_3}\right)\big\}\nonumber\\
 &&\cr
& & \times\left[\frac{1}{\left(q_1^2-m_\pi^2\right)\left(q_3^2-m_\pi^2\right)}+\frac{1}{\left(q_1^2-m_\pi^2\right)\left(q_2^2-m_\pi^2\right)} 
    \right. \cr
& & \left.  +\frac{1}{\left(q_2^2-m_\pi^2\right)\left(q_3^2-m_\pi^2\right)}\right]
\end{eqnarray}

\begin{eqnarray}
i\mathcal{M}^{\left(\mbox{\scriptsize \ref{fig:3N-contralpha}c}\right)} & = &  \alpha\;\frac{g_A^2 m_\pi}{4f_\pi^6}\left(\vec{\sigma}_1\cdot\vec{q}_1\right)\left(\vec{\sigma}_3\cdot\vec{q}_3\right)\tau_2^f \cr 
& & \big\{\epsilon^{adf}\left(\tau_1^b\tau_3^d+\tau_1^d\tau_3^b\right)+\epsilon^{abf}\left(\vec{\tau}_1\cdot\vec{\tau_3}\right)\big\}\nonumber\\
 &&\cr
& &  \times\left[\frac{1}{\left(q_1^2-m_\pi^2\right)\left(q_3^2-m_\pi^2\right)}+\frac{1}{\left(q_1^2-m_\pi^2\right)\left(q_2^2-m_\pi^2\right)}
\right. \cr 
& & \left. +\frac{1}{\left(q_2^2-m_\pi^2\right)\left(q_3^2-m_\pi^2\right)}\right]
\end{eqnarray}
\begin{eqnarray}
i\mathcal{M}^{\left(\mbox{\scriptsize \ref{fig:3N-contralpha}d}\right)} & = &   -\alpha\;\frac{g_A^2m_\pi}{4f_\pi^6}\frac{1}{\left(q_1^2-m_\pi^2\right)\left(q_2^2-m_\pi^2\right)} \cr 
& & \left(\vec{\sigma}_1\cdot\vec{q}_1\right)\left(\vec{\sigma}_3\cdot\vec{q}_3\right)\tau_2^f\nonumber\\
 &&\cr
& & \times\big\{\epsilon^{dbf}\left(\tau_1^a\tau_3^d+\tau_1^d\tau_3^a\right)+\epsilon^{abf}\left(\vec{\tau}_1\cdot\vec{\tau}_3\right)\big\}
\end{eqnarray}
\begin{eqnarray}
i\mathcal{M}^{\left(\mbox{\scriptsize \ref{fig:3N-contralpha}e}\right)} & = &    -\alpha\;\frac{g_A^2m_\pi}{4f_\pi^6}\frac{1}{\left(q_1^2-m_\pi^2\right)\left(q_2^2-m_\pi^2\right)} \cr 
& & \left(\vec{\sigma}_1\cdot\vec{q}_1\right)\left(\vec{\sigma}_3\cdot\vec{q}_3\right)\tau_2^f\nonumber\\
 &&\cr
& & \times\big\{\epsilon^{adf}\left(\tau_1^b\tau_3^d+\tau_1^d\tau_3^b\right)+\epsilon^{abf}\left(\vec{\tau}_1\cdot\vec{\tau}_3\right)\big\}
\end{eqnarray}
\begin{eqnarray}
i\mathcal{M}^{\left(\mbox{\scriptsize \ref{fig:3N-contralpha}f}\right)} & = & -\alpha\;\frac{g_A^2m_\pi}{4f_\pi^6}\frac{1}{\left(q_2^2-m_\pi^2\right)\left(q_3^2-m_\pi^2\right)} \cr 
& & \left(\vec{\sigma}_1\cdot\vec{q}_1\right)\left(\vec{\sigma}_3\cdot\vec{q}_3\right)\tau_2^f\nonumber\\
 &&\cr
& & \times\big\{\epsilon^{dbf}\left(\tau_1^a\tau_3^d+\tau_1^d\tau_3^a\right)+\epsilon^{abf}\left(\vec{\tau}_1\cdot\vec{\tau}_3\right)\big\}
\end{eqnarray}
\begin{eqnarray}
\label{eq:3balphag}
i\mathcal{M}^{\left(\mbox{\scriptsize \ref{fig:3N-contralpha}g}\right)} & = &   -\alpha\;\frac{g_A^2m_\pi}{4f_\pi^6}\frac{1}{\left(q_2^2-m_\pi^2\right)\left(q_3^2-m_\pi^2\right)} \cr 
& & \left(\vec{\sigma}_1\cdot\vec{q}_1\right)\left(\vec{\sigma}_3\cdot\vec{q}_3\right)\tau_2^f\nonumber\\
 &&\cr
& & \times\{\epsilon^{adf}\left(\tau_1^b\tau_3^d+\tau_1^d\tau_3^b\right)+\epsilon^{abf}\left(\vec{\tau}_1\cdot\vec{\tau}_3\right)\}
\end{eqnarray}
Based on these expressions, it is easy to convince oneself 
that the sum of the expressions Eqs.~(\ref{eq:3balphaa}) to 
(\ref{eq:3balphag}) cancels. This implies that 
the sum of the diagrams is independent of $\alpha$. 

\section{Explicit expressions for the half-Coulombian three-nucleon diagrams }
\label{app:halfcoul3n}
In this appendix, we summarize the explicit expressions 
for the amplitudes of Fig.~\ref{fig:3N-contralpha}. 
The sum of these diagrams is independent of the parametrization 
of the pion field as discussed in Appendix~\ref{app:alphadep}. 
We refer to these amplitudes as half-Coulombian because 
for most of the diagrams of this group  one of the pion propagators  
is $1 \over {\vec q}^2$. 
They read 
\begin{eqnarray}
i\mathcal{M}^{\left(\mbox{\scriptsize \ref{fig:3N-contralpha}a}\right)} & = &   -\frac{g_A^2m_\pi}{32f_\pi^6}\frac{1}{\left(\vec{q}_1^{\,2}+m_\pi^2\right)\left(\vec{q}_3^{\,2}+m_\pi^2\right)}\nonumber \\[5pt] 
& & \times \left(\vec{\sigma}_1\cdot\vec{q}_1\right)\left(\vec{\sigma}_3\cdot\vec{q}_3\right)\nonumber\\[5pt]
& &\times\big\{\epsilon^{adf}\left(\tau_1^b\tau_3^d+\tau_1^d\tau_3^b\right)+\epsilon^{dbf}\left(\tau_1^a\tau_3^d+\tau_1^d\tau_3^a\right)
\cr 
& & +2\epsilon^{abf}\left(\vec{\tau}_1\cdot\vec{\tau}_3\right)\big\}\tau_2^f
\end{eqnarray}
\begin{eqnarray}
    i\mathcal{M}^{\left(\mbox{\scriptsize \ref{fig:3N-contralpha}b}\right)} & = &  \frac{g_A^2m_\pi}{8f_\pi^6}\;\frac{1}{\vec{q}_2^{\,2}\left(\vec{q}_1^{\,2}+m_\pi^2\right)\left(\vec{q}_3^{\,2}+m_\pi^2\right)}
    \nonumber\\[5pt]
    & & \times \left(\vec{\sigma}_1\cdot\vec{q}_1\right)\left(\vec{\sigma}_3\cdot\vec{q}_3\right)\nonumber\\[5pt]
    & & \times\big\{\epsilon^{dbf}\left(\vec{q}_1^{\,2}\;\tau_1^a\tau_3^d+\vec{q}_3^{\,2}\;\tau_1^d\tau_3^a\right)
    \cr 
    && +\left(\vec{q}_2^{\,2}+m_\pi^2\right)\epsilon^{abf}\left(\vec{\tau}_1\cdot\vec{\tau_3}\right)\big\}\tau_2^f
\end{eqnarray}
\begin{eqnarray}
i\mathcal{M}^{\left(\mbox{\scriptsize \ref{fig:3N-contralpha}c}\right)} & = &  \frac{g_A^2m_\pi}{8f_\pi^6}\;\frac{1}{\vec{q}_2^{\,2}\left(\vec{q}_1^{\,2}+m_\pi^2\right)\left(\vec{q}_3^{\,2}+m_\pi^2\right)} \nonumber \\[5pt]
& & \times \left(\vec{\sigma}_1\cdot\vec{q}_1\right)\left(\vec{\sigma}_3\cdot\vec{q}_3\right)\nonumber\\[5pt]
& & \times\big\{\epsilon^{adf}\left(\vec{q}_1^{\,2}\;\tau_1^b\tau_3^d+\vec{q}_3^{\,2}\;\tau_1^d\tau_3^b\right) \cr 
& & +\left(\vec{q}_2^{\,2}+m_\pi^2\right)\epsilon^{abf}\left(\vec{\tau}_1\cdot\vec{\tau_3}\right)\big\}\tau_2^f
\end{eqnarray}
\begin{eqnarray}
i\mathcal{M}^{\left(\mbox{\scriptsize \ref{fig:3N-contralpha}d}\right)} & = &  \frac{g_A^2m_\pi}{16f_\pi^6}\;\frac{1}{\vec{q}_2^{\,2}\left(\vec{q}_1^{\,2}+m_\pi^2\right)}\left(\vec{\sigma}_1\cdot\vec{q}_1\right)\nonumber\\
 &&\cr
& & \times\vec{\sigma}_3\cdot\{\epsilon^{dbf}\left(\vec{q}_1\;\tau_1^a\tau_3^d-\vec{q}_3\;\tau_1^d\tau_3^a\right) \cr 
& & +\vec{q}_2\;\epsilon^{abf}\left(\vec{\tau}_1\cdot\vec{\tau}_3\right)\}\tau_2^f
\end{eqnarray}
\begin{eqnarray}
i\mathcal{M}^{\left(\mbox{\scriptsize \ref{fig:3N-contralpha}e}\right)} & = & 
\frac{g_A^2m_\pi}{16f_\pi^6}\;\frac{1}{\vec{q}_2^{\,2}\left(\vec{q}_1^{\,2}+m_\pi^2\right)}\left(\vec{\sigma}_1\cdot\vec{q}_1\right)\nonumber\\
 &&\cr
& &  \times\vec{\sigma}_3\cdot\{\epsilon^{adf}\left(\vec{q}_1\;\tau_1^b\tau_3^d-\vec{q}_3\;\tau_1^d\tau_3^b\right) \cr 
& & +\vec{q}_2\;\epsilon^{abf}\left(\vec{\tau}_1\cdot\vec{\tau}_3\right)\}\tau_2^f
\end{eqnarray}
\begin{eqnarray}
i\mathcal{M}^{\left(\mbox{\scriptsize \ref{fig:3N-contralpha}f}\right)} & = & 
 \frac{g_A^2m_\pi}{16f_\pi^6}\;\frac{1}{\vec{q}_2^{\,2}\left(\vec{q}_3^{\,2}+m_\pi^2\right)}\left(\vec{\sigma}_3\cdot\vec{q}_3\right)\nonumber\\
 &&\cr
& & \times\vec{\sigma}_1\cdot\{\epsilon^{dbf}\left(-\vec{q}_1\;\tau_1^a\tau_3^d+\vec{q}_3\;\tau_1^d\tau_3^a\right) \cr 
& & +\vec{q}_2\;\epsilon^{abf}\left(\vec{\tau}_1\cdot\vec{\tau}_3\right)\}\tau_2^f
\end{eqnarray}
\begin{eqnarray}
i\mathcal{M}^{\left(\mbox{\scriptsize \ref{fig:3N-contralpha}g}\right)} & = &  \frac{g_A^2m_\pi}{16f_\pi^6}\;\frac{1}{\vec{q}_2^{\,2}\left(\vec{q}_3^{\,2}+m_\pi^2\right)}\left(\vec{\sigma}_3\cdot\vec{q}_3\right)\nonumber\\
 &&\cr
& & \times\vec{\sigma}_1\cdot\{\epsilon^{adf}\left(-\vec{q}_1\;\tau_1^b\tau_3^d+\vec{q}_3\;\tau_1^d\tau_3^b\right) \cr 
& & +\vec{q}_2\;\epsilon^{abf}\left(\vec{\tau}_1\cdot\vec{\tau}_3\right)\}\tau_2^f
\end{eqnarray}
The definition of the momenta can again be read off from the 
figures. For these expressions, we have assumed $\alpha=0$,
which corresponds to the $\sigma$-gauge.

\section{PW decomposition of the two-nucleon operators}
\label{app:2Npw}

In this appendix, we briefly summarize the PW decomposed 
expressions for the pertinent two-nucleon operators. 

The PW states for the NN system read 
\begin{equation}
| \alpha \rangle \equiv | p \, (l\, s) \, j m \rangle
\end{equation}
where $p$ is the magnitude of the NN relative momentum, 
$l$ and $s$ are the corresponding orbital angular momentum 
and NN spin and $j$,$m$ are the total angular momentum and its projection. 
We abbreviate this set of quantum numbers by $\alpha$. 
Here we only consider $\pi^-$-$^2$H scattering. Therefore, 
the two nucleons are in an $|t=0 \, m_{t}=0\rangle$ isospin state
and the cartesian components $a,b=1...3$ of the pion are 
$(1/\sqrt{2},-i/\sqrt{2},0)$. The pertinent isospin matrix elements are 
\begin{eqnarray}
& & \langle t=0 m_{t}=0 |  \delta^{ab}\left(\vec{\tau}_1\cdot\vec{\tau}_2\right)   | t=0 m_{t}=0 \rangle = -3 \nonumber \\[5pt]
& & \langle t=0 m_{t}=0 |  \tau_1^b\tau_2^a+\tau_1^a\tau_2^b  | t=0 m_{t}=0 \rangle = -2 \cr 
& & 
\end{eqnarray}

The spin-orbital part can be expressed in terms of 
integrals 
\begin{eqnarray}
g^n_{k}(p'p) & = & 2\pi (-)^k \sqrt{2k+1} \, \int_{-1}^1 dx \, 
P_{k}(x) {1 \over {q}^n} \nonumber \\[5pt]
\tilde g^f_{k}(p'p) & = & 2\pi (-)^k \sqrt{2k+1} \, \int_{-1}^1 dx\,  P_{k}(x) {1 \over {q}^f} {q^2 \over (q^2 + m_{\pi}^2)^2}  \nonumber \\[5pt]
\end{eqnarray} 
where $q=\sqrt{p^2+{p'}^2-2pp'x} $ is an internal momentum transfer
and $P_{k}(x)$ is the degree $k$ Legendre polynomial. 

With these definitions the PW decomposition 
of the amplitudes Eqs.~(\ref{eq:m2a}), (\ref{eq:m2bc}) and 
(\ref{eq:m2triple}) read 
\begin{eqnarray}
\langle \alpha' |  \mathcal{M}^{\left(\mbox{\scriptsize \ref{fig:2N-ops}a}\right)}| \alpha  \rangle & = & - {m_{\pi}^2 \over 8 \pi^3 f_{\pi}^4} 
\delta_{ll'}\delta_{ss'}\delta_{jj'}\delta_{mm'} \   {(-)^l \over \sqrt{2l+1}} g^2_{l}(p'p) \cr & & 
\end{eqnarray}
\begin{eqnarray}
& & \langle \alpha' |  \mathcal{M}^{\left(\mbox{\scriptsize \ref{fig:2N-ops}bc}\right)}| \alpha  \rangle  = 
 {g_{A}^2 m_{\pi}^2 \over  32 \pi^3  f_{\pi}^4}
  \sum_{f\, k} 
\sum_{\mu_{1}+\mu_{2}=f} (2f+1)^{2} 
 \left\{ \begin{array}{ccc} 
 1 & 1  & 0 \cr 1 & 1 & 0 \cr f & f & 0 
 \end{array}   
 \right\}  \nonumber \\[5pt]
 & & \quad (11f,000) \sqrt{(2f+1)! \over (2\mu_{1})! (2\mu_{2})!} \
 {p'}^{\mu_{1}} (-p)^{\mu_{2}} \ \tilde g_{k}^f (p'p) \nonumber \\[5pt] 
 & & \quad  (2k+1) \left\{ \begin{array}{ccc} 
 k & k  & 0 \cr \mu_{1} & \mu_{2} & f \cr l' & l & f 
 \end{array}   
 \right\} \ (k \mu_{1} l',000) \ (k \mu_{2} l,000) \ (-)^l \nonumber \\[5pt] 
& & \quad 18 \sqrt{(2s+1)(2s'+1)(2j+1)}
 \left\{ \begin{array}{ccc} 
 s' & s  & f \\[2pt] {1\over 2} & {1\over 2} & 1 \\[2pt] {1\over 2} & {1\over 2} & 1  
 \end{array}   
  \right\}  
 \left\{ \begin{array}{ccc} 
 l' & l  & f \cr s' & s & f \cr j' & j & 0  
 \end{array}   
  \right\} \nonumber \\[5pt]
  & & 
\end{eqnarray}
\begin{eqnarray}
& & \langle \alpha' |  \mathcal{M}^{\left(\mbox{\scriptsize \ref{fig:2N-triple}}\right)}| \alpha  \rangle \cr 
& & =   { 1 \over 2 } 
\left( { m_{\pi} \over 4 \pi  f_{\pi}^2 }\right)^3 
\delta_{ll'}\delta_{ss'}\delta_{jj'}\delta_{mm'} \   {(-)^l \over \sqrt{2l+1}} g^1_{l}(p'p) \cr & &  
\end{eqnarray}
for $\pi^-$-$^2$H scattering in terms of 9j-coefficients 
and Clebsch-Gordan coefficients 
$(j_{1}j_{2}j_{3},m_{1}m_{2}m_{3})$. Here, we normalize 
the momentum eigenstates such that the 
expectation values read 
\begin{eqnarray}
& & \langle \mathcal{M} \rangle \cr 
& & =  \sum_{lsj} \sum_{l's'j'} \int_{0}^\infty dp \, p^2 \, dp' \,  {p'}^2 
   \, \psi^*_{l's'j'}(p') \langle \alpha' | \mathcal{M} | \alpha \rangle 
   \psi_{lsj}(p) \cr & & 
\end{eqnarray}
for wave functions normalized to 
\begin{eqnarray}
\sum_{lsj} \int_{0}^\infty dp \, p^2 \,
   \, | \psi_{lsj}(p) |^2  & = & 1 \ . 
\end{eqnarray}
\bibliographystyle{lit-database/epj}
\bibliography{lit-database/lit}

\end{document}